\documentclass[fleqn,usenatbib]{mnras}

\usepackage{newtxtext,newtxmath}

\usepackage[T1]{fontenc}

\DeclareRobustCommand{\VAN}[3]{#2}
\let\VANthebibliography\thebibliography
\def\thebibliography{\DeclareRobustCommand{\VAN}[3]{##3}\VANthebibliography}

\usepackage{graphicx}	%
\usepackage{amsmath}	%

\usepackage{multirow}   

\usepackage{ulem}

\defcitealias{Farias2017}{Paper~I}
\defcitealias{Farias2019}{Paper~II}
\newcommand\paperI{\citetalias{Farias2017}}
\newcommand\paperII{\citetalias{Farias2019}}

\newcommand\Msun{\ensuremath{M_{\odot}}}
\newcommand\gpcm{\ensuremath{{\rm g\:cm}^{-2}}}
\newcommand\kms{\ensuremath{\rm km\:s^{-1}}}

\newcommand\mcl{\ensuremath{M_{\rm cl}}}
\newcommand\Rcl{\ensuremath{R_{\rm cl}}}

\newcommand\krho{\ensuremath{k_{\rho}}}
\newcommand\sigmacl{\ensuremath{\sigma_{\rm cl}}}
\newcommand\Sigmacl{\ensuremath{\Sigma_{\rm cloud}}}
\newcommand\tcross{\ensuremath{t_{\rm cr}}}
\newcommand\trelax{\ensuremath{t_{\rm relax}}}
\newcommand\tff{\ensuremath{t_{\rm ff}}}
\newcommand\fbound{\ensuremath{f_{\rm bound}}}

\newcommand\rhb{\ensuremath{r_{\rm h,b}}}
\newcommand\sfeff{\ensuremath{\epsilon_{\rm ff}}}
\newcommand\sigmas{\ensuremath{\sigma_{s}}}
\newcommand\tsf{\ensuremath{t_*}}
\newcommand\bmin{\ensuremath{b_{\rm min}}}
\newcommand\mt{\ensuremath{m_{\rm t}}}

\newcommand\Nint{\ensuremath{N_{\rm int}}}
\newcommand\Qb{\ensuremath{Q_{\rm b}}}

\newcommand\smallCloudL{\texttt{m300L}}
\newcommand\mediumCloudL{\texttt{m3000L}} 
\newcommand\largeCloudL{\texttt{m30000L}}
\newcommand\smallCloudH{\texttt{m300H}}
\newcommand\mediumCloudH{\texttt{m3000H}}
\newcommand\largeCloudH{\texttt{m30000H}}

\newcommand\smallClouds{\texttt{m300}}
\newcommand\mediumClouds{\texttt{m3000}}
\newcommand\largeClouds{\texttt{m30000}}

\title[Across the mass spectrum]{Star Cluster Formation from Turbulent Clumps. III.\\
        Across the mass spectrum}

\author[J. P. Farias and J. C. Tan]{
Juan P. Farias$^{1,2}$\thanks{E-mail: juan.farias@austin.utexas.edu}
and
Jonathan C. Tan$^{1,3}$
\\
$^{1}$ Department of Space, Earth \& Environment, Chalmers University of Technology, Gothenburg SE-41293, Sweden \\
$^{2}$ Department of Astronomy, Univerisity of Texas at Austin, TX 78712, USA\\
$^{3}$ Department of Astronomy, University of Virginia, Charlottesville, VA 22904, USA 
}

\date{Accepted XXX. Received YYY; in original form ZZZ}

\pubyear{2023}

\begin{document}
\label{firstpage}
\pagerange{\pageref{firstpage}--\pageref{lastpage}}
\maketitle

\begin{abstract}
We study the formation and early evolution of star clusters that have a wide range
of masses and background cloud mass surface densities, $\Sigma_{\rm cloud}$, which
help set the initial sizes, densities, and velocity dispersions of the natal gas
clumps.  Initial clump masses of 300, 3,000 and 30,000~$M_\odot$ are considered,
from which star clusters are born with an assumed 50\% overall star formation
efficiency and with 50\% primordial binarity. This formation is gradual, i.e., with
a range of star formation efficiencies per free-fall time from 1\% to 100\%, so
that the formation time can range from 0.7~Myr for low-mass, high-$\Sigma_{\rm
cloud}$ clumps to $\sim30$~Myr for high-mass, low-$\Sigma_{\rm cloud}$ clumps.
Within this framework of the Turbulent Clump model, for a given $\Sigma_{\rm
cloud}$, clumps of higher mass are of lower initial volume density, but their
dynamical evolution leads to higher bound fractions and causes them to form much
higher density cluster cores and maintain these densities for longer periods. This
results in systematic differences in the evolution of binary properties, degrees of
mass segregation and rates of creation of dynamically ejected runaways. We discuss
the implications of these results for observed star clusters and stellar
populations.
\end{abstract}

\begin{keywords}
methods: numerical -- galaxies: star clusters
\end{keywords}

\section{Introduction}

Most stars appear to form in clusters (or at least initially clustered
associations) inside molecular clouds \citep[e.g.,][]{Gutermuth2009}. A wide range
of scales is involved, including the broad distribution of cluster masses that make
up the initial cluster mass function (ICMF). For masses $\gtrsim100\:M_\odot$ and
up to at least $\sim10^5\:M_\odot$, the ICMF appears to follow a power law of the
form $d N / d {\rm log} M \propto M^{-1}$ \citep[e.g.,][]{Lada2003,Dowell2008}, so
that there is an equal mass contributed by clusters in each decade of the mass
spectrum. Thus considering a broad range of cluster masses is needed when
understanding the origin of galactic stellar populations.

Star cluster formation itself is a very complex process that involves the interplay
of many physical processes, including fragmentation of self-gravitating, turbulent,
magnetised molecular clouds, protostellar outflow feedback from accreting stars
\citep[e.g.,][]{Nakamura2007,Nakamura2014,Cunningham2011,Hansen2012,Federrath2014,Geen2015},
other feedback processes from already
formed, especially massive, stars
\citep[e.g.,][]{Peters2010,Peters2011,Rogers2013,Dale2015}, and dynamical
evolution of the stellar population, including dynamical ejection of runaway stars
\citep[e.g.,][]{Bannerjee2012,Oh2016,Gavagnin2017}. All these processes have their own spatial regimes
and timescales over which they are important.

It is not currently possible to include all the above processes in a unified
simulation to model star cluster formation. Our approach, developed in a series of
papers of which this is the third, explores star cluster formation within the
paradigm of the Turbulent Core/Clump Model \citep{mt03} with approximate
implementation of the birth of stars via their gradual introduction into
simulations that follow the $N$-body dynamical evolution of the system. The overall
goal is to explore how the stellar population, including realistic binary
properties, is processed dynamically during the formation phase of a star cluster,
and how this processing may be affected by model parameters. There are two basic
parameters describing the initial star-forming clumps: the clump mass, $M_{\rm
cl}$, and the mass surface density of the surrounding ``cloud'' environment,
$\Sigma_{\rm cloud}$, which sets the bounding pressure of the clumps and thus their
radii, $R_{\rm cl}$. High $\Sigma_{\rm cloud}$ environments have high pressures,
i.e., due to the self-gravity of the cloud, which means that clumps of a given mass
are denser in such environments. The formation phase of the cluster also involves
two main parameters: the star formation efficiency per free-fall time, $\sfeff$,
and the overall star formation efficiency of the clump, $\epsilon$.

In \cite{Farias2017} (hereafter \paperI), we first explored an extreme version of
this scenario in which the star clusters are formed instantaneously from their
parent clumps. While instantaneous formation appears to be an unrealistic case, we
note that this has been the standard practice in almost all such similar studies so
far \citep[with the notable exception of][]{Proszkow2009}. In our second work of
this series, \citet{Farias2019} (hereafter \paperII), we implemented gradual
formation of stars, which enabled us to explore a wide range of formation
timescales (achieved via a range of values of $\sfeff$ and a fixed, fiducial value
of $\epsilon=0.5$). We showed that such timescales strongly influence the dynamical
evolution of the clusters in both the embedded phase (i.e., when gas is still
present) and during the subsequent gas-free phase, including the rate and amount of
expansion, the fraction of stars that remain bound, the frequency of ejection
events, the establishment of age-radius gradients and the degree of processing of
binaries.

However, in these previous papers we limited the studies to a fixed parent clump
mass of 3,000\,\Msun.  It is not immediately obvious how our results would scale
with mass (at a fixed $\Sigmacl$), since there are several coupled processes at
play with various different timescales and dependencies.  Thus, our goal in this
paper is to present a series of $N$-body simulations that explore different clump
masses using the same framework as in \paperII. These simulations will help us to
elucidate how the various dynamical processes, described above, combine to control
the dynamical evolution of clusters across the mass spectrum.

\section{Theoretical Background}

\subsection{Background gas model}
\label{sec:gasmodel}

We perform star cluster formation simulations following the methods presented in
\paperI\ and \paperII. In these models star clusters are assumed to be forming from
gravitationally bound, initially starless gas clumps within
giant molecular clouds (GMCs), partially supported by magnetic fields and
turbulence. The structure of the parent clumps is described
following the turbulent core/clump model of \cite{mt03}, i.e., they are polytropic
spheres in virial and pressure equilibrium with their surroundings. The density
profile of such clumps is modeled as:
\begin{eqnarray}
        \label{eq:dens}
        \rho_{\rm cl} (r) &=& \rho_{\rm s,cl} \left(\frac{r}{\Rcl} \right)^{-\krho},
\end{eqnarray}
and the velocity dispersion profile as: 
\begin{eqnarray}
\sigma_{\rm cl}(r) &=& \sigmas \left( \frac{r}{\Rcl} \right)^{(2-\krho)/2},
\label{eq:sigmar}
\end{eqnarray}
where $\rho_{\rm s,cl}$ and $\sigmas$ are the density and velocity dispersion at
the surface of the clump, respectively, $\Rcl$ is the clump radius, i.e., where its
boundary is located, and we adopt $\krho=1.5$ as the fiducial power law of the
density distribution \citep[e.g.,][]{Butler2012}. One important feature to note is
that the velocity dispersion increases with radius \citep[see][]{mt03}, which is a
general feature of interstellar turbulence. We refer the reader to \paperI\ and
\paperII, where we discuss the dynamical implications of such a characteristic for
the formation and evolution of our model star clusters.

Using our fiducial parameters for the structure of the parent clump, the
characteristic radius and velocity dispersion at the clump surface are controlled
by the surrounding cloud's mass surface density, $\Sigmacl$, and are given by:
\begin{eqnarray}
\label{eq:rcl}
\Rcl & = 0.365
         \left( \frac{M_{\rm cl}}{3000 \Msun}  \right)^{1/2} 
         \left( \frac{\Sigmacl}{1 {\rm ~g~cm^{-2}}} \right)^{-1/2}\:{\rm pc},
\end{eqnarray}
and
\begin{eqnarray}
\sigmas &= 3.04 \left(\frac{M_{\rm cl}}{3000~\Msun} \right)^{1/4}
 \left( \frac{\Sigmacl }{1~{\rm g~cm^{-2}}} \right)^{1/4}\:{\rm km~s^{-1}}.
\label{eq:sigmas} 
\end{eqnarray}
Following our previous works, we model clumps in two different cloud environments:
the high-$\Sigma$ case with $\Sigmacl=1.0\,\gpcm$ and the low-$\Sigma$ case with
$\Sigmacl=0.1\,\gpcm$. Such a range is likely to be relevant for a large
fraction of the star-forming systems of our Galaxy \citep{Tan2014}: for
example, large portions of the samples of the IRDC clumps of \cite{Butler2012},
of the high-mass star-forming clumps of \cite{Mueller2002}, and the massive
clumps of \cite{Ginsburg2012} are in or near this region of parameter space.
Then, given $\Rcl$, defined by
$M_{\rm cl}$ and $\Sigmacl$, the density at the surface of the clump is:
\begin{eqnarray}
        \rho_{\rm s,cl}&=& \frac{(3-\krho)M_{\rm cl}}{4\pi\Rcl^3}.
\end{eqnarray}

In \paperII\ we introduced models of gradual formation of star clusters, i.e., in
which natal gas is still present while stars are being formed. The influence of the
natal gas in the evolution of the forming star cluster is modeled as a
time-dependent background potential derived from Eq.~\ref{eq:dens}, i.e.:
\begin{eqnarray}
        \label{eq:pot}
        \Phi_{\rm gas} (r,t) &=&  \left\{
        \begin{array}{lr}
                \dfrac{ GM_{\rm cl}(t) }{(2-\krho) \Rcl } \left[ \left(
                \dfrac{r}{\Rcl} \right)^{2-\krho}\hspace{-20pt} - 3 + \krho \right] & (r\leq \Rcl)\\
                &\\
                -\dfrac{GM_{\rm cl}(t)}{r} & (r>\Rcl)
        \end{array}
        \right.,
\end{eqnarray}
where $G$ is the gravitational constant and $M_{\rm cl}(t)$ the time-dependent
clump gas mass. Note the radius of the clump is truncated at $R_{\rm cl}$ and no
additional gas mass is modeled beyond this radius, i.e., no further contributions
to the potential are made from the surrounding cloud.

We keep our previous assumption of a constant star formation rate (SFR) defined
using the {\it initial} parameters of the clump, i.e.,
\begin{eqnarray}
        \label{eq:mdot}
        \dot{M}_* &=& \frac{\sfeff M_{\rm cl,0} }{t_{\rm ff,0}},
\end{eqnarray}
where the initial free fall time of the clump, $t_{\rm ff,0}$, is also defined by
$\Sigmacl$ and $M_{\rm cl,0}$. Using the fiducial clump parameters, it is given by:
\begin{eqnarray}
        \label{eq:tff0}
        t_{\rm ff,0} &=& 0.069 \left( \frac{M_{\rm cl, 0}}{3000\,{\Msun}}
        \right)^{1/4}\left( \frac{\Sigmacl} {1\,{\rm g\,cm^{-2}}} \right)^{-3/4}
        \, {\rm Myr}.
\end{eqnarray}

We assume that there is a local star formation efficiency, $\epsilon$, i.e.,
defined as the ratio between the stellar mass formed and the total mass required to
form such a stellar mass. The fiducial value of $\epsilon=0.5$ with such a value
being typical of expectations of local star formation efficiency from individual
pre-stellar cores due to protostellar outflow feedback \citep[e.g.,][]{Matzner2000,Tanaka2017}.
For simplicity, the gas that does not make it into a star is assumed to be
instantaneously lost from the clump.  We assume star formation proceeds in this way
until all the gas from the clump is exhausted.  Therefore, the time-evolution of
the global gaseous mass of the clump is given by:
\begin{eqnarray}
        \label{eq:mgas}
        M_{\rm cl}(t) &=& \left\{  
        \begin{array}{lr}
                M_{\rm cl,0} - \dfrac{ \dot{M}_* }{ \epsilon } t & (t \leq \tsf) \\
                & \\
                0 & (t > \tsf),\\
        \end{array} 
        \right.
\end{eqnarray}
where $\tsf$ is the time at which gas is exhausted. Since we assume a constant SFR,
the formation time is given by :
\begin{eqnarray}
        \tsf = \dfrac{\epsilon}{\sfeff}  t_{\rm ff,0} 
        &\propto \dfrac{\epsilon}{\sfeff} \left( \dfrac{\mcl}{\Sigmacl^3}
        \right)^{1/4}
        \label{eq:tstar}
\end{eqnarray}

\subsection{Scaling of clump properties with mass}
\label{sec:masstheory}

\begin{figure}
        \centering
        \includegraphics[width=\columnwidth]{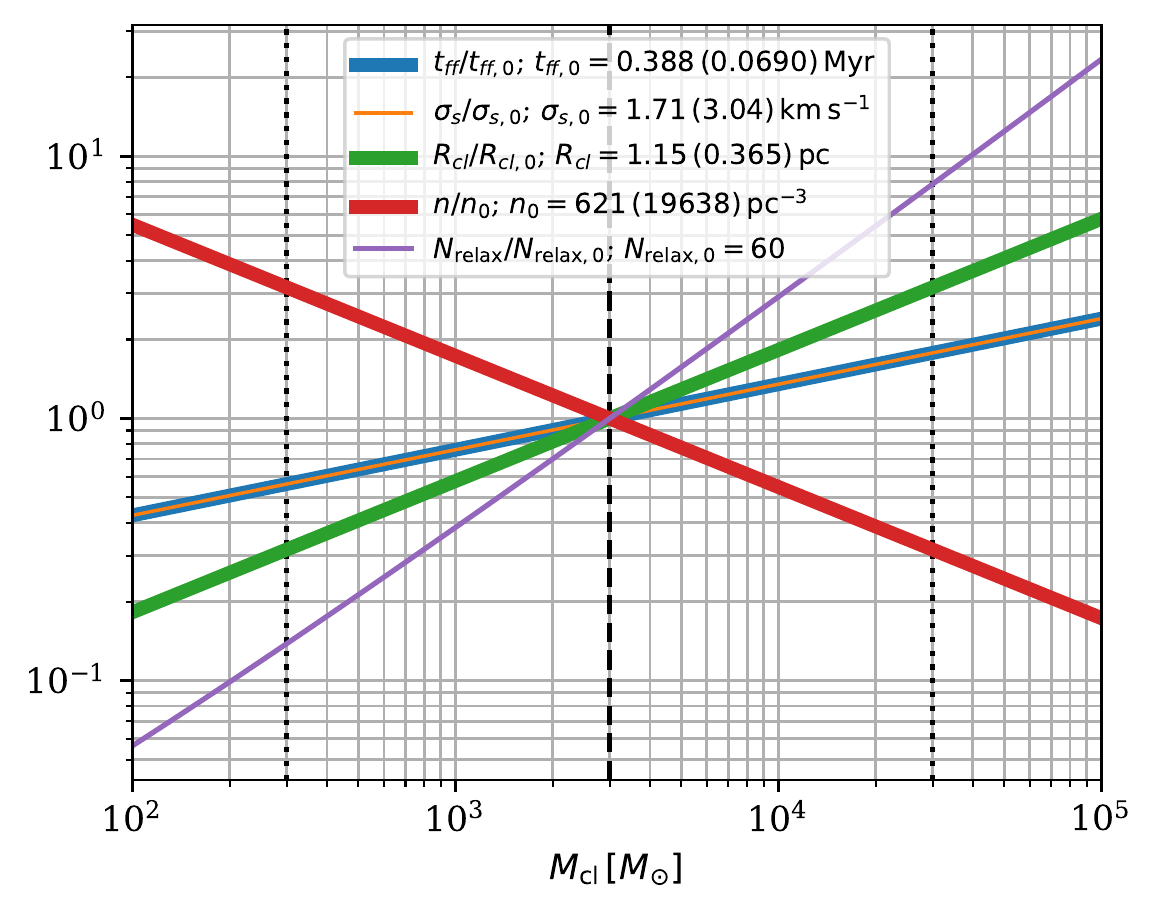}
        \caption{Variation of global clump parameters with mass, normalised with
                respect to clumps with $\mcl=3,000\,\Msun$. Legend shows fiducial
                values for this case with $\Sigmacl=0.1\,\gpcm$ and 
                $\Sigmacl=1\,\gpcm$ in parenthesis. Dotted vertical
                lines show the masses of the new models introduced in this paper
                with 300 and 30,000$\,\Msun$.  }
        \label{fig:MDtheory}
\end{figure}

In this work, we explore how the formation and early evolution of star clusters
depends on the initial mass of the clump, $\mcl$.  Within the context of the
Turbulent Clump Model, several important parameters and features of the clumps and
clusters vary with clump mass, which we overview in Figure~\ref{fig:MDtheory}. In
particular, this figure shows how several clump properties vary with \mcl, while
keeping the bounding cloud mass surface density, $\Sigmacl$, constant. Values are
normalised relative to the $\mcl=3,000\,\Msun$ case (and numerical values shown in
the legend of the figure apply for the model with $\Sigmacl=0.1\,\gpcm$).
Naturally, increasing the mass of the clump requires a larger clump radius to keep
mass surface density constant \citep[there is a one-to-one relation of clump mass
surface density and surrounding cloud mass surface density,][]{mt03}, with
$\Rcl\propto\mcl^{1/2}$ (thick green line).  From Eq.~\ref{eq:rcl}, we see that the
one dimensional velocity dispersion at the surface of the clump, $\sigma_{s}$,
scales with mass as $\sigma_{s}\propto \mcl^{1/4}$.  Thus the crossing time varies
as $\tcross\propto \Rcl/\sigma_s \propto \mcl^{1/4}$. The same scaling applies to
the free-fall time $\tff\propto\mcl^{1/4}$ (see Eq.~\ref{eq:tff0}). 

For fixed star formation efficiency, the number density of stars that would be
initially contained in the volume of the clump scales as $n_* \propto \mcl / \Rcl^3
\propto \mcl^{-1/2}$.  Defining $N_{\rm relax}$ as the relaxation time, $\trelax$,
divided by the crossing time, it is known that $N_{\rm relax}\propto \mcl/\ln \mcl$
\citep[see][]{BT1987}.  A star cluster needs to evolve for about $N_{\rm relax}$
crossing times for the individual stars to lose information of their
initial orbits and reach a near dynamical equilibrium state. Thus a more
massive cluster takes longer (both in terms of number of crossing times and in
terms of absolute time) to reach an equilibrium configuration. This is relevant
since the initial state of the stars that are formed from the Turbulent Clump Model
are not in the equilibrium configuration of a gas-free stellar cluster.  Finally,
for fixed values of $\epsilon$ and $\sfeff$, the formation time scales as
$\tsf\propto\tff \propto \mcl^{1/4}$.

\section{Methods}\label{methods}
\begin{table*}
\centering
\caption{Simulations parameters}
\label{tab:masspar}
\begin{tabular}{r|cccccccc}
Set Name & \sfeff& \Sigmacl & \mcl & $\langle N_* \rangle$ & $\tsf$ & $\tff$ & 
\Rcl & $\sigma_{s}$\\ 
& & [\gpcm] & [\Msun]  &   & [Myr] & [Myr] & [pc] & [$\kms$] \\ \hline
\multirow{ 5 }{*}{ \smallCloudL } 
  & 0.01  & 0.1 & 300    & 400    & 10.91 & 0.22  & 0.36 & 0.96 \\
  & 0.03  & 0.1 & 300    & 400    & 3.64  & 0.22  & 0.36 & 0.96 \\
  & 0.1   & 0.1 & 300    & 400    & 1.09  & 0.22  & 0.36 & 0.96 \\
  & 0.3   & 0.1 & 300    & 400    & 0.36  & 0.22  & 0.36 & 0.96 \\
  & 1.0   & 0.1 & 300    & 400    & 0.11  & 0.22  & 0.36 & 0.96 \\\cline{2-9}
\multirow{5}{*}{\mediumCloudL}
  & 0.01 & 0.1 & 3,000  & 4,000  & 19.40 & 0.39  & 1.15 & 1.71 \\
  & 0.03 & 0.1 & 3,000  & 4,000  & 6.47  & 0.39  & 1.15 & 1.71 \\
  & 0.1  & 0.1 & 3,000  & 4,000  & 1.94  & 0.39  & 1.15 & 1.71 \\
  & 0.3  & 0.1 & 3,000  & 4,000  & 0.65  & 0.39  & 1.15 & 1.71 \\
  & 1.0  & 0.1 & 3,000  & 4,000  & 0.19  & 0.39  & 1.15 & 1.71 \\\cline{2-9}
\multirow{5}{*}{\largeCloudL} 
  &  0.01 & 0.1 & 30,000 & 40,000 & 34.50 & 0.69  & 3.65 & 3.04 \\
  &  0.03 & 0.1 & 30,000 & 40,000 & 11.50 & 0.69  & 3.65 & 3.04 \\
  &  0.1  & 0.1 & 30,000 & 40,000 & 3.45 & 0.69  & 3.65 & 3.04 \\
  &  0.3  & 0.1 & 30,000 & 40,000 & 1.15 & 0.69  & 3.65 & 3.04 \\
  &  1.0  & 0.1 & 30,000 & 40,000 & 0.34 & 0.69  & 3.65 & 3.04 \\
\hline
\multirow{ 5 }{*}{\smallCloudH} 
 &  0.01   & 1.0 & 300    & 400    & 1.94  & 0.039  & 0.115 & 1.71 \\
 &  0.03   & 1.0 & 300    & 400    & 0.65  & 0.039  & 0.115 & 1.71 \\
 &  0.1    & 1.0 & 300    & 400    & 0.19  & 0.039  & 0.115 & 1.71 \\
 &  0.3    & 1.0 & 300    & 400    & 0.06  & 0.039  & 0.115 & 1.71 \\
 &  1.0    & 1.0 & 300    & 400    & 0.02  & 0.039  & 0.115 & 1.71 \\\cline{2-9}
\multirow{5}{*}{\mediumCloudH} 
 &  0.01 & 1.0 & 3,000  & 4,000  & 3.45  & 0.069  & 0.365 & 3.04 \\
 &  0.03 & 1.0 & 3,000  & 4,000  & 1.15  & 0.069  & 0.365 & 3.04 \\
 &  0.1  & 1.0 & 3,000  & 4,000  & 0.35  & 0.069  & 0.365 & 3.04 \\
 &  0.3  & 1.0 & 3,000  & 4,000  & 0.12  & 0.069  & 0.365 & 3.04 \\
 &  1.0  & 1.0 & 3,000  & 4,000  & 0.03  & 0.069  & 0.365 & 3.04 \\\cline{2-9}
\multirow{5}{*}{\largeCloudL} 
 &  0.01 & 1.0 & 30,000 & 40,000 & 6.14  & 0.123  & 1.154 & 5.41 \\
 &  0.03 & 1.0 & 30,000 & 40,000 & 2.05  & 0.123  & 1.154 & 5.41 \\
 &  0.1  & 1.0 & 30,000 & 40,000 & 0.61  & 0.123  & 1.154 & 5.41 \\
 &  0.3  & 1.0 & 30,000 & 40,000 & 0.20  & 0.123  & 1.154 & 5.41 \\
 &  1.0  & 1.0 & 30,000 & 40,000 & 0.06  & 0.123  & 1.154 & 5.41 \\
\end{tabular}
\end{table*}
\subsection{Gradual formation of stars}

In this paper, matching the examples of Paper II, we will follow the formation and
early evolution of star clusters for up to about 20~Myr. This involves a formation
phase, i.e., when gas is still present, and then a post-formation, gas-free phase.
During the formation phase, as the background gas model evolves, stars are
gradually introduced in the simulations according to the previously calculated
constant SFR following the same phase-space distribution of the gas. As introduced
in \paperII, we include this prescription in a modified version of the direct
$N$-body code \texttt{Nbody6++} \citep{Aarseth2003,Wang2015}, where we are able to
introduce stars, including primordial binaries, at arbitrary times during runtime.
The minimum number of stars we can model with this code is on the order of 150,
which is the initial number of stars all models start with. The primordial binary
fraction in a given simulation is held constant in time, so that if no binaries are
disrupted, then the total binary fraction would, on average, remain constant during
star formation.

As in \paperI\ and \paperII, stellar mass loss from stellar evolution was included
in the simulations using the analytical models developed by
\cite{Hurley2000,Hurley2002} implemented in \texttt{Nbody6++}, including mass
transfer between binaries and close interacting stars, so that the full stellar
evolutionary path is not simply defined by the initial mass and metallicity, but
can change with the dynamical history of the stars.  We note that these stellar
evolutionary models do not include the pre-main sequence phase, when stars are
generally larger than their main sequence sizes, so that such interactions will
tend to be underestimated somewhat. However, in general, close interactions between
stars occur only very rarely in our simulations and this limitation is not expected
to influence the overall results significantly. The models also include velocity
kicks for neutron stars (but not black holes) that are formed from asymmetrical
supernovae ejections. The magnitude of the kicks follows a Maxwellian velocity
distribution with $\sigma = 265$ km/s, based on proper motion observations of
runaway pulsars \citep{Hobbs2005}. 

\begin{figure*}
  \includegraphics[width=\textwidth]{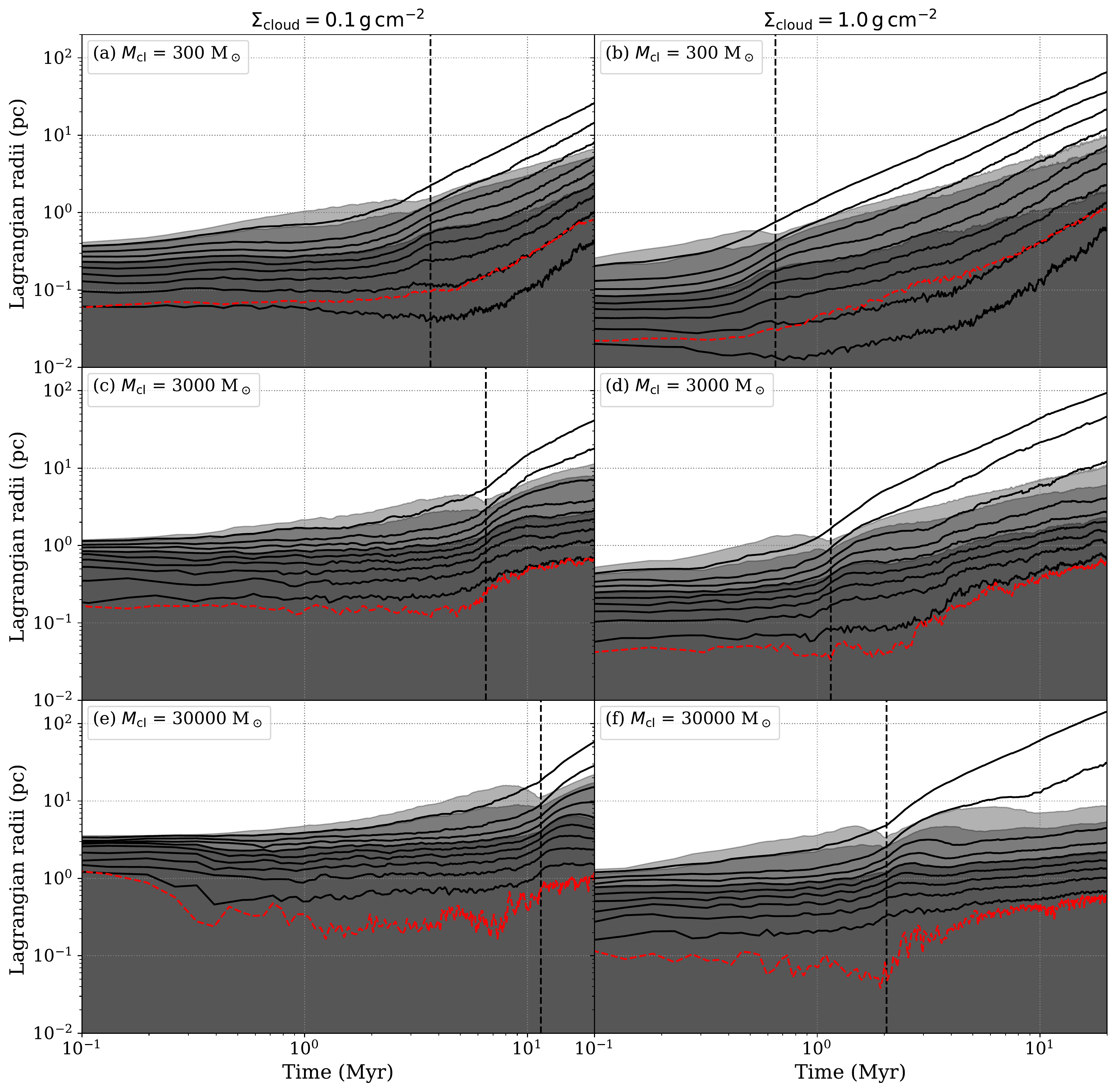}\\
  \caption{%
          Time evolution of Lagrangian radii enclosing 10, 20, 30, 40, 50, 60, 70,
          80 and 90\% of the stellar mass for star clusters with different $\mcl$
          (top, middle and bottom rows, as labelled), formed with $\epsilon=0.5$
          and $\sfeff=0.03$.  Left panels shows the low-\Sigmacl\ case and right
          panels the respective simulations for the high-\Sigmacl\ case.  The lines
          show the averages over all the simulations performed in each set.
          Background shaded areas shows the average Lagrangian radii for the 50, 80
          and 90\% bound stellar masses. Dashed red lines shows the average core
          radii. Vertical dashed lines show the respective gas exhaustion time 
          $t_*$.
        }
  \label{fig:lradmass}
\end{figure*}

\subsection{Primordial stellar population}
\label{sec:icbinary}

In this work we aim to isolate the effects of the different parent clump masses and
so we use a stellar population that is identical to the fiducial set of simulations
in our previous works. This uses a canonical initial mass function (IMF)
\citep{KroupaIMF} with 50\% binaries in circular orbits. We construct the binary
population from a log-normal period distribution with a mean of $P=293\,$yr and
standard deviation of $\sigma_{\log P} = 2.28$ according to observations of
\cite{Raghavan2010}. The mass ratio distribution follows the form $dN/dq\propto
q^{0.7}$ as observed in young star clusters \citep{Reggiani2011}. The binary
population is constructed from the full set of individual stars (binary members and
singles) that follows the adopted IMF. We note that this construction implies that
in general low-mass stars end up with slightly higher binary fractions than more
massive stars. This is because once a primary star is selected, the companion,
which has a lower mass by construction, is selected according to the mass ratio
distribution.  Therefore low mass stars have higher chances of being selected to be
part of a binary system. In our scheme, brown dwarfs then have primordial binary
fractions of 60\%, while stars above 0.4\,\Msun\ have 40\% primordial binary
fractions resulting in an average of 50\%. We note that this disagrees with
observations where most massive stars tend to have higher multiplicity fractions
\citep{Offner2022}. However, it is possible that such a trend develops dynamically
after the formation phase, which we will assess in a future work in this series.

\begin{figure*}
  \includegraphics[width=\textwidth]{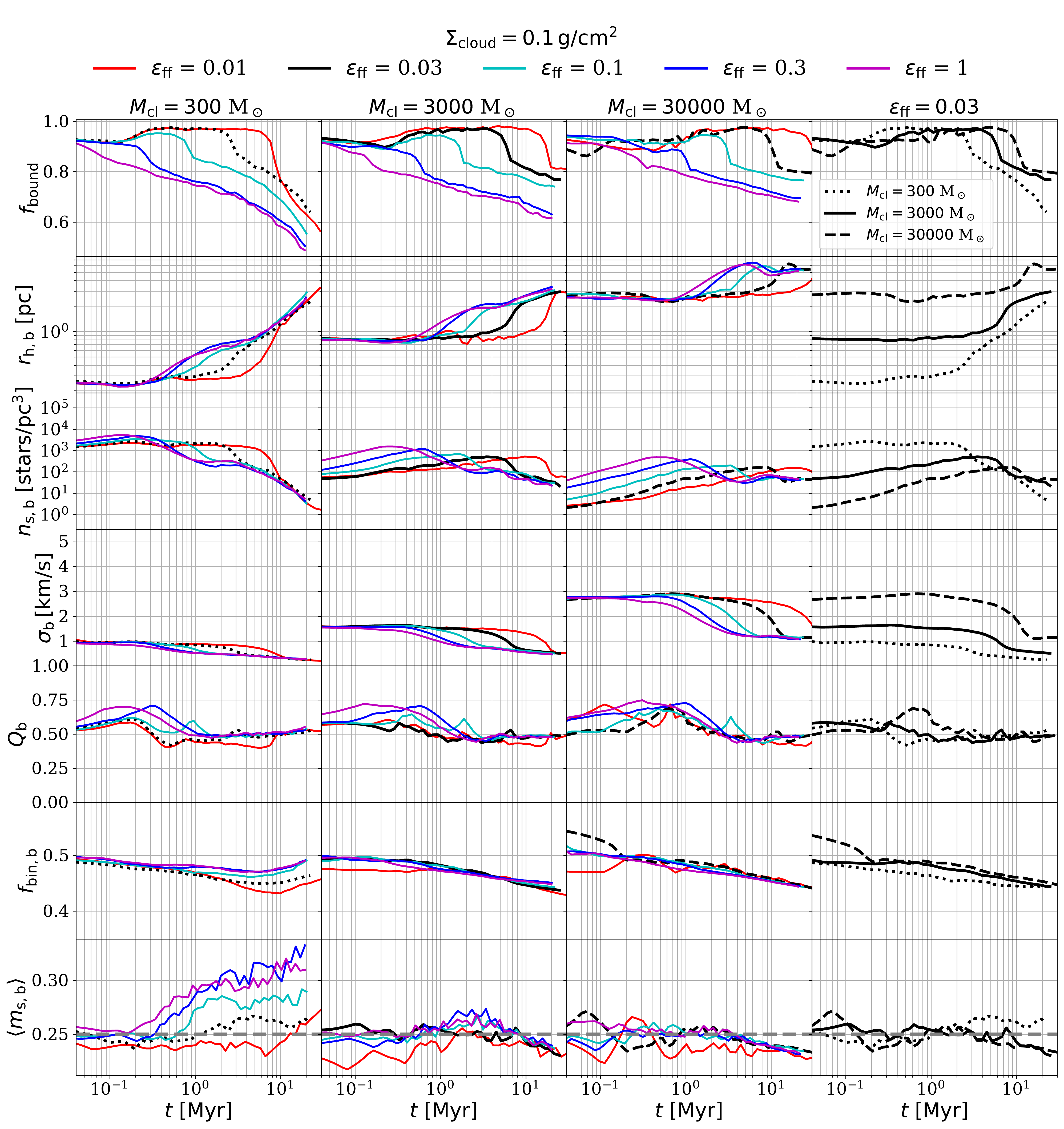}\\
  \caption{
           Time evolution of various properties of star clusters formed within a
           mass surface density environment of $\Sigmacl=1.0\,\gpcm$, global
           $\epsilon=0.5$ and different values of $\sfeff$ (see legend). The lines
           in each panel show median values calculated from all the simulations of
           each set. First, second and third columns show the cases of $\mcl=300$,
           3,000 and 30,000\,\Msun, respectively, while the fourth column shows a
           comparison of all masses for the fiducial choice of $\sfeff=0.03$. Top
           row shows the fraction of bound mass in the cluster relative to the
           instantaneous total formed stellar mass. Second row shows the evolution
           of the half mass radius $\rhb$ for the bound stars. Third row shows the
           average number density of systems ($n_{\rm s,b}$), i.e., singles and
           binaries, measured inside the volume defined by $r_{\rm h,b}$. Fourth
           row shows the evolution of the velocity dispersion measured inside
           $r_{\rm h,b}$. Fifth row shows the evolution of the virial ratio of the
           bound stellar component (\Qb). Sixth row shows the evolution of the
           bound binary fraction ($f_{\rm bin,b}$). Bottom row show the average
           system mass (singles and binaries) for stellar systems with primaries
           less massive than 7\:\Msun, where horizontal gray dashed line shows the
           expected average value given the input IMF.
  }
  \label{fig:evolmass}
\end{figure*}

\begin{figure*}
  \includegraphics[width=\textwidth]{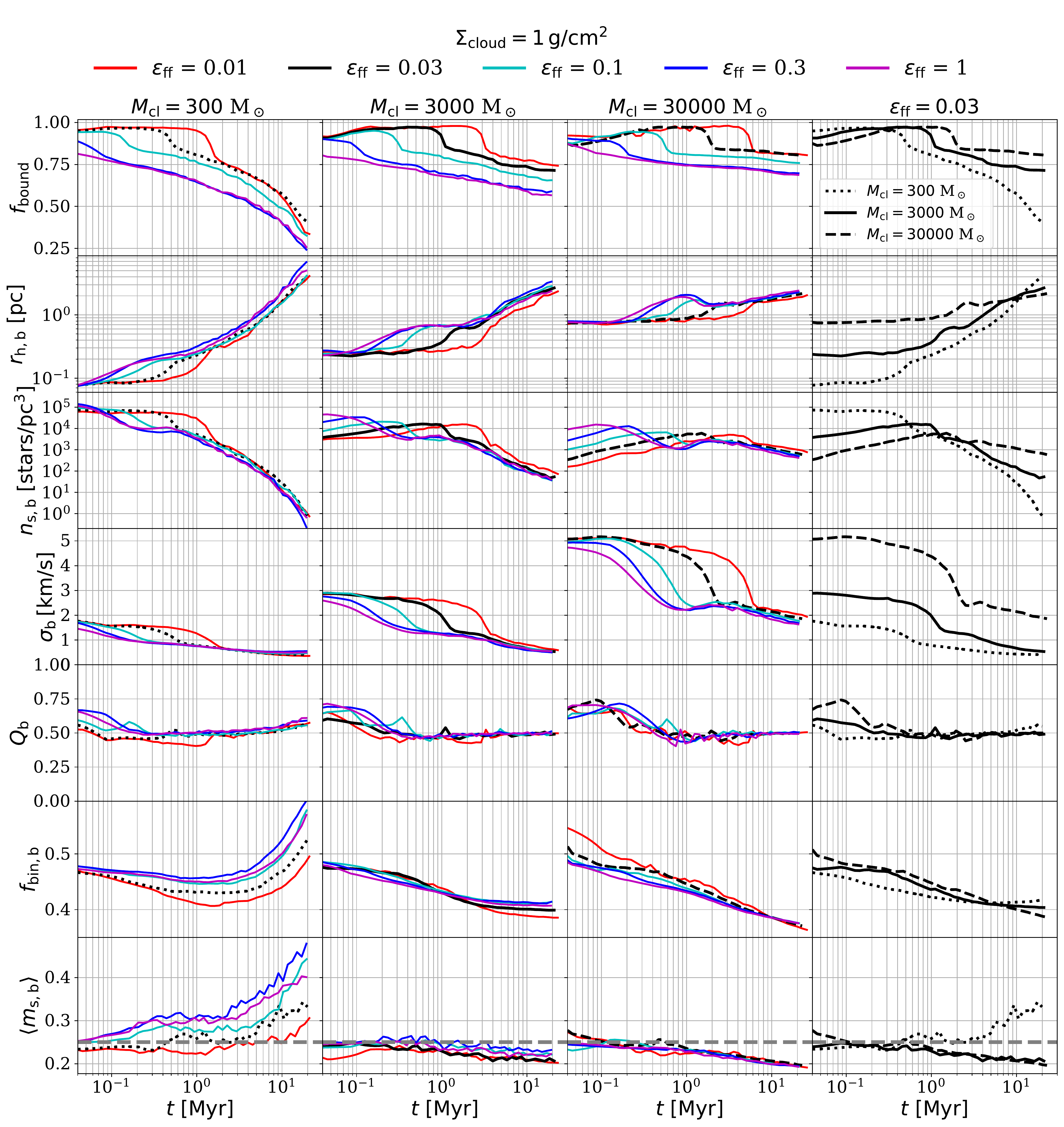}\\
  \caption{%
 Same as Fig.~\ref{fig:evolmass}, but for $\Sigmacl=1.0\,\gpcm$.
          }
  \label{fig:evolmassss1}
\end{figure*}

\subsection{Different mass models}

We perform two sets of simulations, i.e., with clump masses ten times greater and
ten times smaller than the clumps of \paperII\, which had $\mcl=3,000\,\Msun$. To
make the sets statistically comparable, we carry out 200 simulations with
$\mcl=300\,\Msun$ and 2 simulations with $\mcl=30,000\,\mcl$ for each value of
$\Sigmacl=0.1$ and 1\,\gpcm\ (hereafter low and high $\Sigmacl$ cases,
respectively).  We compare these to 20 simulations of \paperII\ for each $\Sigmacl$
case. All the simulations used for this comparison have a global star formation
efficiency, $\epsilon=$ 50\%. We also explore a range  star formation efficiency
per free fall time with the same values as in \paperII, i.e., $\sfeff=0.01$, 0.03,
0.1, 0.3 and 1.  Table~\ref{tab:masspar} shows the simulation parameters for the
different simulations performed. For the most massive clusters, we have utilised
GPUs to run the simulations to be able to access greater computational resources
and run the calculations more efficiently. This large set of simulations was
scheduled using the automated Simulation Monitor for Computational Astrophysics,
SiMon \citep{Quian2017}. 

\section{Results}

\subsection{Evolution of global structure and kinematics}\label{sec:evol}

Figure~\ref{fig:lradmass} shows the evolution of the Lagrangian radii of the star
clusters with reference to all the stars in the system (solid lines), along with
the bound stellar component (shaded regions), for our fiducial choice of
$\sfeff=0.03$. We see that during the formation stage, when the gas is still
present, the clusters tend to be confined by its gravitational potential. This
behaviour was already noted for the $\mcl=3,000\:M_\odot$ case in Paper II. After
star formation is completed, then the clusters expand more quickly. This phase
begins earlier for lower-mass and higher-density clusters (see
Table~\ref{tab:masspar}). Figure~\ref{fig:lradmass} also shows the evolution of the
cluster core radii, discussed in more detail below.

We next consider the effect of varying \sfeff\ on the evolution of the clusters.
Figures~\ref{fig:evolmass} and \ref{fig:evolmassss1} show the evolution of the
different parameters for the low and high $\Sigma_{\rm cloud}$ cases, respectively.
In each figure, the first, second and third columns show results for star clusters
forming from clumps with $\mcl=300$, 3,000, and 30,000\,\Msun\, respectively, while
the fourth column shows the three cases together for the fiducial value of
$\sfeff=0.03$.

The top rows of Figures~\ref{fig:evolmass} and \ref{fig:evolmassss1} show the
evolution of the bound mass fraction, \fbound. The values of \fbound\ of 
the various models are quite similar at the end of the formation time, which is 
determined mainly by $\Sigmacl$ and $\sfeff$, but also by $\mcl$ to a lesser degree 
(see Equation~\ref{eq:tstar}).

In the post formation phase, cluster dissolution and
evaporation effects then occur. The rates of these processes are mostly driven by
the rate of cluster relaxation, with lower mass clusters evolving more quickly to
smaller bound fractions. For example, by 20~Myr in the high $\Sigmacl$ case, the
clusters formed from $\mcl=300\:\Msun$ clumps have bound fractions of only about
0.3, i.e., these are very low-mass clusters with bound stellar masses of only $\sim
50\:\Msun$. In the low $\Sigmacl$ case, these low-mass clusters have higher bound
fractions at 20~Myr, with values of $\sim 0.5$, mostly because their formation took
longer and the post formation phase is a smaller fraction of the 20~Myr evolution.
Considering the $\mcl=3,000\:\Msun$ cases, the bound fractions at 20~Myr are
higher, i.e., $\sim 0.6$ in the low density models and $\sim 0.7$ in the high
density models, but with some dispersion caused by $\sfeff$. These higher bound
fractions are caused, at least in part, by the cluster relaxation times being
significantly longer. These general trends continue up to the $\mcl=30,000\:\Msun$
cases, which retain the highest bound fractions at 20~Myr of $\sim 0.8$ for the
fiducial $\sfeff=0.03$ case. This corresponds to a star cluster of mass $\sim
12,000\:\Msun$.

The second rows of Figures~\ref{fig:evolmass} and \ref{fig:evolmassss1} show the
evolution of bound cluster half-mass radii, $\rhb$, while the third rows show the
evolution of the average number density of stars, evaluated inside these radii. We
see that $\rhb$ remains quite constant during the formation phase, and then
undergoes expansion once the gas has been exhausted. The low-mass clusters end
their formation with radii of $\rhb \sim 0.1$ to 0.3~pc. The clusters forming
relatively quickly, i.e., with $\sfeff\gtrsim 0.1$, have a chance to enter a
``post-formation stabilization'' (PFS) phase, when \rhb\ stays at a nearly constant
level, i.e., $r_{\rm h,b,PFS}$, that is about a factor of 2 to 3 greater than
during formation. After this the clusters undergo very dramatic expansion, driven
by dynamical relaxation. Note, the slow-forming models do not have a chance to
enter the PFS phase, since they are still forming when dynamical relaxation starts
to drive their expansion. The low-mass clusters reach sizes of about 2~pc in the
low density case and about 5~pc in the high density case, which thus, in fact,
achieve the lowest number density of stars of any of our models, i.e., only
$\sim10\:{\rm pc}^{-3}$, after a decline of about a factor of $10^4$. The effects
of $\sfeff$ are relatively modest on the values of $\rhb$ reached by 20~Myr, with
the main differences occurring at earlier times around $\sim1\:$Myr due to the
different durations of the formation phases and whether or not they have a chance
to enter the PFS phase. 

These general trends continue for the $\mcl=3,000\:\Msun$ cases, though with the
variation in sizes due to different onsets of the PFS phases shifted to somewhat
later times, ranging from about 0.5 to 2~Myr with $r_{\rm h,b,PFS}\simeq 0.7\:$pc
in the high density environments and about 2 to 5~Myr with $r_{\rm h,b,PFS}\simeq
2\:$pc in the low density environments. Dynamical relaxation drives subsequent
expansion, but at a much slower rate than in the low-mass clusters. Again, the
slow-forming models do not have a chance to enter the PFS phase. We note that by
20~Myr the clusters forming from $\mcl=3,000\:M_\odot$ clumps reach sizes of
$\rhb\sim3\:$pc, with this being quite insensitive to $\Sigmacl$ and $\sfeff$, even
though they can reach this size with quite different evolutionary histories,
especially for the low \sfeff\ and low \Sigmacl\ cases. 

The $\mcl=30,000\,\Msun$, $\Sigmacl=1\:\gpcm$ case produces even more compact
clusters with $\rhb\sim 2\:$pc at 20~Myr, independent of $\sfeff$. Again, the
evolution to this state involves a phase in which the cluster expansion is
essentially halted (even with a ``bounce'' for $\sfeff\gtrsim 0.1$) at $r_{\rm
h,b,PFS}\simeq 1.5\:$pc within the first few Myr (depending on \sfeff), before the
cluster relaxation expansion phase, which here occurs at a very slow rate compared
to the lower-mass clusters. In the $\Sigmacl=0.1\:\gpcm$ case, the massive clusters
stop expanding at $r_{\rm h,b,PFS}\simeq 6\:$pc (and with $n_{\rm s,b}\sim
100\:{\rm pc}^{-3}$), although the slowest forming model with $\sfeff=0.01$ does
not have time to reach this state within $\sim20\:$Myr.  Furthermore, these
massive, low-density clusters do not have time to exhibit significant expansion
driven by dynamical relaxation during the duration of the simulations investigated
here, i.e., up to $\sim 20\:$Myr.

\begin{figure*}
 \includegraphics[width=\linewidth]{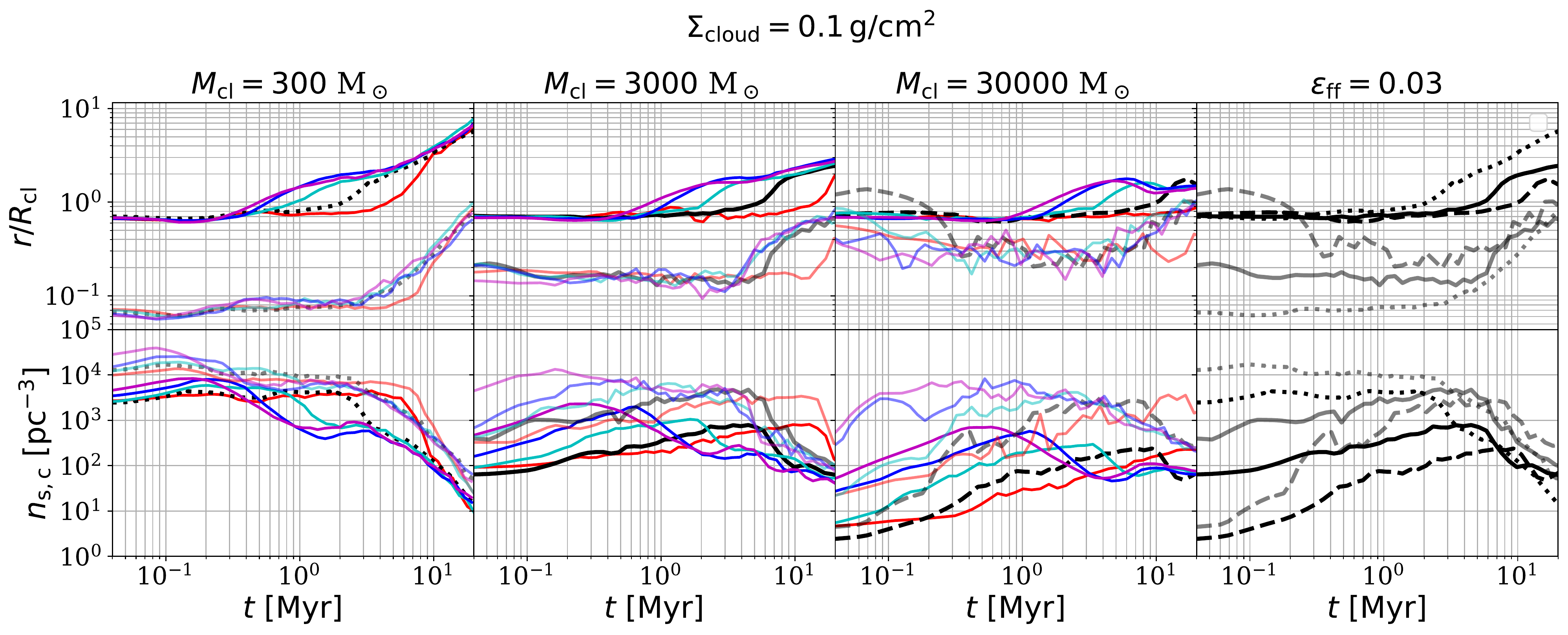}
 \includegraphics[width=\linewidth]{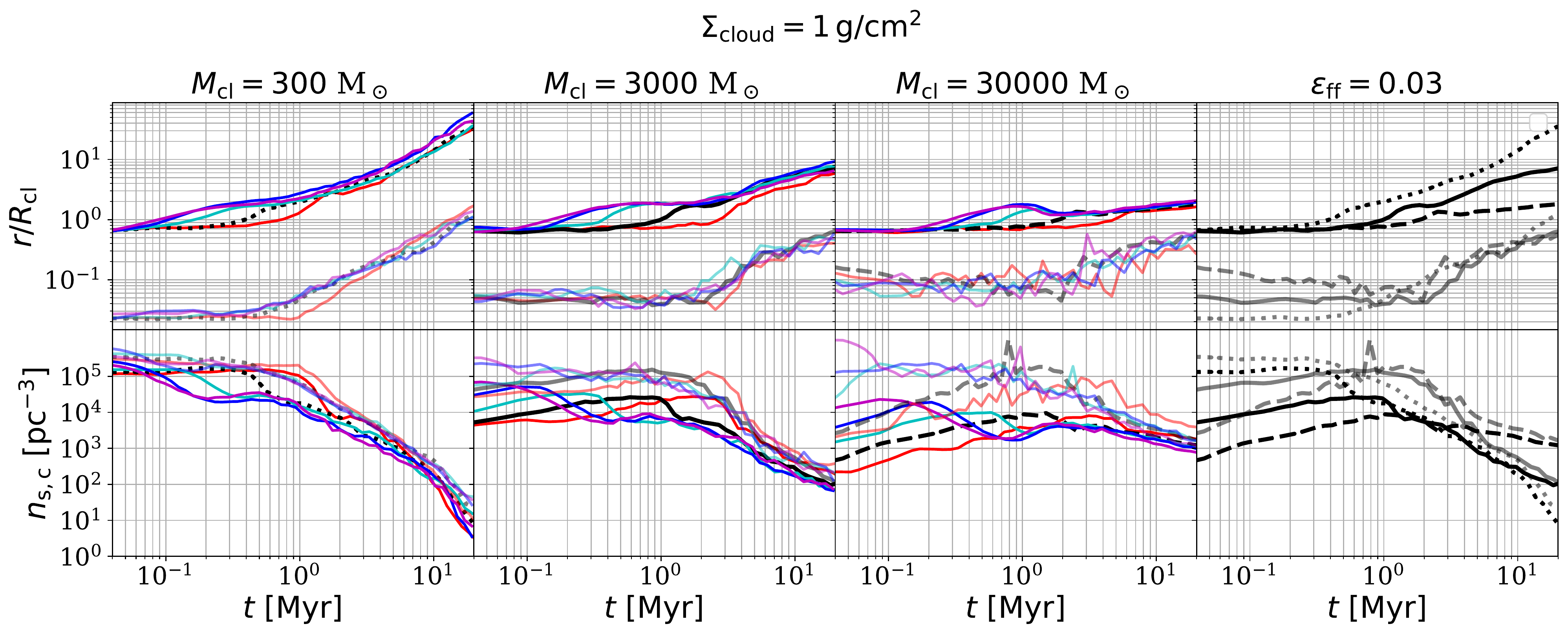}
 \caption{%
          Evolution of average size and number density for all simulations in this
          work. Top and bottom set of panels shows simulations with
          $\Sigmacl=0.1\:\gpcm$ and $\Sigmacl=1.0\:\gpcm$, respectively. Different
          colors show the adopted $\sfeff$ with the same color-scheme as in
          Figures~\ref{fig:evolmass} and \ref{fig:evolmassss1}. The first three
          columns group simulations with the same $\mcl$, while the fourth column
          compares models with different values of $\mcl$ for the $\sfeff=0.03$
          case. Solid lines in the top rows show the evolution of the average bound
          half-mass radii, $\rhb$, normalized by the initial clump radius, \Rcl.
          Semi transparent lines show the respective normalized core radii. The
          bottom rows show the evolution of number densities within the bound
          half-mass radii, $n_{\rm s,b}$, and within the core radii, $n_{\rm s,c}$.
         }\label{fig:rhdens}
\end{figure*}

To more fully illustrate the evolution of cluster sizes, in Figure~\ref{fig:rhdens}
we show in the top rows of the top and bottom set of panels the time evolution of
\rhb\ normalized by the initial clump radius. We see that $r_{\rm h,b,PFS}$ is
about a factor of 1.5 and 2 times larger than $\Rcl$ for the low and high-\Sigmacl\
cases, respectively. Then, by about 20~Myr, clusters have typically been able to
expand by factors of 2 (for large clusters with $\mcl=30,000\:\Msun$) to 40 (for
small clusters with $\mcl=300\:\Msun$) compared to the size of their natal gas
clumps. 

Another important radial scale is the cluster core radius, defined as the
``density''-weighted average distance of the stars from the density center in the
cluster, where the ``density'' of each star is estimated using the mass in a sphere
containing the six nearest neighbors \citep{Casertano1985,Aarseth2003}.  The time
evolution of the core radii, normalised by $R_{\rm cl}$, are also shown in
Figure~\ref{fig:rhdens}. These cluster core radii are relatively constant during
the formation phase and are systematically larger for the more massive clusters. In
addition, we see that core radius evolution appears to be independent of $\sfeff$,
with the exception of the $\sfeff=0.01$ case. In all models we see that the main
expansion phase of the core radius begins at about the same time, i.e., after about
one crossing time of the region. The PFS phase ends as part of this core radius
expansion phase, i.e., the half-mass radius expands in step with the core radius.
The case of $\sfeff=0.01$ is different because the core radius is held in place by
the background potential, delaying the expansion of the cluster and not going
through a PFS phase since the core is already relaxed. We see that 
at about 20~Myr these cluster core radii, although still expanding, have evolved to
be quite similar to the initial clump radii. 

The third rows of Figures~\ref{fig:evolmass} and \ref{fig:evolmassss1} show the
time evolution of the average number densities of the stars inside $\rhb$. These
respond accordingly to the evolution of $\fbound$ and $\rhb$. In general, in our
models, lower-mass clusters form from denser clumps and so during the formation
phase have higher number densities of stars than more massive clusters. However,
given that they start expanding earlier, this situation reverses during the first
few Myr. The slower forming clusters take longer to build up their stellar
densities, but retain these levels for longer periods of time. We will see later
that this affects their overall efficiency at producing runaway stars via dynamical
ejections. However, we note that it is the number densities in the densest part of
the clusters, i.e., in their cores, which are important for production of most
close interactions leading to dynamical ejections. Thus, in Figure~\ref{fig:rhdens}
(bottom rows of each set of panels) we also show the time evolution of $n_{\rm
s,c}$, i.e., the average number density of stellar systems within the core radius.
We see that the number densities in the core regions can be many times larger than
that averaged over the half-mass scale, especially for the most massive clusters.
Thus, in general, the full density profile of a cluster needs to be considered for
estimation of quantities, such as interaction rates, that depend on local
densities.

\begin{figure*}
 \centering
 \includegraphics[width=\linewidth]{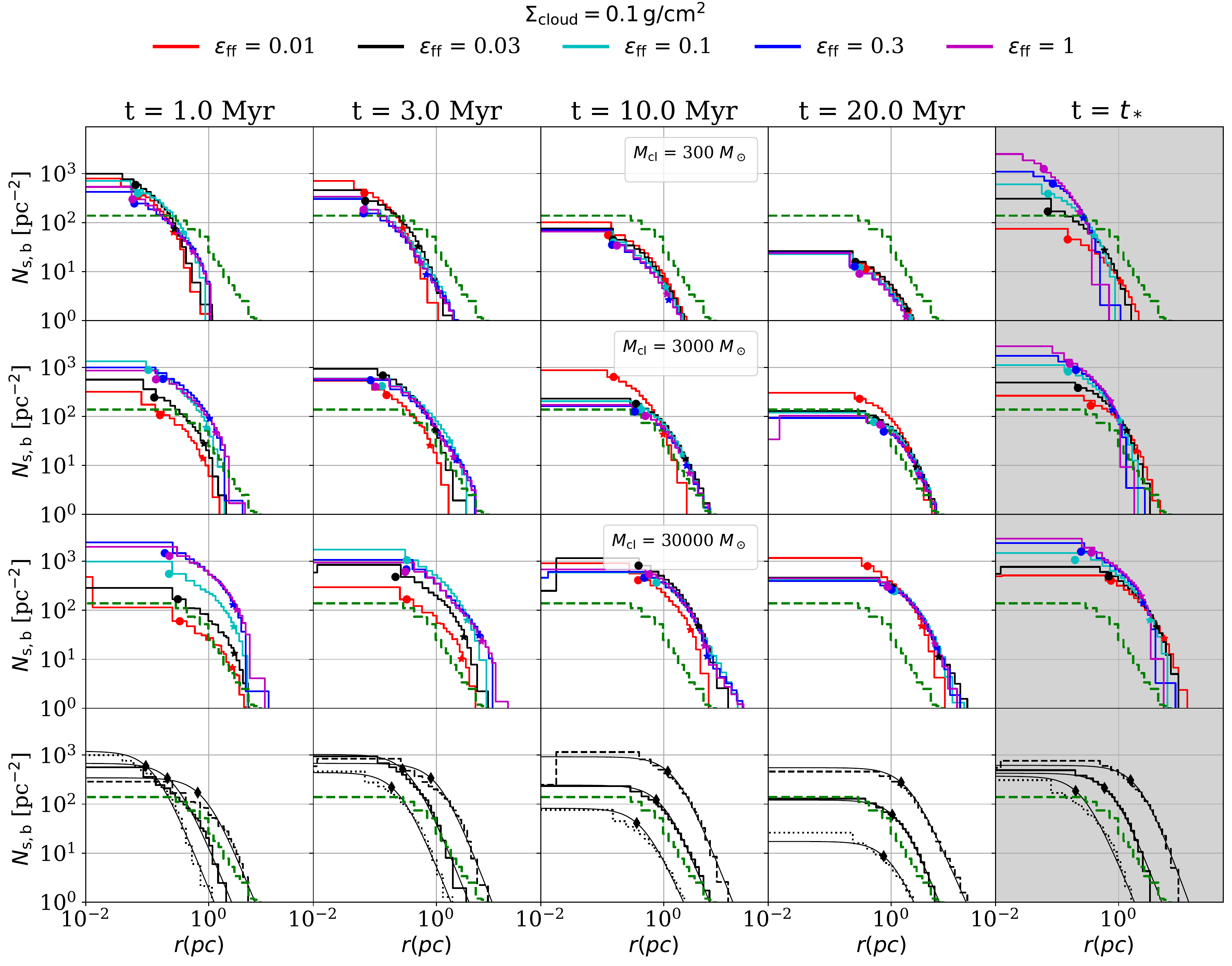} 
 \includegraphics[width=0.5\linewidth]{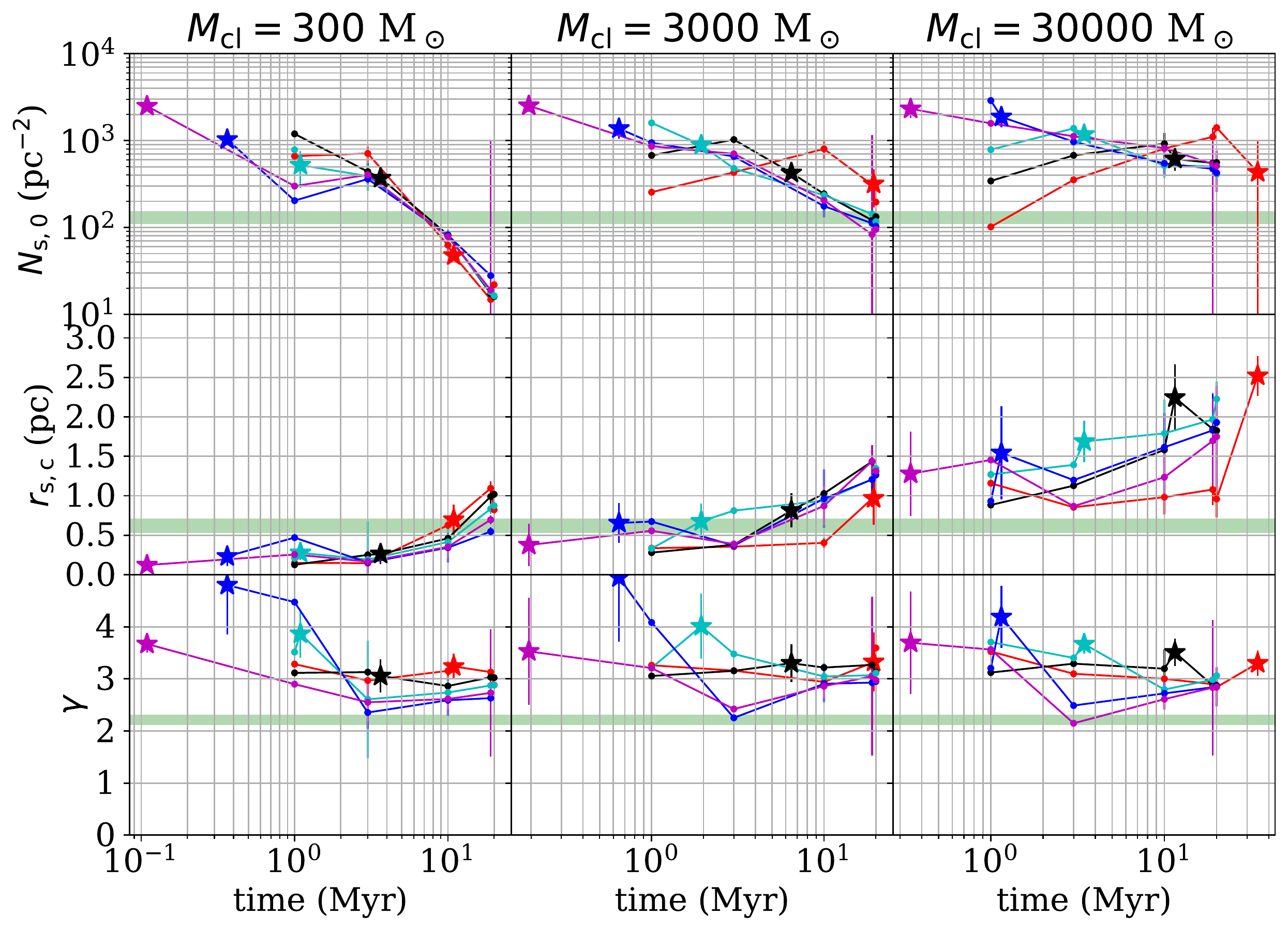}
 \caption{%
         {\it (a) Top:} Projected number density profiles for simulations with
         $\Sigmacl=0.1\gpcm$. Profiles are measured at $t = 1$, 3, 10, 20 Myr and
         when star formation is finished at $t=t_*$ (columns, left to right).
         Filled circle, diamond and star symbols shows the positions of the core
         radius ($r_{\rm c,b}$), fit scale radius ($r_{\rm 0,b}$) and half mass
         radius ($r_{\rm h,b}$), respectively. The first three rows present the
         cases for $M_{\rm cl}=$ 300, 3,000 and 30,000~\Msun, i.e., to allow easy
         visualization of the effects of $\epsilon_{\rm ff}$. The fourth row
         compares the $\epsilon_{\rm ff}=0.03$ cases for the different masses, with
         dotted, solid and dashed histograms showing $\mcl=$ 300, 3,000 and
         30,000~\Msun, respectively. Thin solid lines in the fourth row show the
         best fits of Equation~\ref{eq:nsprofile}. Radial binning is constructed so
         that each bin has the same number of stars. Green dashed lines show the
         density profile of the ONC based on the membership list by
         \protect\cite{DaRio2016}. {\it (b) Bottom:} Time evolution of fitted
         structural parameters of Equation~\ref{eq:nsprofile}, measured at the same
         times as in the above profiles. Star symbols show the results at
         $t=\protect\tsf$. Green horizontal bands show the values of these
         parameters that are estimated for the ONC from the data of
         \protect\cite{DaRio2016}.%
         }\label{fig:structure0.1}
\end{figure*}

\begin{figure*}
\centering
\includegraphics[width=\linewidth]{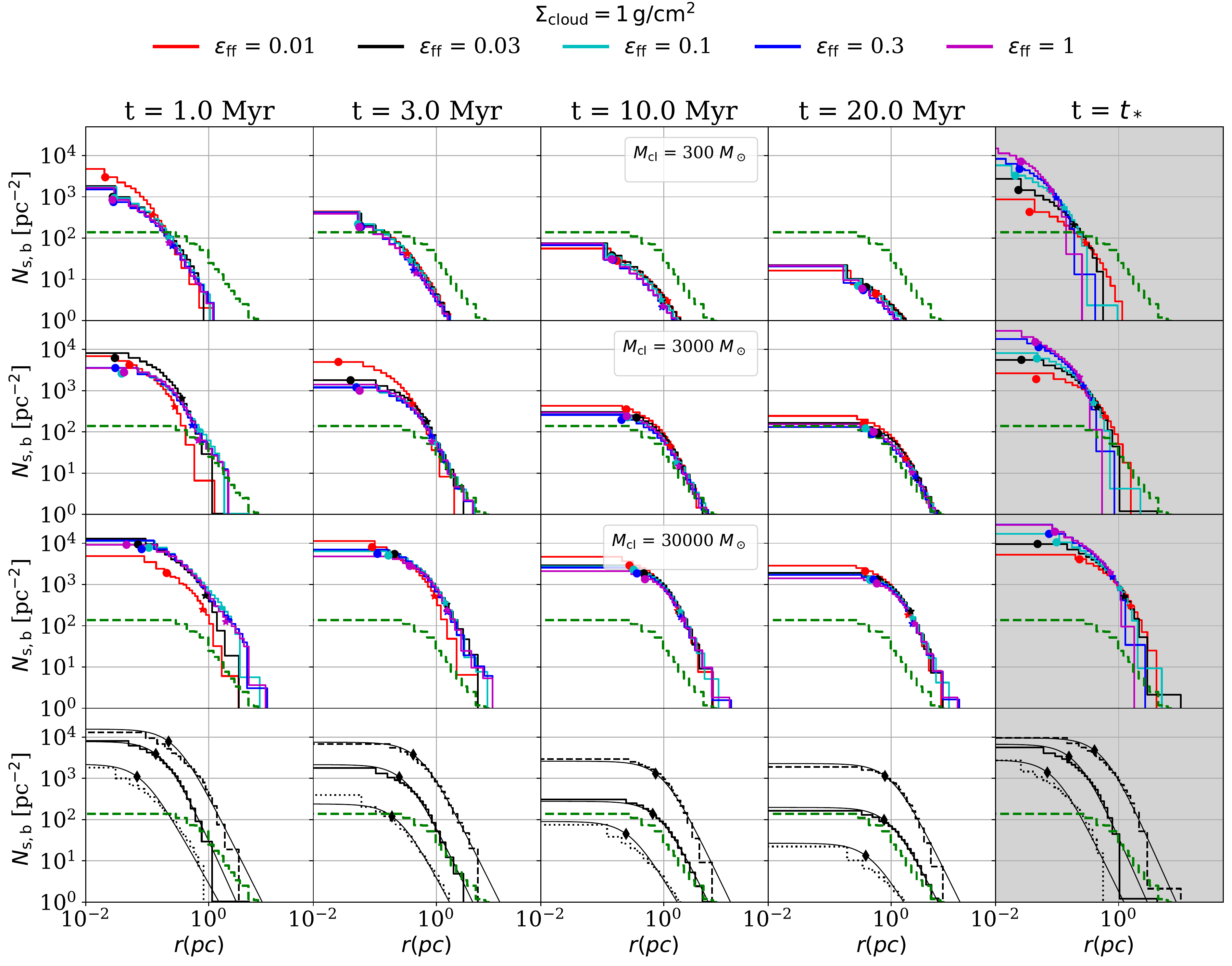} \\
\includegraphics[width=0.5\linewidth]{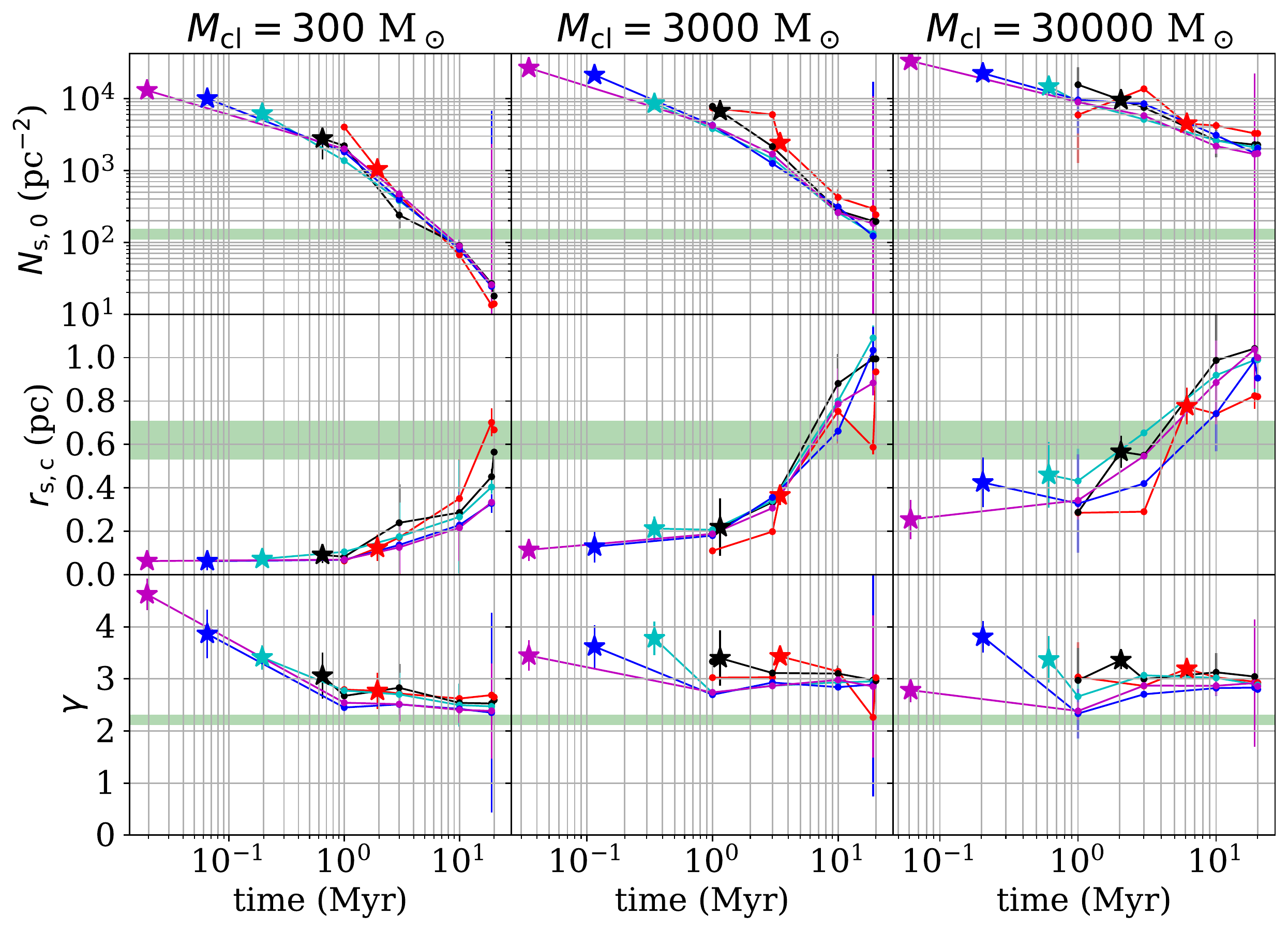}
\caption{%
  As Fig.~\ref{fig:structure0.1}, but for $\Sigmacl=1.0\:\gpcm$. 
        }
\label{fig:structure1}
\end{figure*}

\begin{figure}
  \includegraphics[width=\columnwidth]{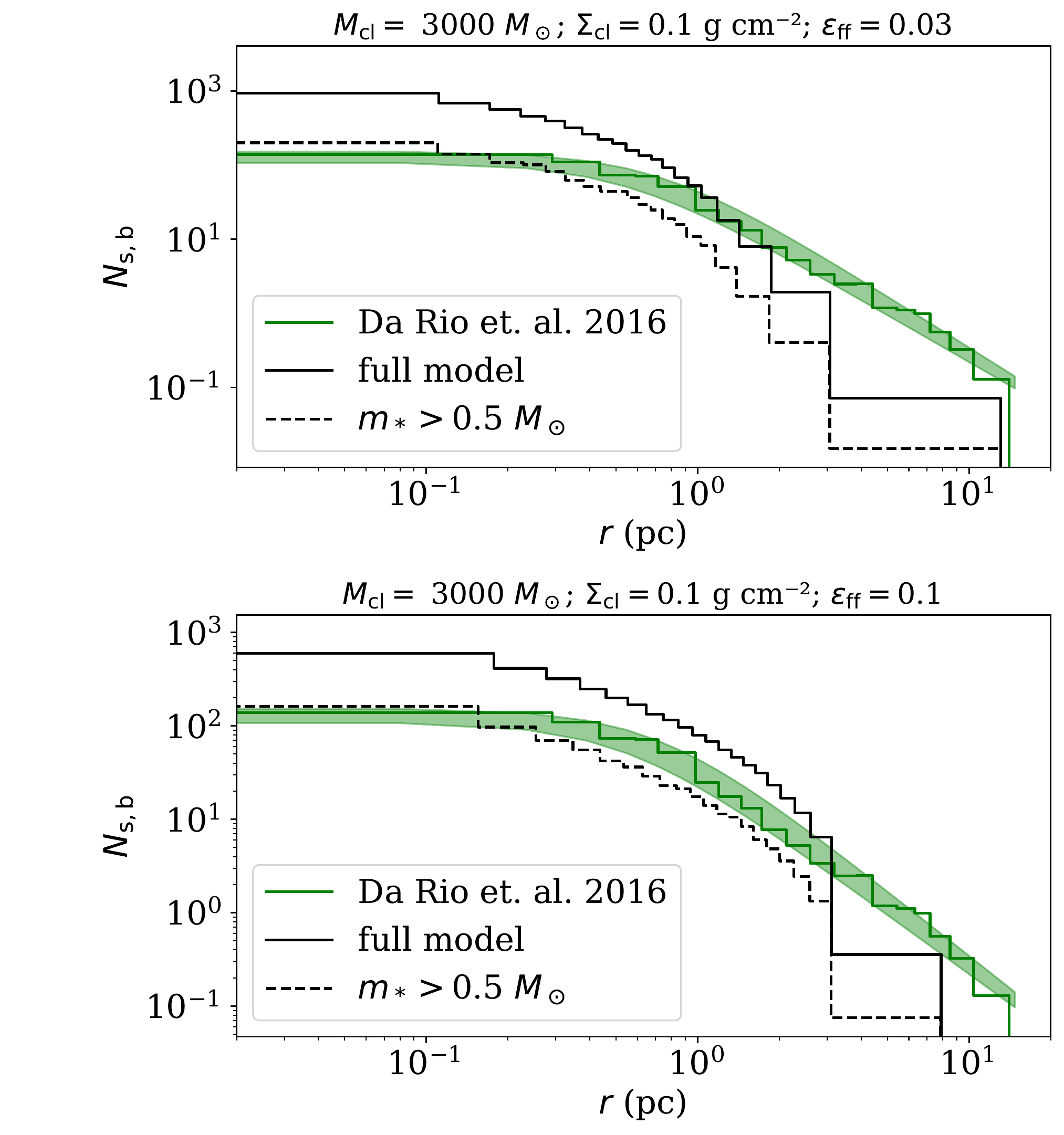}
  \caption{%
          In each panel, the green line shows the ONC projected number density
          profile derived from the membership list provided by
          \protect\cite{DaRio2016}.  Green shaded area shows the range of best
          fitting ONC profiles according to Equation~\ref{eq:nsprofile}.  The ONC
          data are compared to two of our models with $\Sigmacl=0.1\:\gpcm$ and
          $\mcl=3000\:\Msun$ measured at 3 Myr (black solid lines). Top panel
          compares to our fiducial model with $\sfeff=0.03$, and bottom panel to a
          model with $\sfeff=0.1$.  Dashed lines show the same modeled profiles,
          but excluding stars with masses below 0.5\:\Msun, which is a simple way
          to illustrate the effects of potential incompleteness. 
          }
  \label{fig:ONC}
\end{figure}

For the purposes of comparing to observed star clusters, where the true 3D
structure is hard to measure, it is better to consider the projected surface number
density profiles, i.e., $N_s(r)$, with this being the number of stellar systems
(singles, binaries and higher-order multiples) per unit projected area.
Figures~\ref{fig:structure0.1} and \ref{fig:structure1} show these projected radial
profiles of the bound clusters as they evolve during the simulations. These
profiles are averages of all sets of the same model at a given time. 

Our simulated clusters tend to have a similar radial structure.  We characterize
this using the model of \cite{Elson1987}, which was developed to describe
luminosity profiles of young star clusters. Then the surface number density
profiles are described via: \begin{equation} N_{s}(r) = N_{\rm s,0} \left( 1 +
\frac{r^2}{r_{\rm s,0}^2}\right)^{\gamma/2}, \label{eq:nsprofile} \end{equation}
where $N_{\rm s,0}$ is the central surface number density, $r_{\rm s,0}$ is a scale
radius and $\gamma$ is a power law exponent.  The best fit parameters at each
measured time are shown in the lower set of panels in Figures~\ref{fig:structure1}
and~\ref{fig:structure0.1}. 

Equation~\ref{eq:nsprofile} typically provides a good description for the models
presented here. At early stages most models have steep power law halos, but as star
clusters relax they tend to converge to a shallower distribution with $\gamma\simeq
2.5$ for the low-mass clusters and $\simeq 3$ for the more massive ones. The scale
radius tends to be between the core and half mass radii.

As an example comparison with an observed young star cluster, we have constructed
the number density profile of the ONC based on the membership list compilation
provided by \cite{DaRio2016}. We have selected stars flagged as members by any of
the methods described by Da Rio et al., working within a 2 degree radius around the
ONC, which yields a total of 1,464 sources. We have transformed the coordinates of
the stars to parsecs using an ONC distance of 403~pc \citep{Kuhn2019} and
constructed the projected number density profile using 20 bins, each with
approximately equal number of stars. The obtained best fit parameters of
Equation~\ref{eq:nsprofile} to these data are:
\begin{eqnarray}
        N_{\rm s,0} &=& 132 \pm 22 \text{ pc}^{-2}\\ \nonumber
        r_{\rm s,0} &=& 0.62 \pm 0.09 \text{ pc} \\ \nonumber
        \gamma &=& 2.2 \pm 0.1. \nonumber 
\end{eqnarray}
We show the profile defined by these values with the green dashed lines and green
shaded areas in Figures~\ref{fig:structure0.1} and \ref{fig:structure1}, as well as
in Figure~\ref{fig:ONC}.

While our numerical models have not been specifically tailored to the ONC
properties, we see that our derived the fitting parameters, especially of the
low-\Sigmacl\ cases, are typically quite similar to those shown by the ONC in its
current state. For instance, at the age of the ONC (i.e., $\sim3$ Myr), the closest
models to the ONC in terms of total mass are clusters with $\mcl = 3000\,\Msun$. At
$\sim3$~Myr, the low-density clusters reproduce the measured scale radius $r_{\rm
s,c}$. However, the ONC's central density, $N_{\rm s,0}$, is rather low in
comparison with our models. One potential mitigating factor is that the
observational sample of \cite{DaRio2016} is incomplete in the brown dwarf regime
and its incompleteness may be relatively higher in the central regions due to
effects of higher extinction, nebulosity and crowding compared to outer regions.
Figure~\ref{fig:ONC} shows more detailed comparisons of some of our model clusters,
including the effects of incompleteness below $0.5\:\Msun$, with the observed
surface number density profile of the ONC.  We see here that the ONC has a
relatively shallower outer projected density distribution, i.e., with $\gamma=2.2$,
compared to our simulated clusters, i.e., with $\gamma=2.5$. While this could be a
real physical discrepancy, i.e., indicating a limitation of the model, it could also be caused by contamination by false
positive members in the outskirts of the ONC.

In summary, we see that our modeled star clusters develop a surface density profile
that is quite similar to that exhibited by the ONC. However, further work on
simulated clusters that are more specifically tailored to this and other observed
clusters, including effects of observational incompleteness, are needed before one
would be able to constrain model parameters of $\mcl$, $\Sigmacl$ and $\sfeff$.

\subsection{Evolution of kinematics and dynamics}

The fourth rows of Figures~\ref{fig:evolmass} and \ref{fig:evolmassss1} show the
time evolution of the 1-D velocity dispersions of the bound members of the
clusters, $\sigma_b$. The clusters start with velocity dispersions given by their
parental gas clumps (see Table~\ref{tab:masspar}). At first, during the formation
phase, these remain relatively constant, although in the high $\Sigma_{\rm cloud}$
cases $\sigma_b$ declines slowly even during this phase. Following the formation
phase, the velocity dispersions decline at a faster rate as the clusters expand and
lose mass from the bound component. By $\sim 20~$Myr some clusters, e.g., the most
massive clusters forming from low density environments, have a chance to reach a
relatively stable level of $\sigma_b$, just larger than $1\:\kms$. 

The fifth rows of Figures~\ref{fig:evolmass} and \ref{fig:evolmassss1} show the
evolution of the virial ratio of the bound stellar system defined, in its most
general form as:
\begin{equation}
        \Qb \equiv  \frac{ E_{\rm kin,b} }{E_{\rm grav,b}},
\end{equation}
where $E_{\rm kin,b}$ is the total kinetic energy of the bound stars and $E_{\rm
grav,b}=\sum_{i=1}^{N_b} \vec{F_i}\cdot \vec{r_i}$ is the gravitational energy of
the bound stars. In this calculation, binaries are treated as single unresolved
systems.

A cluster in virial equilibrium has $\Qb = 0.5$. As introduced in previous papers
in this series, the star clusters formed in our framework do so from an initially
supervirial state since the natal clump has a significant surface pressure applied
to it from its surroundings. For all simulations in this paper, the initial global
$Q\approx1$. As the stellar systems relax from their initial configurations, they
will achieve approximate equilibrium on a timescale of the order of one relaxation
time, $\trelax \simeq (N/\ln N) \tcross$.

As seen in Paper II, the ratio of the timescale over which $Q$ relaxes into
equilibrium, \trelax, compared to the formation time, $t_*$, is important because
by the end of formation, when the background gas is exhausted, the different
\sfeff\ models can then start their gas-free stage from different dynamical states.
For example, a star cluster that forms quickly (e.g., $\sfeff = 1$) does not have
time to relax and is still supervirial by the time gas is exhausted. On the other
hand, a star cluster that forms slowly ($\sfeff = 0.01$) has enough time to relax
and starts its gas free evolution closer to virial equilibrium. 

Figures~\ref{fig:evolmass} and \ref{fig:evolmassss1} show that this trend persists
at different clump masses. In general, the crossing time (and thus also the
relaxation time) in the formation phase is shorter than after gas is gone. Then,
slow forming models are able to relax even earlier than fast forming models that
already lost their gas mass at supervirial states. Thus, slow forming star clusters
are able to be already near virial equilibrium long before star cluster formation
is finished.

\subsection{Evolution of binary properties}\label{sec:binary}
\begin{figure*}
 \centering
 $\begin{array}{lr}
 \includegraphics[width=0.485\textwidth]{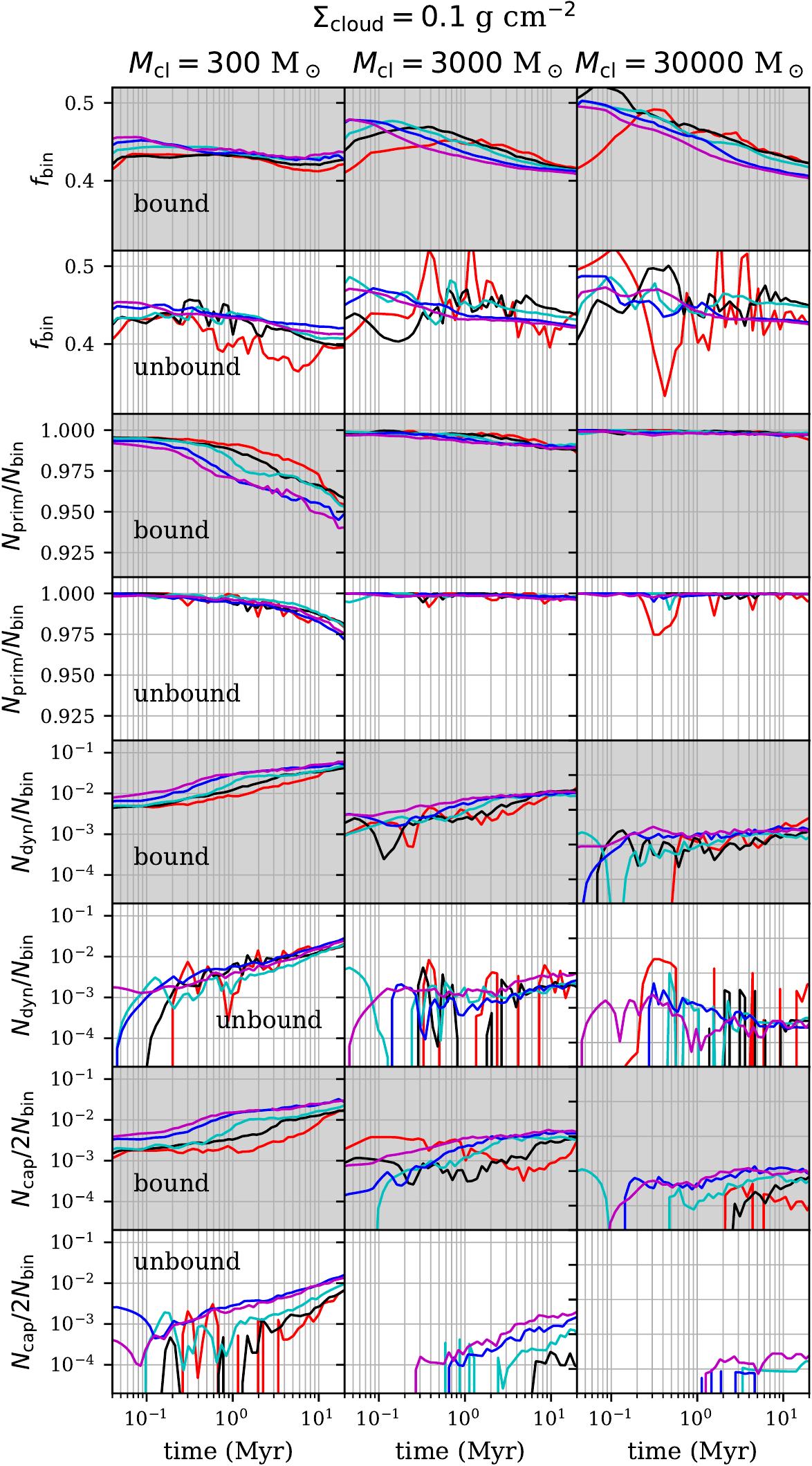} &
 \includegraphics[width=0.485\textwidth]{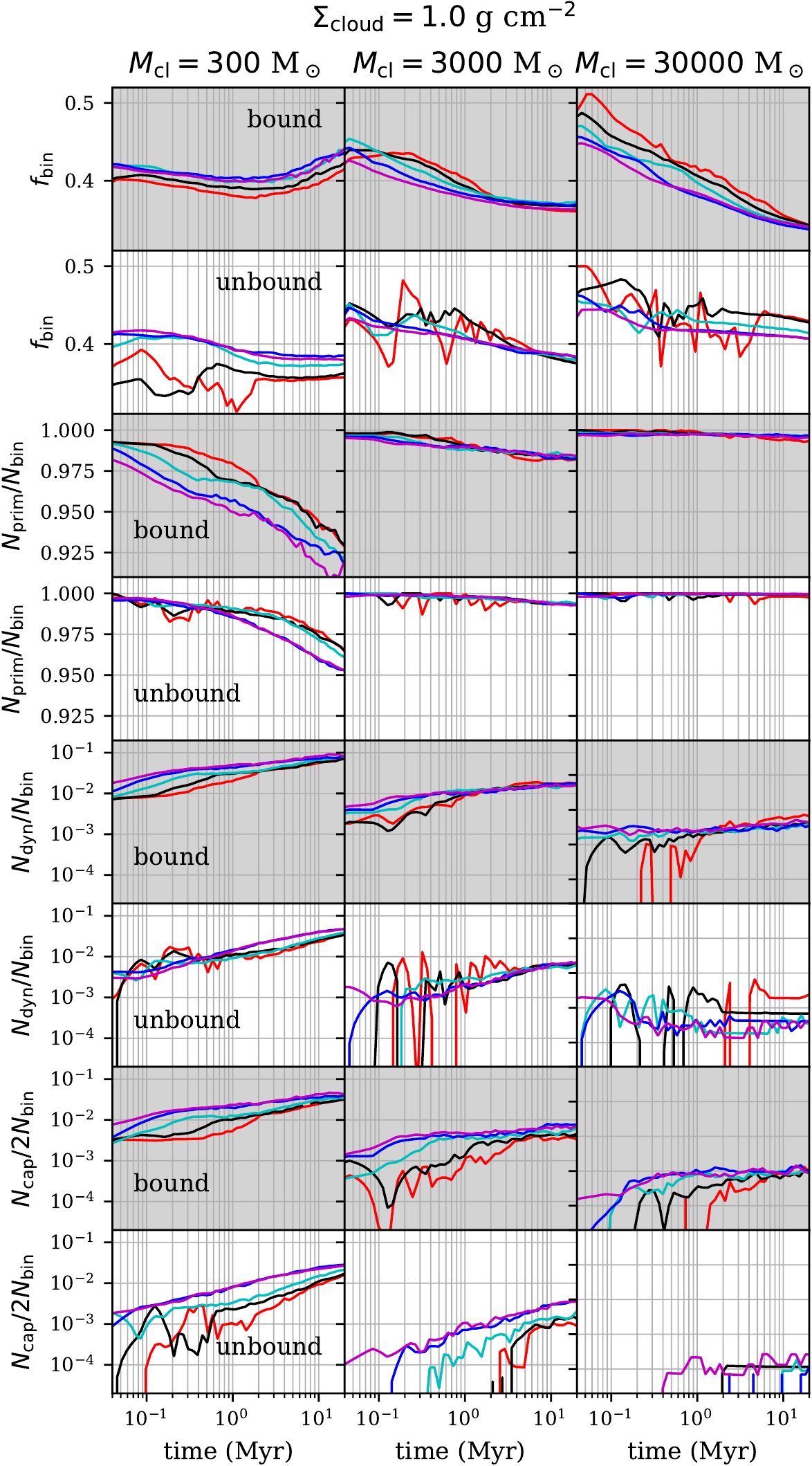}
 \end{array}$ 
 \caption{%
          Detailed evolution of binary properties, shown separately from the bound
          (gray background panels) and unbound (white background panels)
          populations of stars, for simulations with $\Sigmacl=0.1\:\gpcm$ (left)
          and $1.0\:\gpcm$ (right). Color scheme of the lines is the same as in
          previous figures denoting the adopted $\sfeff$.  From top to bottom, the
          first two rows show the binary fraction ($f_{\rm bin}$) of the bound and
          unbound populations.  The third and fourth rows show the fraction of
          binaries that are primordial, i.e., $N_{\rm prim}/N_{\rm bin}$, where
          $N_{\rm prim}$ and $N_{\rm bin}$ are the number of primordial and total
          number of binaries, respectively. The fifth and sixth rows show the
          fraction of binaries formed dynamically, with $N_{\rm dyn}$ as their
          total number.  The seventh and eight rows show the fraction of binary
          members that were originally single stars, i.e., the number of captured
          singles $N_{\rm cap}$ divided by the total number of binary members,
          i.e., $2N_{\rm  bin}$.
 }\label{fig:fbin}
\end{figure*}

The sixth rows of Figures~\ref{fig:evolmass} and \ref{fig:evolmassss1} show the
evolution of the binary fractions in the bound clusters, $f_{\rm bin,b}$. Note that
the stars are formed, statistically, with an average binary fraction of 0.5. There
is a gradual decline seen due to disruption of binaries, which can occur via
dynamical interactions and also as a result of stellar evolution, i.e., supernova
explosions. The evolution of the binary fraction shows significant differences
depending on \mcl, \Sigmacl\ and \sfeff. Small clusters of $\mcl=300\,\Msun$
quickly process binaries during their formation phase, especially at low
$\sfeff=0.01$ where $f_{\rm bin,b}$ reaches a minimum of 0.4 in the high-\Sigmacl\
case. This minimum is quite sensitive to \sfeff, given the longer formation time
compared to the local crossing time of the regions. After reaching this minimum, a
few Myr after the start of formation, there then follows a significant increase in
$f_{\rm bin,b}$. 

In Figure~\ref{fig:fbin} we present a more detailed exploration of the evolution of
the binary fraction, separating those for the bound (gray panels) and unbound
(white panels) populations. The rise in $f_{\rm bin}$ at late times in the low-mass
simulations mostly happens within the bound clusters and is caused by the creation
of new ``dynamically-formed'' binaries. The fifth and sixth rows in this figure
show the fraction of dynamically-formed binaries in the bound and unbound
populations, respectively. These dynamically formed binaries include those that
were primordial but later exchanged one of their members with other binaries or
singles stars. We see that in the low-mass simulations by 20 Myr about $\sim6-8\%$
of binaries in the bound cluster are formed dynamically, while it is about half of
this level ($\sim3-4\%$) in the unbound population.

In the higher \mcl\ cases we see that the binary fractions in the bound clusters
decrease to lower values, e.g., reaching close to 0.3 after about 20~Myr in the
$\mcl = 30,000\:\Msun$, $\Sigma_{\rm cloud}=1.0\:{\rm g\:cm}^{-2}$ case. This is
caused by there being more time for disruption of binaries by close encounters with
other stars in these clusters, which retain a high bound fraction over this period.
These clusters also have a smaller fraction of dynamically-formed binaries, which
is not enough to change the global binary fractions, unlike in the lower-mass
cases.

\subsubsection{Binary Population}
\begin{figure*}
 \centering
 \includegraphics[width=\linewidth]{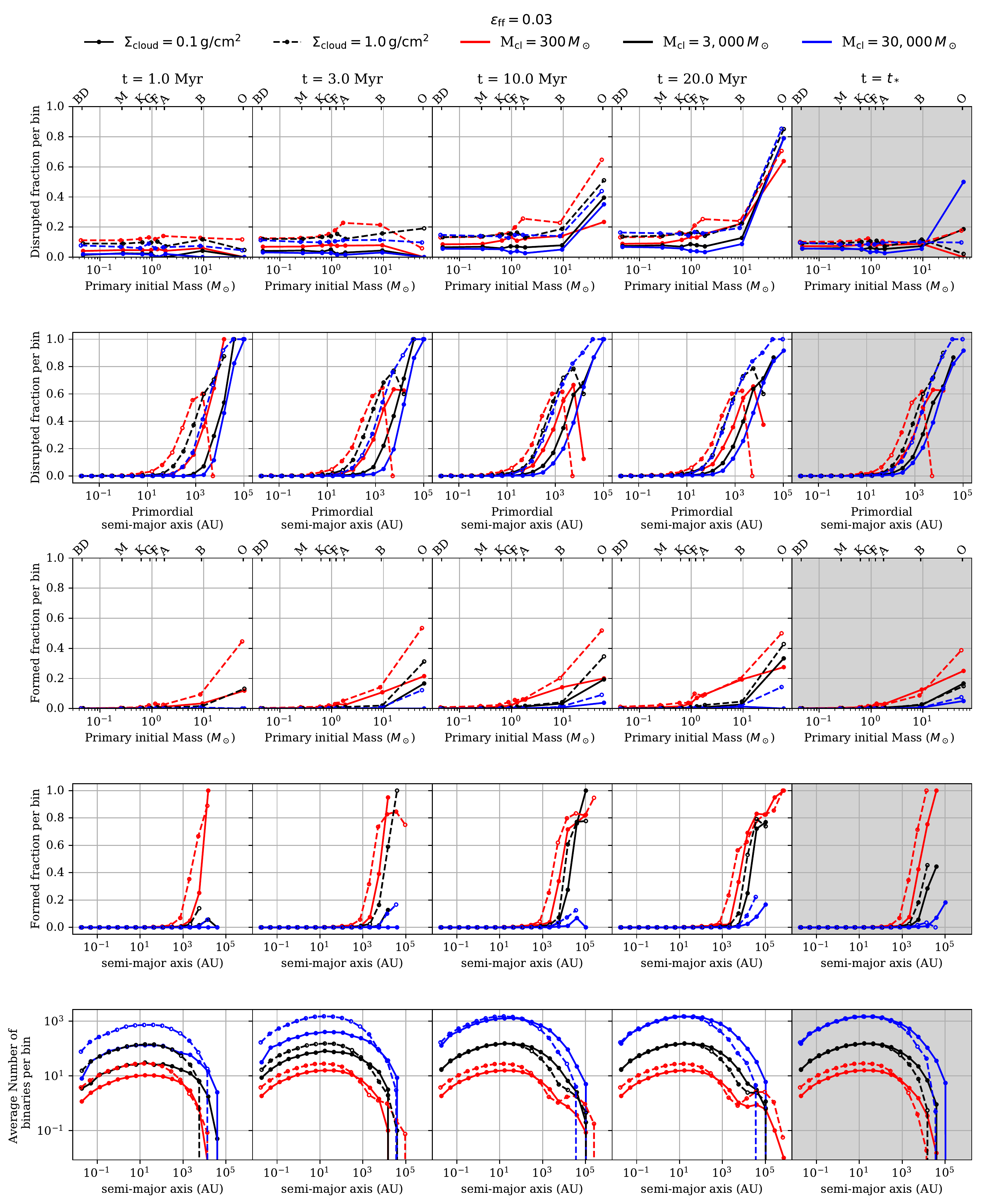}
 \caption{%
          Distribution of primary mass and semi-major axis for different kinds of
          binaries, measured at (from left to right columns) $t=$ 1, 3, 10, 20 Myr
          and when star formation stops ($t_*$). This figure shows the results for
          simulations with $\sfeff=0.03$ and $\mcl = 300$ (red), 3,000 (black) and
          30,000 \Msun\ (blue), and for simulations with $\Sigmacl = 0.1$ (solid
          lines) and 1.0 \gpcm\ (dashed lines). First and second row shows the
          fraction of disrupted primordial binaries per primary stellar type (first
          row) and primordial semi-major axis bin (second row). Third and fourth
          rows shows the fraction of binaries that are formed dynamically separated
          by stellar type (third row) and current semi-major axis. Fifth row shows
          the total distribution of semi-major axes averaged by simulation set.
 }\label{fig:Bdisrupted}
\end{figure*}

Here we examine how the binary population is processed in the different models. We
have seen in \paperI\ and \paperII\ that there was relatively little processing of
the primordial binary population. However, we have also seen from
Figure~\ref{fig:evolmassss1} that in the clusters presented in this paper there is
a significant reduction in the fraction of binaries by 20~Myr in the most massive,
high-density model: about a third of the initial binaries have been disrupted in
this case. This change appears to be driven by dynamical processing, rather than as
a consequence of stellar evolution.

We examine the properties of the populations of binaries, both the ones that have
been disrupted and the ones formed later during the evolution of the cluster. We
note that we have also looked for higher order multiples in the models (as defined
by the Nbody6++ code), however these are found to form in negligible numbers, on
the order of one per simulation in most models and a maximum typical number of four
in the \largeCloudH\ models. Such small number of multiples is expected given that
our initial conditions did not include them, and we see that forming stable
multiples by capture is a rare event in these models.

In Figure~\ref{fig:Bdisrupted} for simulations with $\sfeff=0.03$ we show the
average fraction of binaries that are disrupted at different times in the
evolution, i.e., from left column to right, at $t=$ 1, 3, 10, 20~Myr and at
$t=t_*$, with $\mcl =300\:M_\odot$ in red, $3000\:M_\odot$ in black and
$30,000\:M_\odot$ in blue, and with low- and high-\Sigmacl\ cases with solid and
dashed lines, respectively. The first row shows the fraction of disrupted binaries
as a function of initial primary mass. We see that at early times the mass of the
primary star has little influence on binary disruption. However, by 10~Myr the
disrupted fraction has risen for more massive stars, which is due to stellar
evolution, especially core collapse supernovae to neutron stars that then receive
high kick velocities.

The second row of panels in Figure~\ref{fig:Bdisrupted} shows the disrupted
fraction of binaries as a function of initial semi-major axis ($a$). The typical
semi-major axis in our models is around 20~AU. Below this value, most binaries
survive across the models, which is expected since these are relatively hard
binaries.  Wider binaries (i.e., with $a \gtrsim 100\:AU$) are the most affected,
with disruption fractions that depend sensitively on the environment, e.g., between
10\% to 80\% of binaries with $a\sim1000\:$AU are disrupted depending on the model,
where the main factor is the density of the environment as parameterized by
$\Sigmacl$. For the range above 100\:AU the disruption fractions are clearly
defined by density and parent clump mass. The most massive clusters show lower
disruption fractions within the same $\Sigmacl$, since these clusters have lower
initial number densities (see~\S\ref{sec:masstheory}). In general low $\sfeff$
results in a larger disruption fraction, with \sfeff\ having a larger effect on
low- and medium-mass models (see Appendix~\ref{ap:binaries}), with a variation of
25--40\% at $a=1000$\:AU. In the high-mass clusters with $\mcl=30,000\:\Msun$,
variations in disruption fractions are less than 5\% between different \sfeff\
cases. Most binary disruption happens early in the evolution, so that by 1 Myr most
of these features are already set.

We also explore the details of the dynamically formed binaries in the clusters. As
shown in Figure~\ref{fig:fbin}, up to 7\% of binaries in clusters with
$\mcl=300\:\Msun$ are formed dynamically, where most of these binaries are part of
the bound cluster component. The third row of Figure~\ref{fig:Bdisrupted} shows the
fraction of binaries that are dynamically formed as a function of primary initial
mass.  A clear trend appears where the more massive stars tend to capture other
stars more efficiently. This trend is strongest in the lowest mass clusters, which
undergo the highest degree of dynamical processing, and the fact in these low-mass
clusters, A, F and even G-type stars, can be the most massive stars in the cluster,
and be the ones segregating to the centre. Furthermore, since the overall velocity
dispersion is lower, then gravitational focusing is favored for less massive stars
in these environments. Again, the \sfeff\ parameters appear to play only a minor
role in the formation of binaries, as can be seen in the formation fractions at 20
Myr for other \sfeff\ (see Appendix~\ref{ap:binaries}). Most of the dynamically
formed binaries are wide binaries with semi-major axes larger than 1000 AU, as can
be seen in the fourth row of panels in Figure~\ref{fig:Bdisrupted}. Here we can
also see that higher primordial density favors the formation of tighter binaries,
since harder binaries are able to be perturbed allowing interchange of their
members.

Note that the results shown in the first to fourth rows in
Figure~\ref{fig:Bdisrupted} are fraction of binaries in each bin. The most affected
types of binaries, i.e., the wider and more massive ones, are in fact the less
populated parts of the binary distribution, representing only a small fraction of
the total number of binaries in the system. In the fifth row of
Figure~\ref{fig:Bdisrupted} we show the full distribution of binaries as a function
of semi-major axis as an average per simulation. Then after all stars are formed,
the average numbers in the \largeClouds\ case is 10 times larger than in the
\mediumClouds\ case and 100 times than in the \smallClouds\ case. We see that in
the \mediumClouds\ and \largeClouds\ models, dynamical disruption of binaries
causes a steeper decrease of frequency for binaries with large $a$, where the
fractional decrease is shown in the second row of panels. However, for
\smallClouds\ the dynamical formation of binaries is considerable ($\sim7\%$) and
concentrated in the high end of the distributions ($a>1000$\:AU), producing a
second peak at $a \sim 10^5$\:AU, with the strength of this peak being higher for
the larger \Sigmacl\ case.

\subsection{Evolution of the stellar mass function}\label{sec:massfunction}

The stellar mass function is expected to evolve within the clusters due to a
combination of mass segregation, binary formation/disruption, ejection of
walkaway/runaway stars and stellar evolution. The bottom rows of
Figures~\ref{fig:evolmass} and \ref{fig:evolmassss1} show the average system mass,
i.e., single stars, binaries and higher order multiples (although the latter are
negligible), in the bound cluster populations for systems with primary masses below
$7\Msun$ (i.e., so that these are not significantly affected by stellar evolution
during the period considered). For the assumed IMF and binary sampling methods,
this average system mass has an expected value of 0.25~\Msun, shown by a horizontal
gray dashed line in each of the panels. 

As the clusters evolve we see that small clusters show the largest deviation from
the expected value. Models \smallClouds\ show a remarkable variation in the average
system mass, i.e., rising by a factor of $\gtrsim 1.6$ in the high $\Sigma_{\rm
cloud}$ cases with $\epsilon_{\rm ff}\gtrsim 0.1$. This dramatic change is related
to the fact that these clusters evolve to have the smallest bound mass fractions,
i.e., $f_{\rm bound}\sim 0.3$ and undergo the most significant dynamical
processing, including significant formation of new binaries (see
\S\ref{sec:binary}). The variations in average system mass are more modest in the
\mediumClouds\ and \largeClouds\ models and move in the opposite direction, i.e.,
decreasing to lower values. We attribute this behavior to the fact that these
clusters retain high bound mass fraction and tend to destroy their primordial
binaries without forming significant numbers of new binaries.

\begin{figure*}
 \includegraphics[width=0.8\textwidth]{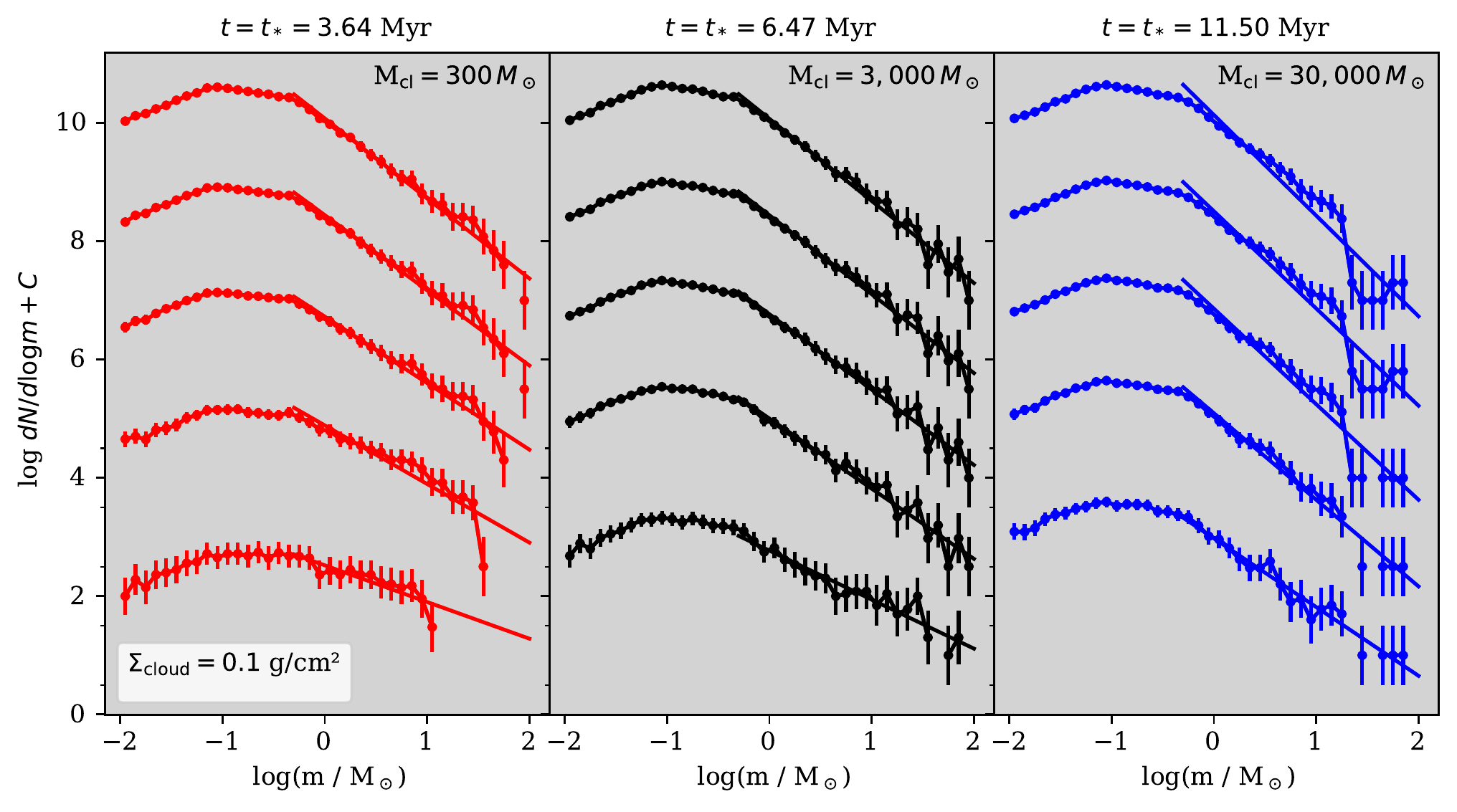} %
 \includegraphics[width=0.8\textwidth]{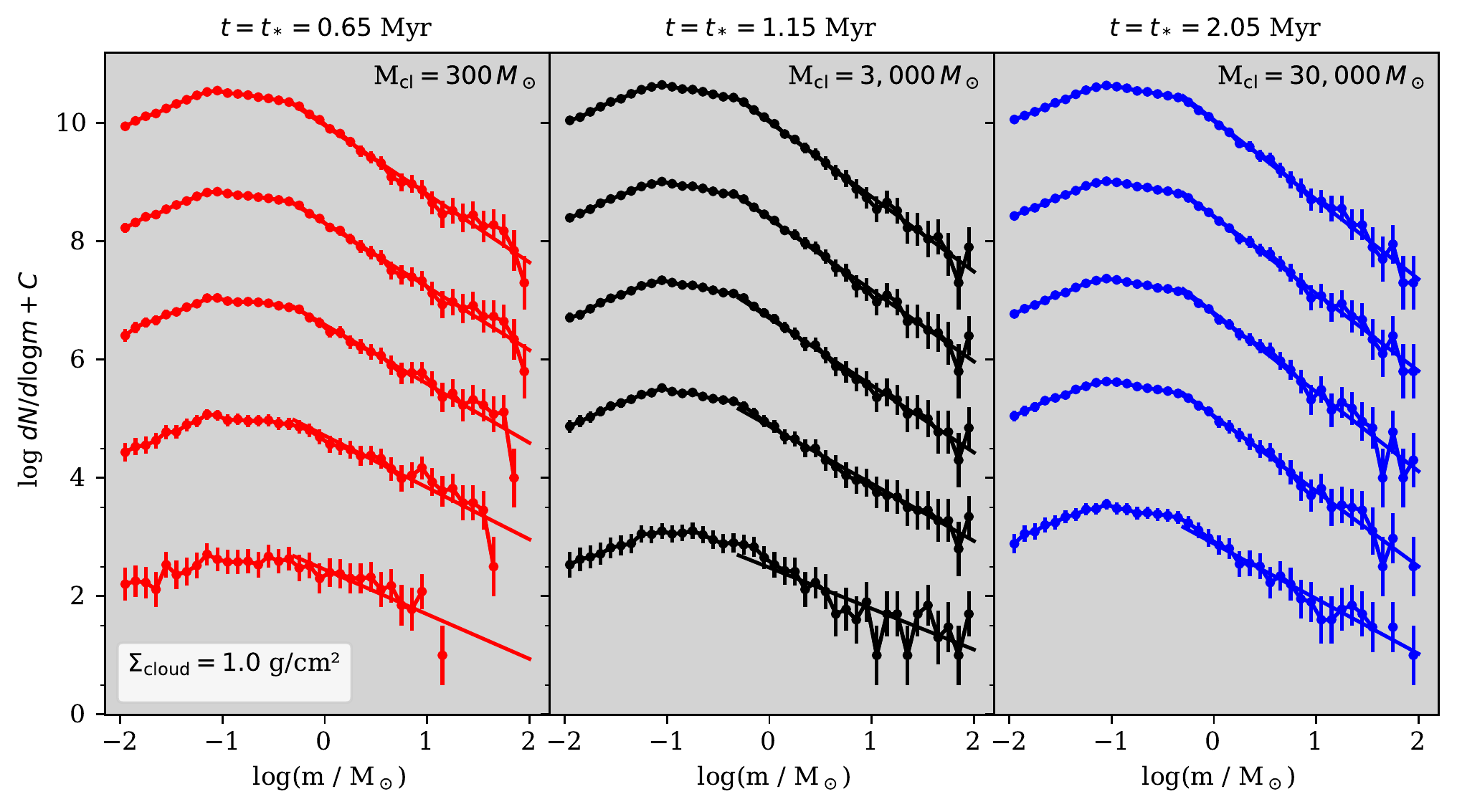}
 \includegraphics[width=0.9\textwidth]{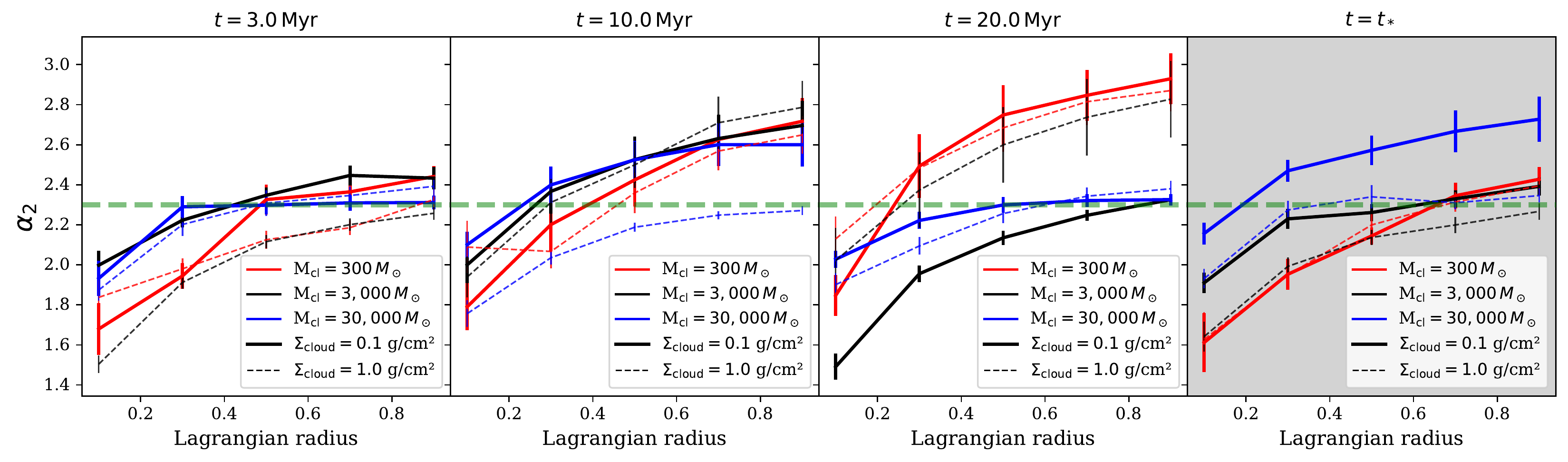}
 \caption{%
         Mass functions measured at different Lagrangian radii and times for star
         clusters with $\mcl=300\,\Msun$ (red), 3,000\,\Msun\ (black) and
         $30,000\mcl$ (blue). First and second rows show the MFs measured at the
         end of star formation ($\tsf$). The different MFs in each panel are
         measured within 0.1, 0.3, 0.5, 0.7 and 0.9 Lagrangian radii (from bottom
         to top). Bottom row shows linear fits to the range between 0.5-100\,\Msun
         measured at $t=$ 3, 10, 20 Myr and at $t=\tsf$. The horizontal green
         dashed line shows the input value of $\alpha_2=2.3$ from
         \protect\cite{KroupaIMF}.
 }
 \label{fig:mf}
\end{figure*}

\begin{figure}
 \includegraphics[width=\linewidth]{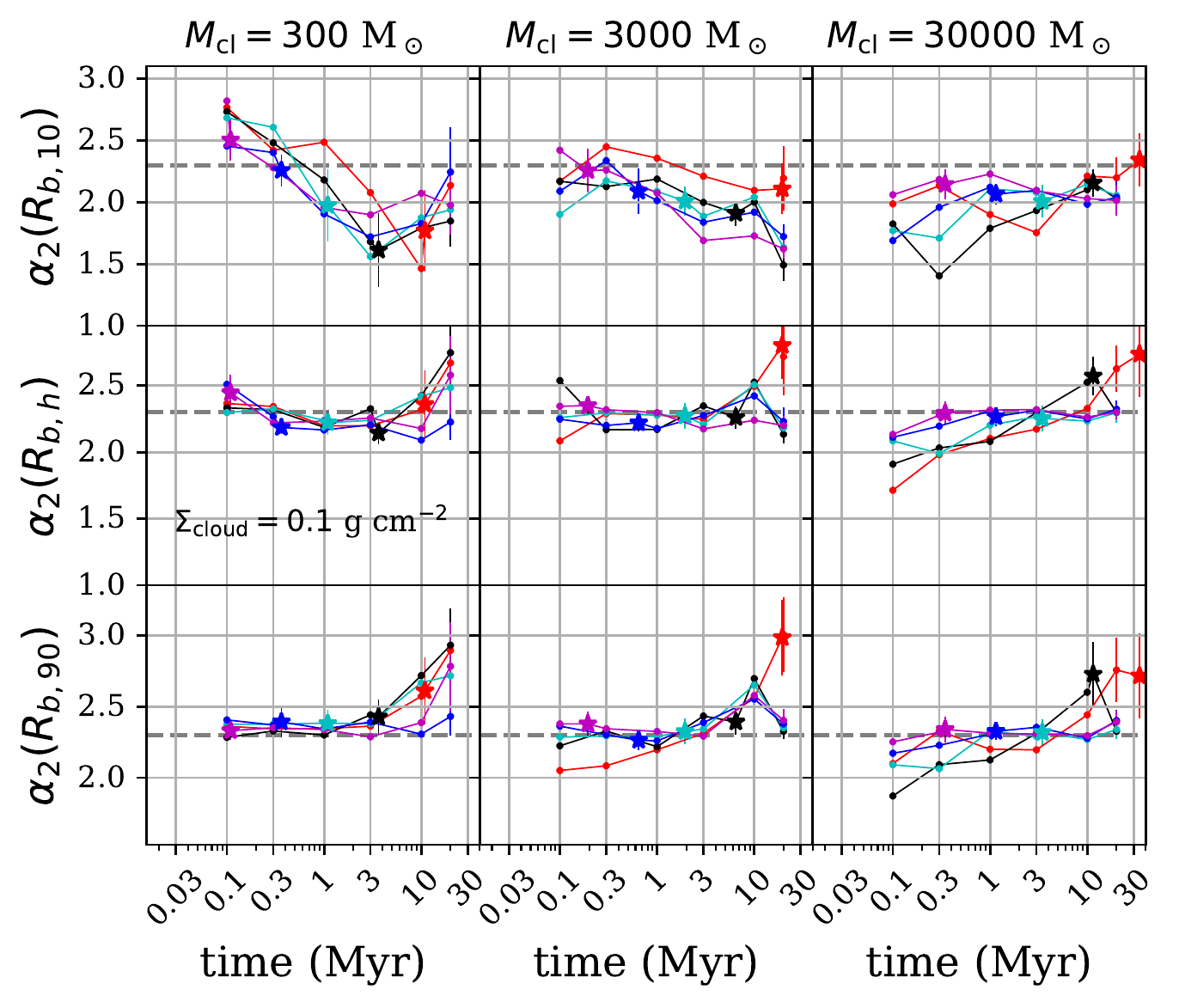}\\
 \includegraphics[width=\linewidth]{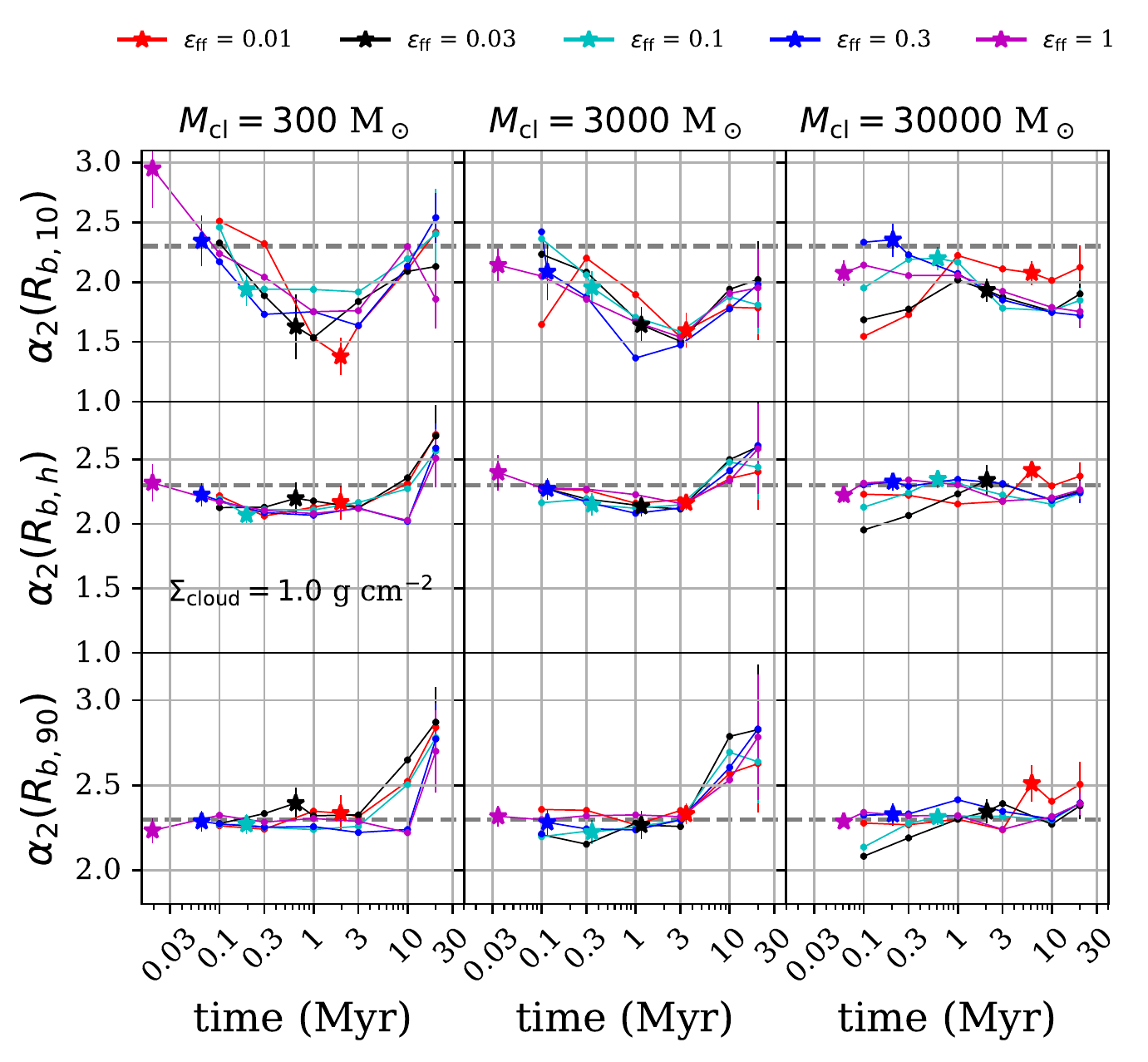}
 \caption{%
          Evolution of $\alpha_2$, i.e., the fitted mass function power law
          parameter for the mass range between 0.5--100\,\Msun. Top and bottom sets
          of panels show the results for the low and high-\Sigmacl\ cases,
          respectively, with different initial $\mcl$ cases shown in the three
          columns. The top row in each set shows $\alpha_{2,10}$, i.e., the
          $\alpha_2$ parameter measured within 10\% mass radius. Middle and bottom
          rows shows the same parameter measured at the 50\% ($\alpha_{2,50}$) and
          90\% ($\alpha_{2,90}$) mass radius respectively, The color scheme of the
          lines and points is the same as in previous figures representing the
          various \sfeff\ cases. Gray dashed line in each panel shows the input
          mass function in this range of $\alpha_2=2.3$.  Star symbols mark the
          moment when star formation is finished and background gas is exhausted. 
 }\label{fig:mfevol}
\end{figure}

Next, we examine signatures of mass segregation by considering the evolution of the
mass function slopes in the mass range above $\sim$1\,\Msun. We measure the stellar
mass function (MF) at different stages during the evolution of the modeled
clusters, using only the bound stars and excluding neutron stars and black-holes.
Figure~\ref{fig:mf} shows the resulting MFs when measured for stars within
different Lagrangian radii at different times for our fiducial models with
$\sfeff=0.03$. However, comparison between the models is complicated by the large
differences in formation and dynamical timescales for these clusters that have
orders of magnitude differences in mass and density. For instance, stellar
evolution plays a different role in each case when the formation and relaxation
times are comparable to the stellar evolution timescales of the most massive stars.
Effects due to ejection events derived from stellar evolution, i.e., velocity kicks
of neutron stars or binary breaking, are especially important.

In Figure~\ref{fig:mf} we first show the MFs at the end of the formation stage
($t=\tsf$), when all stars have formed and the clusters start their gas free-phase.
We pay special attention to the evolution of the high mass end of the stellar mass
function, i.e., the range between 0.5-100\,\Msun, which by construction we have
modeled with a canonical initial index of $\alpha_2 = 2.3$ \citep{KroupaIMF}. We
have performed linear fits to this range in logarithmic space as can be seen in
Figure~\ref{fig:mf} as solid lines within the fitting range.  The fourth panel of
the bottom row shows a comparison between the different $\alpha_2$ values obtained
at the different radii, $\mcl$ and $\Sigmacl$ for the case of $\sfeff=0.03$. The
same procedure was performed at 3, 10 and 20 Myr, where the corresponding fits are
shown in the first three panels of the bottom row. 

From the values of $\alpha_2$ as a function of enclosing Lagrangian radius, we see
that, by the end of star formation (fourth panel), the MFs tend to be more top
heavy in the central regions of the cluster. However, in the \largeCloudL\ model,
which has $\tsf$ of 11.5\,Myr, the population is already affected significantly by
stellar evolution at this time (these clusters contain $\sim$25 stars more massive
than 10\,\Msun). 

At 3~Myr top heavy MF signatures are most pronounced for the lowest mass clusters
and the higher $\Sigmacl$ cases, which, as discussed, have shorter relaxation times
and thus shorter mass segregation times. Note, the low $\Sigmacl$ cases are all
still forming stars at 3~Myr. In particular, the \largeCloudL\ model is at about
25\% of its $\tsf$ and has not yet developed significant mass segregation.
Similarly, the \mediumCloudL\ model is at about $\sim$50\% of $\tsf$ at this time
and also does not show strong mass segregation.

When we consider the MFs at 10 and 20 Myr we see that mass segregation signatures
are maintained and that even though stellar evolution mass loss removes some of the
excess of massive stars in the center, enough intermediate massive stars sink here
to keep the signatures present. In particular, the \mediumCloudL\ model has
developed the strongest top heavy feature at 20~Myr.  Therefore, we see that in all
models, the central regions of the bound systems tend to become top heavy
($\alpha_2 < \alpha_{2,i}$), rather than bottom heavy, regardless of stellar
evolution mass loss.

Figure~\ref{fig:mfevol} shows the evolution of the $\alpha_2$ parameter for all
models in this work measured at 10, 50 and 90\% Lagrangian radius, i.e.,
$\alpha_{2,10}$, $\alpha_{2,50}$, and $\alpha_{2,90}$ respectively. The signature
of mass segregation can be more clearly seen when analysing the 10\% mass radius
(top row of panels in each set). 

The evolution of the $\alpha_{2,10}$ parameter is stronger in the \smallClouds\
models since their crossing times are shorter. Also, due to IMF sampling in small
stellar clusters, the initial value of $\alpha_{2,10}$ is typically relatively
high. As massive stars migrate to the center, $\alpha_{2,10}$ decreases quickly.
The maximum level of mass segregation is reached at the point when the core
radii begin to expand, which does not happen at the end of star formation, but after
about one initial crossing time regardless of \sfeff\ (see \S~\ref{sec:evol}). The
$\alpha_{2,10}$ parameter then stabilizes at the onset of rapid expansion of the
cluster core. Nevertheless, the central region mass functions tend to remain top
heavy compared to the initial mass function. Note that eventually, at later stages,
$\alpha_{2,10}$ begins to increase due to the effects of stellar evolution.

\subsection{High velocity population}\label{sec:hvd}
\begin{figure*}
  \begin{minipage}[b]{0.48\textwidth}
   \centering
   \includegraphics[width=\textwidth]{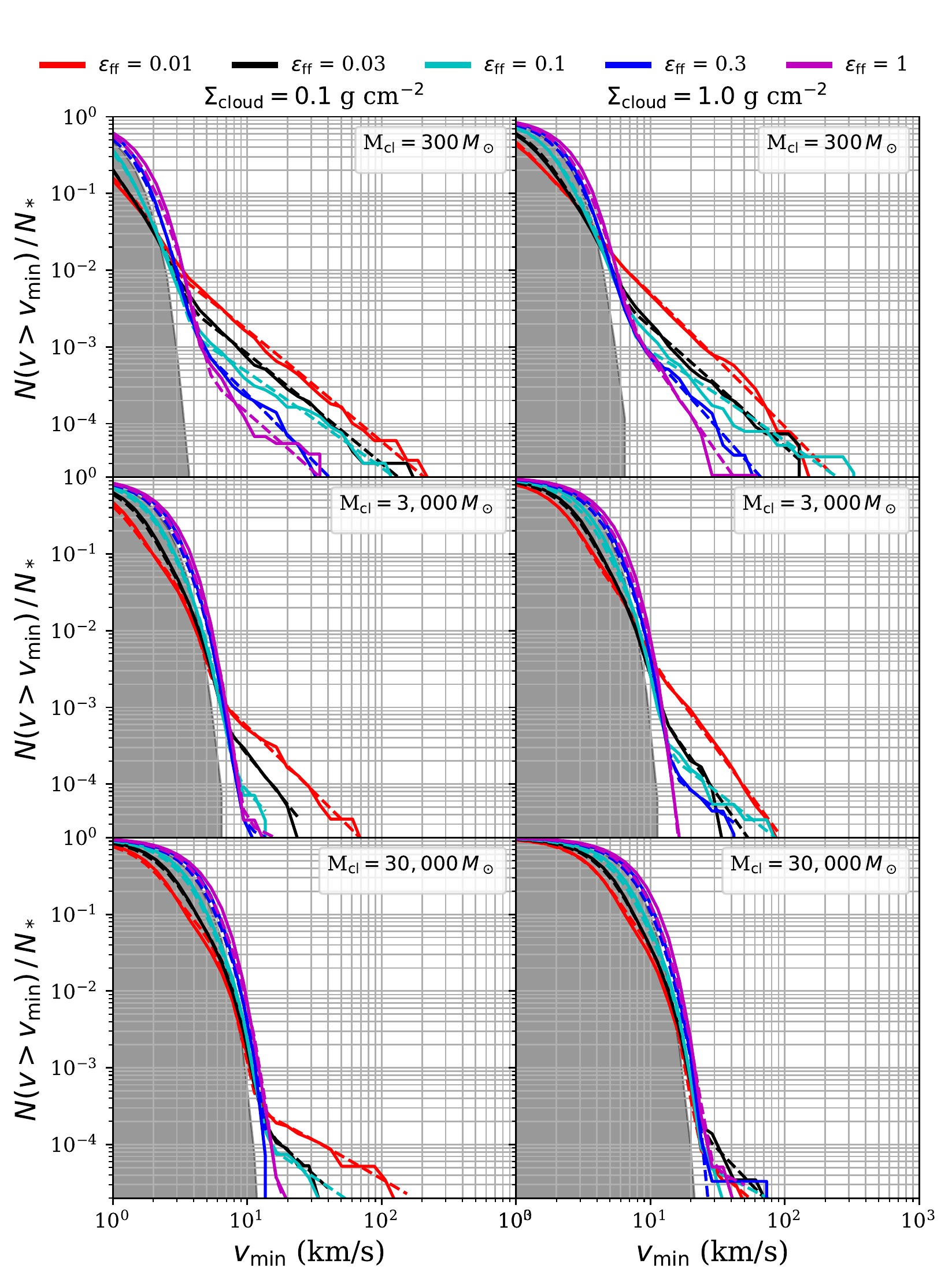}
   \textbf{(a)}
  \end{minipage}%
  \begin{minipage}[b]{0.51\textwidth}
   \centering
   \includegraphics[width=\textwidth]{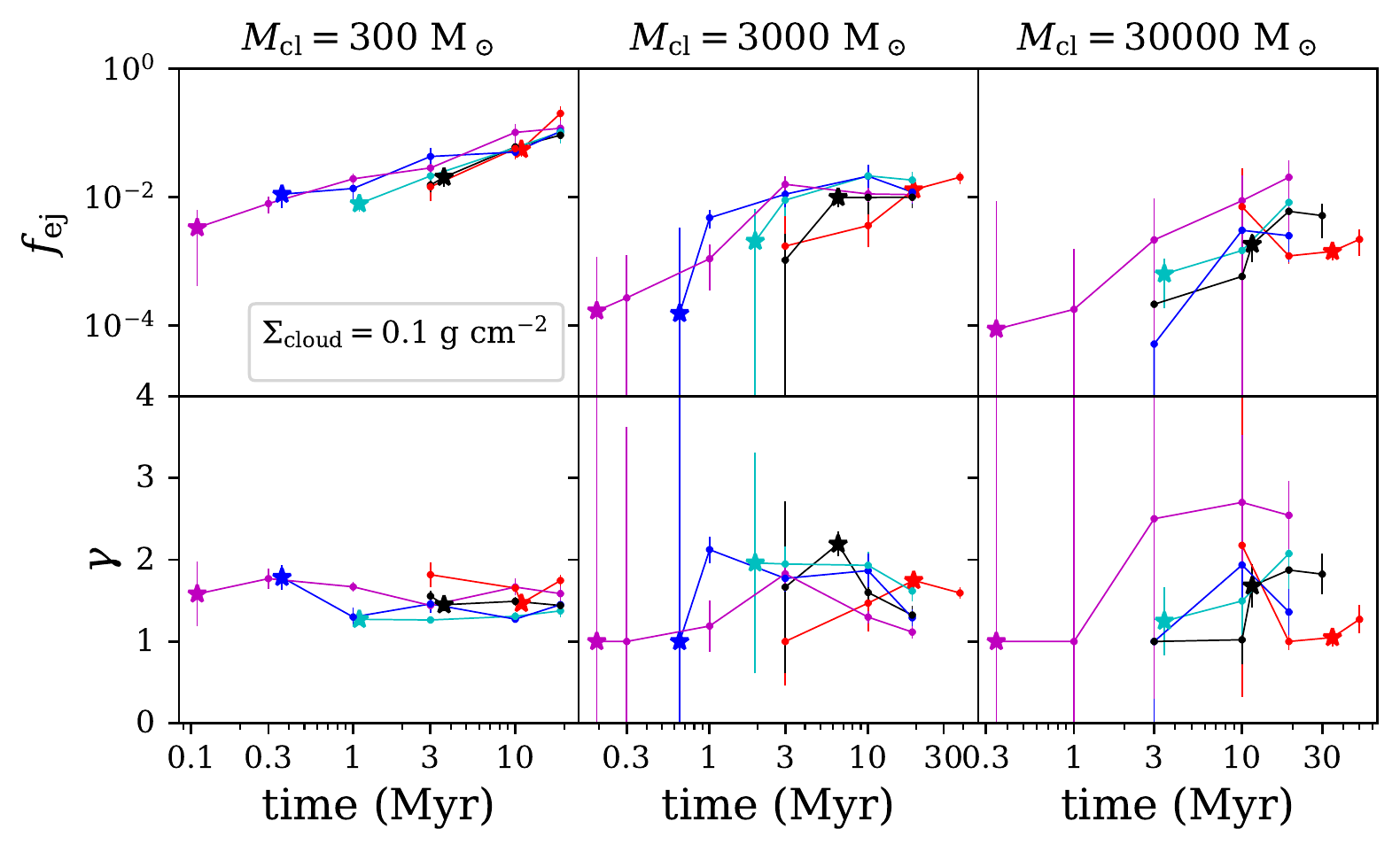}
   \includegraphics[width=\textwidth]{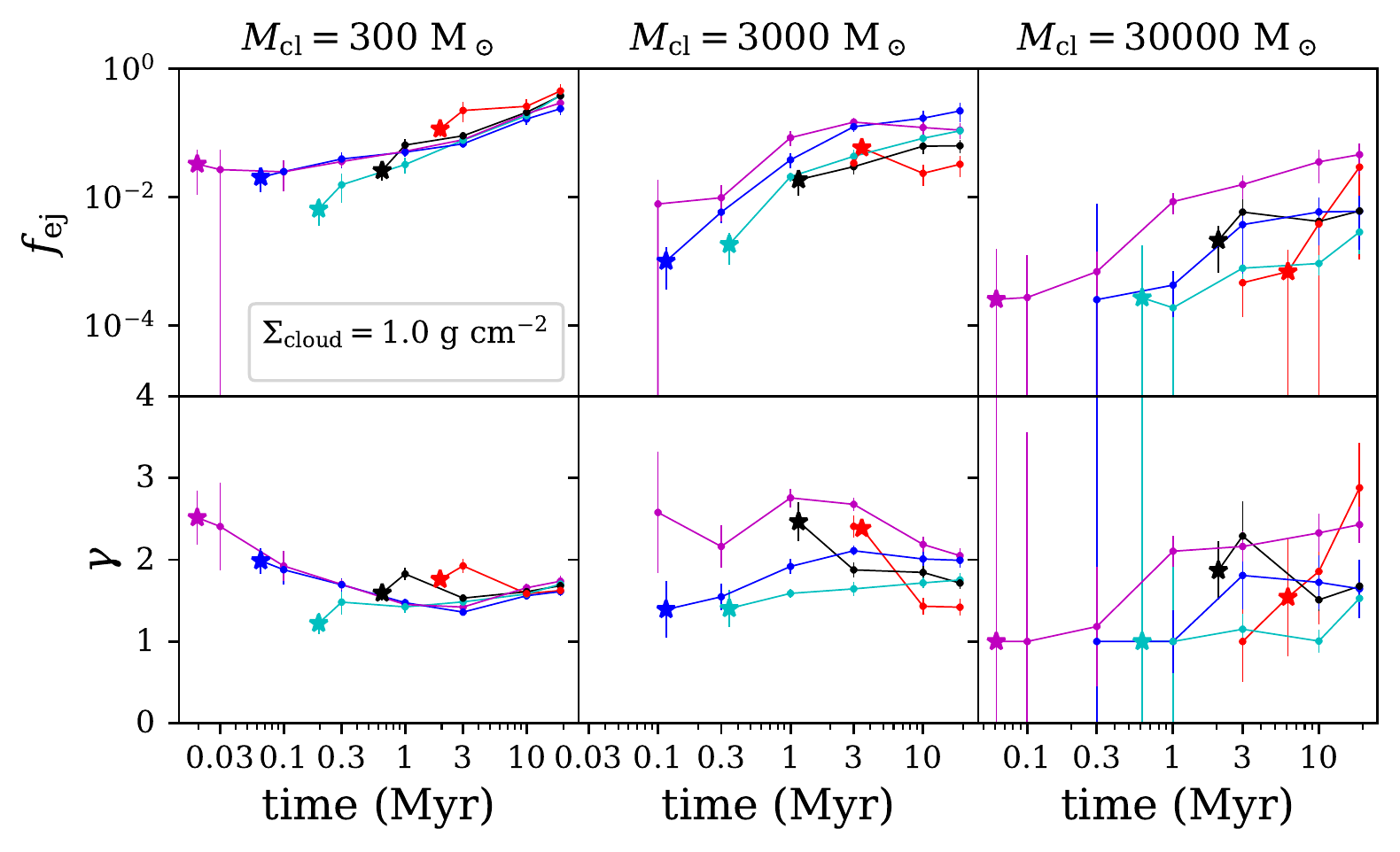}    
   \textbf{(b)}
  \end{minipage}
  \caption{%
          {\bf (a)}: Evolution of the transverse, i.e., 2-D, high velocity
          cumulative distributions normalized by the total number of stars in the
          clusters measured at $t=\tsf$.  Shaded area shows the respective
          Maxwell-Boltzmann distribution with $\sigma$ equal to the initial
          mass-averaged velocity dispersion of the clump.  Rows arrange different
          parent clump masses, \mcl\, while columns arrange the low (left) and high
          (right) \Sigmacl\ cases.  Note, stellar remnants are not included in this
          graph. This figure is equivalent to Figure~10 of
          \protect\cite{Farias2020}, where we make direct comparison of our
          simulations with estimates for the Orion Nebula Cluster. {\bf (b)}:
          Fitting parameters at different times for the excess in the velocity
          distribution of modeled star clusters, $f_{\rm ej}$ (top rows) and
          $\gamma$ (bottom rows) for simulations in the high-\Sigmacl\ (top set of
          panels) and low-\Sigmacl\ (bottom set of panels) regimes.  Star symbols
          show the time at which star formation is finished, where the
          corresponding fitting model is shown in figure (a).  Fitting is made at
          different times starting from $t=\tsf$. Note that the large errorbars
          represent times where fitting the excess is uncertain given that is not
          yet fully developed, e.g., see the high-\Sigmacl\ case of
          $\mcl=3000\,\Msun$ and $\sfeff=1$, i.e., magenta line in the lower right
          panel of figure (a).
        } \label{fig:hvdmass}
\end{figure*}

One important question we have explored during this series of papers is how the
star cluster formation process is linked to the properties of the unbound/ejected
population. We have shown in \paperII\ that slowly forming star clusters tend to
produce more high-mass runaway stars. However, these models have so far only been
for a single mass case of formation from a 3,000~$M_\odot$ clump. Here, we examine
how the high velocity distribution changes with mass and \sfeff\ in the framework
of our models. We expect the results to be useful for interpreting data of runaways
from young clusters, with a first application made for the 3,000~$M_\odot$ models
to the case of the ONC by \cite{Farias2020}.

Figure~\ref{fig:hvdmass}a shows the 2D (plane of sky) velocity distribution for the
low-\Sigmacl\ (left column) and high-\Sigmacl\ (right column) cases. Distributions
are constructed at the time when star formation stops ($\tsf$). The population of
dynamically ejected stars manifests itself as an excess of high velocity stars
relative to the expected Maxwell-Boltzmann distribution for the given velocity
dispersion. The initial expected velocity distribution, given the velocity
dispersion of the natal gas clump, is shown as shaded areas in each panel. As time
advances and gas is ejected, we have seen that the clusters expand and lower their
velocity dispersion. However, the fastest formation models do not have time to
relax and at $t=\tsf$ their velocity dispersion is very similar to the one at
birth. The resulting high velocity excess at $t=\tsf$ can be clearly seen, where
the slowest forming clusters show a more evolved velocity distribution, with a
lower velocity dispersion and greater fraction of high velocity stars. However,
since each cluster has a very different $\tsf$ it is difficult to make a fair
comparison between the models since we need to measure at different points in the
evolution. Below, we develop a simple model to describe the evolution of the
velocity distributions.

\subsubsection{Velocity distribution model}\label{sec:hvdmodel}

Following our analysis in \paperI\ and \paperII, we note that our modeled star
clusters are composed of three kinematially distinct components: (1) bound stars,
which are those with a negative total energy; (2) unbound gently ejected stars,
i.e., those that find themselves unbound given the rapid change in the protocluster
potential; and (3) dynamically ejected stars, which are those that are ejected as a
result of strong dynamical interactions. These groups have distinct velocity
distributions that together compose the total velocity distribution shown in
Figures~\ref{fig:hvdmass}a.  The bound component can be described with a cumulative
2D Maxwell-Boltzmann velocity distribution function
\begin{eqnarray}
        {\rm CDF}(v,a) &= &   1-\exp{\left( - \frac{v^2}{2a^2} \right)},
\end{eqnarray}
with the scale parameter given by
\begin{eqnarray}
        a =  \sqrt{\frac{2}{4-\pi} } \sigma_{b}.
\end{eqnarray}
Note that for practical purposes, given the we are most interested in the high
velocity tail of the distribution where the numbers of stars are low, we instead
use the survival function SF$(v,a) = 1 - $CDF$(v,a)$, which is the function shown
in Figure~\ref{fig:hvdmass}a. The gently ejected component can be modeled with the
same distribution, but with a larger velocity dispersion, which is a remnant of the
dynamical history of the cloud. Therefore, the bound and unbound components are
both described by a SF of the form:
\begin{eqnarray}
        {\rm SF}(v,a) &= &  \exp{\left( - \frac{v^2}{2a^2} \right)}.
\end{eqnarray}

The dynamically ejected stars follow a different distribution, i.e., approximately
a power law tail in the velocity distribution profile with an exponent $\gamma$.
Then, we model the SF of this component as
\begin{eqnarray}
        {\rm SF_{ej}}(v,a) &=& \frac{1}{1+ \left(  \frac{v}{a}  \right)^\gamma}.
        \label{eq:sfej}
\end{eqnarray}

The survival function of the total velocity distribution is thus
\begin{eqnarray}
        {\rm SF}(v) &=& \fbound {\rm SF_{b}}(v,a_{\rm b}) \nonumber\\
        & &+    f_{\rm unbound} {\rm SF_{ub}}(v,a_{\rm ub})  \nonumber \\
        & &+    f_{\rm ej} {\rm SF_{ej}}(v,a_{\rm ub}), \label{eq:hvdmodel}
\end{eqnarray}
where $\fbound + f_{\rm unbound} + f_{\rm ej} = 1$. Note that the bound component
with $\fbound$ and $\sigma_{\rm b}$ is measured directly from the stellar
distribution (see Figures~\ref{fig:evolmass} and \ref{fig:evolmassss1}).  For the
second component, even though we have left $a_{\rm ub}$ as a free parameter, we
have found that this parameter is well represented by the scale parameter obtained
using the velocity dispersion of the parent clump, $\sigmacl$. Then, the fitting
procedure is dominated by the ejected component described in Equation~\ref{eq:sfej}
and its weight, i.e., $\gamma$ and $f_{\rm ej}$. For the scale parameter of this
component we have used the same as for the unbound, $a_{\rm ub}$, since we want the
power law signature to be fully developed at the velocity when the unbound
Maxwell-Boltzmann distribution becomes unimportant.

Figure~\ref{fig:hvdmass}b shows the time evolution of $\gamma$ and $f_{\rm ej}$.
The evolution of $f_{\rm ej}$ shows how the fraction of dynamically ejected stars
grows with time. Small clusters with $\mcl=300\,\Msun$ show larger $f_{\rm ej}$
values, with a similar evolution independent of \sfeff, but mostly dependent of the
age of the clusters and the initial density. At 20 Myr, these small clusters reach
$f_{\rm ej}\sim 0.08-0.2$ for low-\Sigmacl\ models and $0.25-0.45$ in the
high-\Sigmacl\ case.  As $\mcl$ increases, the importance of the ejected population
decreases to a range between $0.01-0.1$ for $\mcl=3000\,\Msun$ and between
$0.001-0.04$ in the most massive clusters.

However, we find no clear trend for the evolution of the power law parameter
$\gamma$, neither with $\mcl$ or \sfeff. Rather than being dependent of global
parameters, $\gamma$ is more likely to depend on the population of binaries as
shown by~\cite{Perets2012}. In our case we obtained an average value of $\gamma =
1.6\pm0.4$. 

\subsection{Runaway stars} \label{sec:runaways}
\begin{figure*}
 \includegraphics[width=\textwidth]{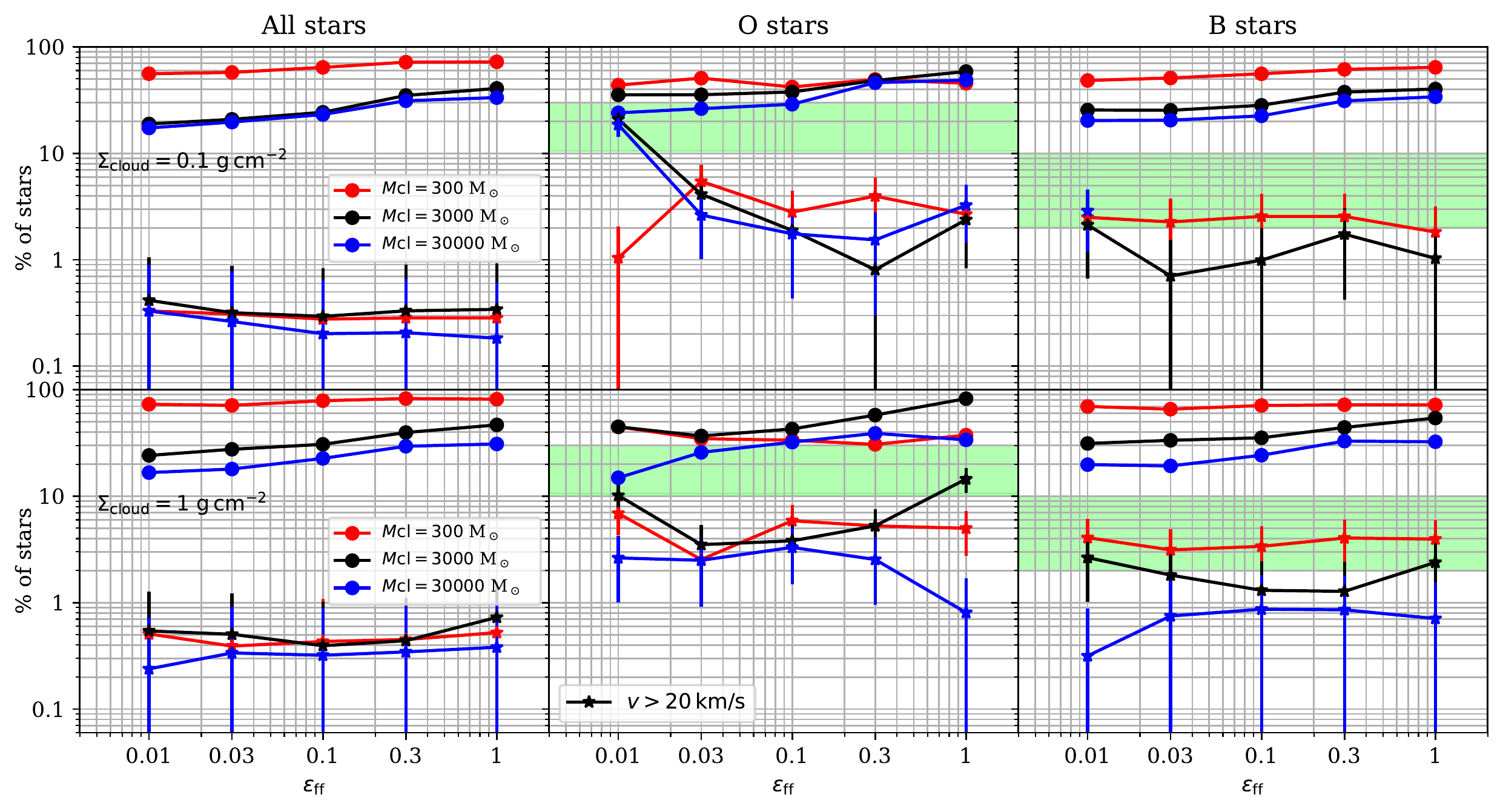}
 \caption{%
          Percentage of ejected stars relative to all stars in each mass range for
          the low (top panels) and high (bottom panels) $\Sigmacl$ cases
          measured at 20 Myr. The first
          column shows the results when using all stars in the set. The second and
          third columns show the fraction of O and B dynamically ejected stars,
          respectively. Shaded areas in these panels show the range of observed
          values (see text). 
 }
 \label{fig:runaways}
\end{figure*}

While the above description is useful at characterizing the different components of
the stellar distribution, compiling such population data is challenging, especially
for the high velocity lower-mass stars that are now far from their origins and thus
hard to find and link to a given population. However, isolated runaway stars are
easier to find, especially O and B stars. Observations of O and B runaway stars
indicate that between 10-30\% of O stars and 2-10\% of B stars are runaways
\citep{Gies1987,Stone1991,Dewit2005}, depending on precise definition of this
class.

Figure~\ref{fig:runaways} shows the percentage of ejected stars per model  without
a velocity cutoff (filled circles), and with velocities above 20\,\kms that we
adopt as a definition for a runaway star. These numbers represent only dynamically
ejected stars via strong interactions or rapid change in the cluster potential,
excluding supernovae related ejections. 

We show the results for three ranges of mass: all stars in the system (left
column); O stars (middle column); and B stars (right column). As found in our
previous work in Paper II, there is a modest increases in the fraction of overall
ejected stars with $\sfeff$, shown in the first column of Figure~\ref{fig:runaways}
as filled circles, which is a result of the increasingly rapid depletion of the
background gas.

The results for high velocity runaway stars appear to be divided into two regimes,
determined by $\Sigmacl$. In the low-$\Sigmacl$ case (top panels), slowly forming
clusters appear to form slightly higher fractions of runaway stars at all masses,
especially for the most massive clusters. However, the differences are modest and
within the uncertainties. 

With the exception of the $\smallCloudL$ case at $\sfeff=0.01$, we find more O
runaway stars when $\sfeff$ approaches to 0.01, as expected given the longer time
stars remain in a dense state during the formation phase, but increase again when
$\sfeff=1$. The former is a consequence of the high peak density reached at the
beginning, given that all stars formed in half a free fall time and collapse into
the center together. 

For B stars the fraction of runaway stars appears to be independent of $\sfeff$,
but with strong dependence on $\mcl$. Small clusters, with higher initial
densities, form higher fractions of B star runaways than the most massive clusters.
If fact, in the low-$\Sigmacl$ case \largeCloudL\ models form no runaways in the B
mass range, except for the slowly forming case with $\sfeff=0.01$.

In the high-$\Sigmacl$ case results appear to be dominated by the high density
environment, and similar fractions of runaway stars are found in each model. In
this case B star runaways are found in all models, but the trend remains the same
as more massive clusters produce smaller fractions of B star runaways.

About 2 and 4\% of B stars are ejected with high velocities for the low and high
$\Sigmacl$ cases, respectively. These figures are consistent with the range of
values found for B stars \citep[see, e.g.,][]{Eldridge2011}. Increasing $\mcl$
brings down the number of high velocity B star runaways down to 0.5-1\% in the
high-$\Sigmacl$ case, and none in the low-$\Sigmacl$ case (with the exception of
the $\sfeff=0.01$ case, where we find 2\%). These results highlight the high
densities reached by the low mass clusters at formation, but the subsequent quick
expansion implies that most of these high velocity ejections happened very early in
the evolution of these systems.

\subsubsection{Interaction rates to produce dynamical ejections}
\begin{figure}
 \centering
 \includegraphics[width=\linewidth]{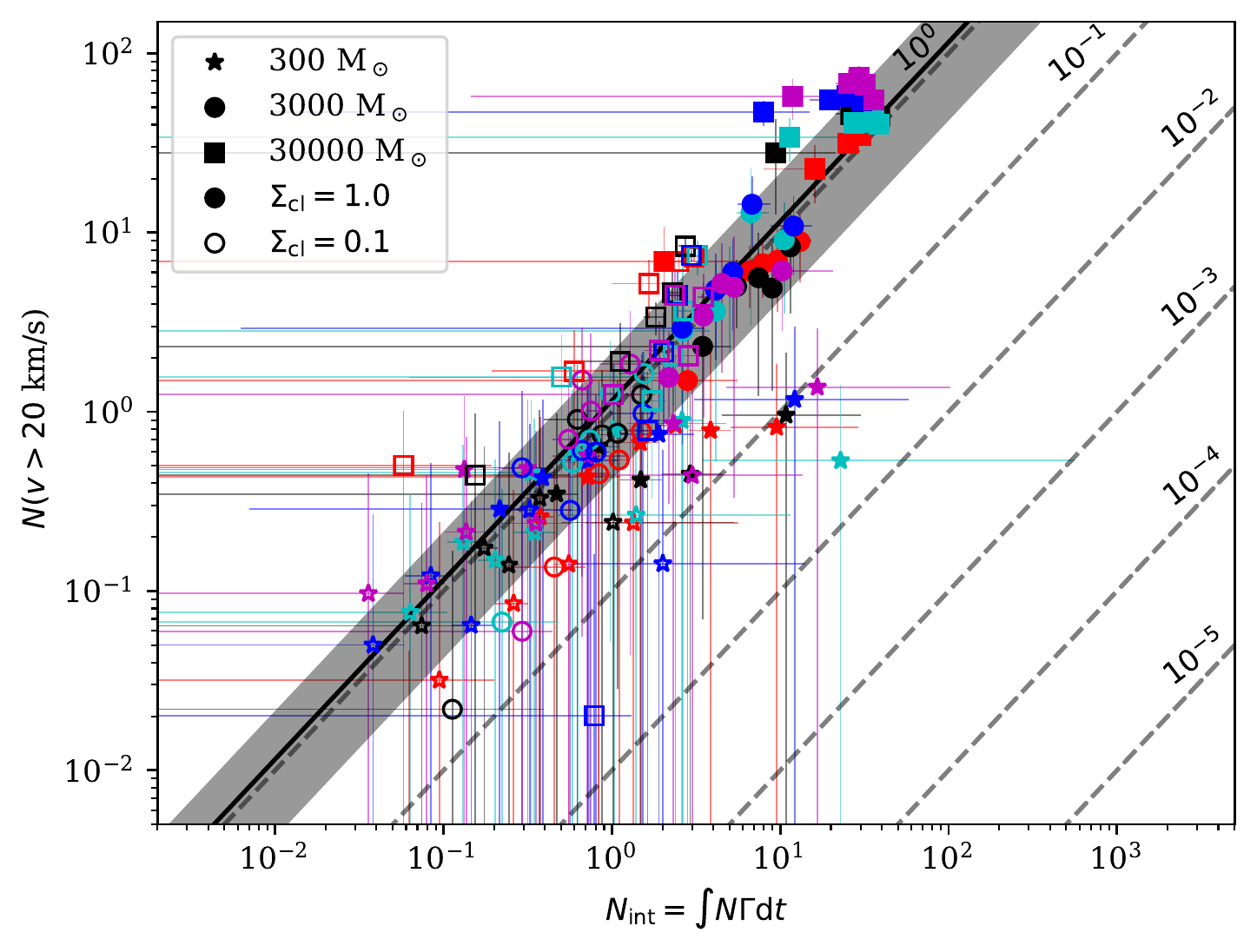}
 \caption{%
          $N_{\rm int}$ versus the number of dynamically ejected high velocity
          stars, collected for all models in this work. Each point represents the
          average number of high velocity stars for measurements with the same
          $\sfeff$ and $\mcl$ that fall in the same $N_{\rm int}$ bin. 
 }\label{fig:Nint}
\end{figure}

We estimate the number of ejected stars we expect at a given time, given the
dynamical history of a star cluster.  For single stars, we can estimate the cross
section $\pi b^2$ of interactions that result in a closest approach of $\bmin$,
where the velocity reaches a value of $v_{\max}$. For two stars approaching from
infinity with relative velocity $\sigma$ and impact parameter $b$, energy and
angular momentum conservation implies that a star reaches a closest approach at a
maximum velocity $v_{\max}$ when:
\begin{eqnarray}
        b &=&  \frac{2G\mt}{ (v_{\max}^2 - \sigma^2) } \frac{v_{\max}}{\sigma},
    \label{eq:b}
\end{eqnarray}
where \mt\ is the total mass of the interacting stars. Note that this result is
only valid for $v_{\max} \geq \sigma$. The $v_{\max}/\sigma$ factor is the
gravitational focusing factor, which increases the impact parameter $b$ in low
velocity environments.  Then, the interaction rate for interactions that can
potentially eject a star, is given by  $\pi b^2 \times n_s \times \sigma$, i.e.,
\begin{eqnarray}
\Gamma &=& 4 \pi n_s\sigma \left[ 
            \frac{G\mt}{ (v_\text{max}^2 - \sigma^2) } \frac{v_\text{max}}{\sigma}
    \right]^2.
\end{eqnarray}

The total number of interactions that will accelerate stars up to $v_\text{max}$,
from $t=0$ to $t=T$ is:
\begin{eqnarray}
   N_{\rm int} = \int_0^T \sum_{i=1}^N \Gamma_i(t) d t, 
   \label{eq:nint}
\end{eqnarray}
where $\Gamma_i$ is the interaction rate of each star in the system.  This number
should be proportional to the number of observed runaway stars with escape
velocities $v>v_{\max}$, i.e., $N_v$. We calculate this proportionality factor in
our models by numerically integrating Equation~\ref{eq:nint} and then comparing to
the number of runaway stars ejected with velocities greater than 20 \kms.
Specifically, we numerically integrate Equation~\ref{eq:nint} over time for each
individual simulation at each snapshot, constructing tracks in the $N_{\rm
int}$-velocity plane.  We calculate $\Gamma_i$ using global values of $\sigma$ and
$n_s$, based on the stellar population within the half mass radius of the system.
We consider as runaway stars any star with $v > v_{\max} = 20\,\kms$ that is not a
stellar remnant. To avoid contamination of stars close to the center of the
potential well, where local escape velocities may be large due episodic close
encounters, we only consider stars beyond 2 stellar half mass radius when counting
the number of runaway stars.  We combine the results of each set of models by
placing all $\Nint$-$v$ data pairs of each simulation at each snapshot on a single
combined set. We divided this set in \Nint\-bins of equal number of data-points and
take the average of $N_v$.  Figure~\ref{fig:Nint} shows these results for all the
different sets of models in this work.  It is expected that $\Nint\propto N_v$ and
this is approximately seen in Figure~\ref{fig:Nint}. From these results we
calculate a proportionality factor $\eta = 1.15^{+1.0}_{0.7}$.  This shows that the
number of runaway stars is a direct constraint on the dynamical history of a star
cluster. For instance, this linear relation indicates that if a star cluster is
characterized with 10 stars with $v > 20\,\kms$, then models trying to reproduce
such a system must reach the same number of strong interactions (on average) during
its age.  

\section{Discussion}

We have expanded our modeling of star cluster formation to cover a wide range of
masses of initial gas clumps and studied their resulting star clusters. This step
is important to eventually produce a comprehensive grid of models that could be
applied to interpret populations of star clusters. Given the assumptions of the
Turbulent Clump Model, certain scalings of properties occur as a function of clump
mass (at fixed mass surface density of surrounding cloud environment). These
scalings, e.g., of density, velocity dispersion, etc., have implications for the
dynamical evolution of the resulting stellar system, for instance, affecting the
relaxation time and degree of binary processing.

The evolution of cluster radii (e.g., half-mass radii of the bound components) and
radial profiles with time are metrics that can be compared to observed systems,
although this will typically be best achieved by converting model mass profiles
into multiwavelength light profiles. In our modeling program, this step is deferred
to a future paper in the series, requiring implementation of pre-main sequence
evolutionary tracks into the simulation framework. However, in principle, a cluster
that is observed to have a given (mean) age, mass, half-mass (or light) radius and
radial mass (or light) profile can be compared with the outputs of the models
presented here. There will be a range of formation parameters ($\mcl$, $\Sigmacl$,
$\epsilon$, $\sfeff$, etc.) that are consistent with a given set of observational
data. The grid of models presented here is a first step in the process of building
tools that will eventually allow constraints to be placed on formation parameters
from observed clusters, which could then be used to estimate initial cluster and
clump mass functions and distributions of $\Sigmacl$ formation environments.

We find variations in cluster sizes with $\sfeff$ for models of the same initial
mass and environmental mass surface density. This is due to the potential of the
natal gas clump restricting the expansion that arises from the initially
supervirial state. However, these differences in sizes are most apparent only for a
relatively short time period similar to the star formation time, since after this
rapid cluster expansion occurs and the various models tend to converge to have
similar sizes at a given age.

Given such degeneracies, direct measures of $\sfeff$, i.e., via measurement of age
spreads, remains important. However, accurately measuring age spreads in young star
clusters is a challenge that involves model-dependent fitting to pre-main-sequence
evolutionary tracks \citep[e.g., ][]{Bressan2012,Tognelli2011,Baraffe2015} and is
also complicated by observational uncertainties in extinction, photometric
variability and unresolved multiplicity \citep[e.g.,][]{DaRio2016}. Associating
individual runaway stars with a given young cluster and/or spreads in kinematic
expansion ages is another, more direct, method for estimating $\sfeff$
\citep[e.g.,][]{Tan2006,Farias2020}.

We have shown that properties that depend on the time-integrated density of the
stellar systems, i.e., amount of binary processing and fraction of dynamically
ejected stars, do have significant sensitivity to $\sfeff$. These tend to show the
strongest variations in small clusters, due to their short relaxation times.
However, testing models via observations of such clusters faces the inherent
problem of small numbers of stars leading to larger sampling uncertainties.
Overcoming this would require observations of large numbers of low-mass clusters.

Independent from $\sfeff$, we have found significant variations of behaviour
between low-mass and high-mass forming clusters, which are mostly due to the
differences in their relaxation times during the formation phase. Even
slowest-forming ($\sfeff\sim0.01$) low-mass clusters evolve to have relatively low
bound mass fractions ($\lesssim 0.6$ for $\Sigma_{\rm cloud}=0.1\:{\rm
g\:cm}^{-2}$; $\lesssim 0.4$ for $\Sigma_{\rm cloud}=1.0\:{\rm g\:cm}^{-2}$) by
$\sim 20\:$Myr and that are continuing to decline quickly, while in the higher-mass
systems $f_{\rm bound}$ can remain as high as $\sim 0.8$ at these times and with
much more gradual rates of decline. Related to this, lower-mass clusters are able
to form a more significant high velocity population of dynamically ejected stars.
While all of the clusters show mass segregation leading to a more top heavy
high-mass end mass function slope in their bound, central components, this effect
is stronger in lower-mass clusters. Finally, the average system mass in the bound
remnants of low-mass clusters shows significant evolution to higher values, partly
driven by significant numbers of dynamically formed binaries. In more massive
clusters, such binary formation is rare and binary processing tends to destroy the
primordial binary population, e.g., from $f_{\rm bin}$ of 1/2 down to as low as
$\sim1/3$ in the most massive, highest density clusters considered. These are
significant variations that may be testable by future observations of young
clusters.

There are a number of caveats and limitations of the models that we have presented.
The protocluster models are globally spherically symmetric and lack spatial and
kinematic substructure that might be expected to arise from interstellar
turbulence. Allowing for such features is planned in future papers in this series.
Furthermore, in the models presented here, higher order multiples were not part of
the initial conditions setup and their formation by capture was negligible. If
significant fractions of triple and higher-order multiple systems are found to
exist in young clusters, then this would indicate a need to incorporate such
systems as part of the primordial population.

\section{Conclusions}
We have presented a set of star cluster formation simulations that span a wide
range of initial clump masses ($M_{\rm cl}$ from 300 to 30,000~$M_\odot$), cloud
environment mass densities ($\Sigma_{\rm cloud}$ from 0.1 to 1.0~$\rm g\:cm^{-2}$)
and star formation efficiencies per free-fall time (\sfeff\ from 0.01 to 1.0). These
simulations, all involving global star formation efficiency of 50\% and all
starting with 50\% primordial binaries, follow the n-body dynamics of the stellar
populations, including evolution of the bound cluster, binary properties, mass
segregation and production of high velocity runaways.

We summarize our main results as:

\begin{itemize}
\item Bound mass fractions at the end of star formation are similar in all
        models, i.e., around 90\% (see \S~\ref{sec:evol}). However, the
        subsequent evolution diverges dramatically depending on $M_{\rm cl}$ and
        $\Sigma_{\rm cloud}$, with low-mass clusters in high-density environments
        retaining the smallest fractions ($f_{\rm bound}\lesssim 0.3$) in their
        remnant bound cores. In general, slowest forming clusters retain higher
        bound fractions.
        
\item The evolution of half-mass radii of the bound clusters also shows large
        differences in behaviour depending on cluster mass and environment.
        Low-mass clusters in high density environments undergo the largest degree
        of expansion during the first 20~Myr of evolution, since they form
        relatively quickly and have short relaxation times that drive this
        dynamical evolution. Variations with
        \sfeff\ are mostly related to the length of the formation phase, during
        which the gravitational potential of the gas clump acts to confine the
        cluster, retarding its expansion. Once the gas is exhausted, clusters can
        enter a post formation stabilization phase, during which they have
        relatively constant sizes (see \S~\ref{sec:evol} and
        Figure~\ref{fig:rhdens}). This phase ends once the cluster has had time to
        undergo dynamical relaxation, which leads to further expansion. This delay
        in expansion means that clusters of a given $M_{\rm cl}$ and $\Sigma_{\rm
        cloud}$ have similar sizes by a time of $\sim 20\:$Myr.

\item The core radius evolution is independent of $\sfeff$ and remains relatively
        constant for about one crossing time. If gas is still present in the
        system, the core radius can remain dense for longer (i.e., $\sfeff <
        0.03$). The expansion of the core radius sets the end of the post formation
        stabilization phase that star clusters undergo after gas expulsion.

\item The above results imply that binary systems are disrupted efficiently in the
        most massive cluster during the initial $\sim$20\:Myr period that has been
        modeled here. However, in lower-mass systems, binary disruption is
        constrained to the formation time only, given their quick
        post-gas-expulsion expansion (see \S~\ref{sec:binary}). Most
        disrupted binaries have semi-major axes greater than 100 - 1000\:AU,
        depending on $\Sigmacl$ (Figure~\ref{fig:Bdisrupted}). Lower-mass
        systems can disrupt harder binaries relative to the most massive clusters,
        given their high initial densities. Binary formation by capture is more
        efficient in lower-mass systems (see Figure~\ref{fig:fbin}). By
        20\:Myr about 6-8\% of binaries are formed by capture in the bound systems.
        This figure drops dramatically for clusters with $\mcl=3,000\:\Msun$, i.e.,
        is below 1\% and practically zero in more massive systems. Binaries formed
        by capture are concentrated at higher end of the semi-major axis
        distribution, showing a noticeable secondary peak in at $a=10^5$\:AU at
        20\:Myr. These binaries are formed mainly after gas-expulsion during the
        expansion of the cluster.

\item Young star clusters develop different levels of central mass segregation
        reaching a peak at the time the core radius begins to expand (see
        \S\ref{sec:massfunction}). The short dynamical timescales of
        clusters with $\mcl=300\:\Msun$ and small IMF sampling, causes these
        systems evolve to have the most top heavy central regions in relation to
        their outskirts. 

\item The fraction of dynamically-ejected stars depends on the initial
        mass of the clump and the mass surface density of its environment
        (\S\ref{sec:runaways}). Low-mass clusters produce greater
        fractions of ejected stars, i.e., ranging from 8 to 20\% in the
        low-\Sigmacl\ case and 25 to 45\% in the high-\Sigmacl\ case. 

\item  The percentage of runaway stars, i.e., dynamically ejected stars, follows
        the same dependence, but differences are modest. B stars, however, show the
        greatest differences across $\mcl$, where low-mass clusters are able to
        reproduce observed percentages, with an average of 2.5\% in the
        low-$\Sigmacl$ case and 4\% in the high density environment.

\end{itemize}

\section*{Acknowledgments}
JPF was supported by NSF Career grant No. 1748571 and NASA grant 80NSSC20K0507.
JPF and JCT acknowledge support from ERC Advanced Grant project MSTAR.

\section*{Data availability}
The data underlying this article will be shared on reasonable request to the corresponding author.

\bibliographystyle{mnras}
\bibliography{references} %

\begin{thebibliography}{}
\makeatletter
\relax
\def\mn@urlcharsother{\let\do\@makeother \do\$\do\&\do\#\do\^\do\_\do\%\do\~}
\def\mn@doi{\begingroup\mn@urlcharsother \@ifnextchar [ {\mn@doi@}
  {\mn@doi@[]}}
\def\mn@doi@[#1]#2{\def\@tempa{#1}\ifx\@tempa\@empty \href
  {http://dx.doi.org/#2} {doi:#2}\else \href {http://dx.doi.org/#2} {#1}\fi
  \endgroup}
\def\mn@eprint#1#2{\mn@eprint@#1:#2::\@nil}
\def\mn@eprint@arXiv#1{\href {http://arxiv.org/abs/#1} {{\tt arXiv:#1}}}
\def\mn@eprint@dblp#1{\href {http://dblp.uni-trier.de/rec/bibtex/#1.xml}
  {dblp:#1}}
\def\mn@eprint@#1:#2:#3:#4\@nil{\def\@tempa {#1}\def\@tempb {#2}\def\@tempc
  {#3}\ifx \@tempc \@empty \let \@tempc \@tempb \let \@tempb \@tempa \fi \ifx
  \@tempb \@empty \def\@tempb {arXiv}\fi \@ifundefined
  {mn@eprint@\@tempb}{\@tempb:\@tempc}{\expandafter \expandafter \csname
  mn@eprint@\@tempb\endcsname \expandafter{\@tempc}}}

\bibitem[\protect\citeauthoryear{Aarseth}{Aarseth}{2003}]{Aarseth2003}
Aarseth S. S.~J.,  2003, {Gravitational N-Body Simulations}.
Cambridge University Press, \mn@doi{10.1017/CBO9780511535246}, \url
  {https://www.cambridge.org/core/product/identifier/9780511535246/type/book}

\bibitem[\protect\citeauthoryear{{Banerjee}, {Kroupa}  \& {Oh}}{{Banerjee}
  et~al.}{2012}]{Bannerjee2012}
{Banerjee} S.,  {Kroupa} P.,   {Oh} S.,  2012, \mn@doi [\apj]
  {10.1088/0004-637X/746/1/15}, \href
  {https://ui.adsabs.harvard.edu/abs/2012ApJ...746...15B} {746, 15}

\bibitem[\protect\citeauthoryear{{Baraffe}, {Homeier}, {Allard}  \&
  {Chabrier}}{{Baraffe} et~al.}{2015}]{Baraffe2015}
{Baraffe} I.,  {Homeier} D.,  {Allard} F.,   {Chabrier} G.,  2015, \mn@doi
  [\aap] {10.1051/0004-6361/201425481}, \href
  {https://ui.adsabs.harvard.edu/abs/2015A&A...577A..42B} {577, A42}

\bibitem[\protect\citeauthoryear{Binney \& Tremaine}{Binney \&
  Tremaine}{2008}]{BT1987}
Binney J.,  Tremaine S.,  2008, {Galactic Dynamics: Second Edition}.
Princeton University Press

\bibitem[\protect\citeauthoryear{Bressan, Marigo, Girardi, Salasnich, {Dal
  Cero}, Rubele  \& Nanni}{Bressan et~al.}{2012}]{Bressan2012}
Bressan A.,  Marigo P.,  Girardi L.,  Salasnich B.,  {Dal Cero} C.,  Rubele S.,
    Nanni A.,  2012, \mn@doi [Mon. Not. R. Astron. Soc.]
  {10.1111/j.1365-2966.2012.21948.x}, 427, 127

\bibitem[\protect\citeauthoryear{Butler \& Tan}{Butler \&
  Tan}{2012}]{Butler2012}
Butler M.~J.,  Tan J.~C.,  2012, \mn@doi [Astrophys. J.]
  {10.1088/0004-637X/754/1/5}, 754, 5

\bibitem[\protect\citeauthoryear{Casertano \& Hut}{Casertano \&
  Hut}{1985}]{Casertano1985}
Casertano S.,  Hut P.,  1985, \mn@doi [Astrophys. J.] {10.1086/163589}, 298, 80

\bibitem[\protect\citeauthoryear{{Cunningham}, {Klein}, {Krumholz}  \&
  {McKee}}{{Cunningham} et~al.}{2011}]{Cunningham2011}
{Cunningham} A.~J.,  {Klein} R.~I.,  {Krumholz} M.~R.,   {McKee} C.~F.,  2011,
  \mn@doi [\apj] {10.1088/0004-637X/740/2/107}, \href
  {https://ui.adsabs.harvard.edu/abs/2011ApJ...740..107C} {740, 107}

\bibitem[\protect\citeauthoryear{{Da Rio} et~al.,}{{Da Rio}
  et~al.}{2016}]{DaRio2016}
{Da Rio} N.,  et~al., 2016, \mn@doi [Astrophys. J.]
  {10.3847/0004-637x/818/1/59}, 818, 59

\bibitem[\protect\citeauthoryear{Dale, Ercolano  \& Bonnell}{Dale
  et~al.}{2015}]{Dale2015}
Dale J.~E.,  Ercolano B.,   Bonnell I.~A.,  2015, \mn@doi [Mon. Not. R. Astron.
  Soc.] {10.1093/mnras/stv913}, 451, 987

\bibitem[\protect\citeauthoryear{Dowell, Buckalew  \& Tan}{Dowell
  et~al.}{2008}]{Dowell2008}
Dowell J.~D.,  Buckalew B.~A.,   Tan J.~C.,  2008, \mn@doi [Astron. J.]
  {10.1088/0004-6256/135/3/823}, 135, 823

\bibitem[\protect\citeauthoryear{Eldridge, Langer  \& Tout}{Eldridge
  et~al.}{2011}]{Eldridge2011}
Eldridge J.~J.,  Langer N.,   Tout C.~A.,  2011, \mn@doi [Monthly Notices of
  the Royal Astronomical Society] {10.1111/j.1365-2966.2011.18650.x}, 414, 3501

\bibitem[\protect\citeauthoryear{Elson, Fall  \& Freeman}{Elson
  et~al.}{1987}]{Elson1987}
Elson R. A.~W.,  Fall S.~M.,   Freeman K.~C.,  1987, \mn@doi [Astrophys. J.]
  {10.1086/165807}, 323, 54

\bibitem[\protect\citeauthoryear{Farias, Tan  \& Chatterjee}{Farias
  et~al.}{2017}]{Farias2017}
Farias J.~P.,  Tan J.~C.,   Chatterjee S.,  2017, \mn@doi [Astrophys. J.]
  {10.3847/1538-4357/aa63f6}, 838, 116

\bibitem[\protect\citeauthoryear{Farias, Tan  \& Chatterjee}{Farias
  et~al.}{2019}]{Farias2019}
Farias J.~P.,  Tan J.~C.,   Chatterjee S.,  2019, \mn@doi [Mon. Not. R. Astron.
  Soc.] {10.1093/mnras/sty3470}, 483, 4999

\bibitem[\protect\citeauthoryear{Farias, Tan  \& Eyer}{Farias
  et~al.}{2020}]{Farias2020}
Farias J.~P.,  Tan J.~C.,   Eyer L.,  2020, \mn@doi [Astrophys. J.]
  {10.3847/1538-4357/aba699}, 900, 14

\bibitem[\protect\citeauthoryear{{Federrath}, {Schr{\"o}n}, {Banerjee}  \&
  {Klessen}}{{Federrath} et~al.}{2014}]{Federrath2014}
{Federrath} C.,  {Schr{\"o}n} M.,  {Banerjee} R.,   {Klessen} R.~S.,  2014,
  \mn@doi [\apj] {10.1088/0004-637X/790/2/128}, \href
  {https://ui.adsabs.harvard.edu/abs/2014ApJ...790..128F} {790, 128}

\bibitem[\protect\citeauthoryear{Gavagnin, Bleuler, Rosdahl  \&
  Teyssier}{Gavagnin et~al.}{2017}]{Gavagnin2017}
Gavagnin E.,  Bleuler A.,  Rosdahl J.,   Teyssier R.,  2017, \mn@doi [Mon. Not.
  R. Astron. Soc.] {10.1093/MNRAS/STX2222}, 472, 4155

\bibitem[\protect\citeauthoryear{{Geen}, {Hennebelle}, {Tremblin}  \&
  {Rosdahl}}{{Geen} et~al.}{2015}]{Geen2015}
{Geen} S.,  {Hennebelle} P.,  {Tremblin} P.,   {Rosdahl} J.,  2015, \mn@doi
  [\mnras] {10.1093/mnras/stv2272}, \href
  {https://ui.adsabs.harvard.edu/abs/2015MNRAS.454.4484G} {454, 4484}

\bibitem[\protect\citeauthoryear{{Gies}}{{Gies}}{1987}]{Gies1987}
{Gies} D.~R.,  1987, \mn@doi [\apjs] {10.1086/191208}, \href
  {https://ui.adsabs.harvard.edu/abs/1987ApJS...64..545G} {64, 545}

\bibitem[\protect\citeauthoryear{Ginsburg, Bressert, Bally  \&
  Battersby}{Ginsburg et~al.}{2012}]{Ginsburg2012}
Ginsburg A.,  Bressert E.,  Bally J.,   Battersby C.,  2012, \mn@doi
  [Astrophys. Journal, Lett.] {10.1088/2041-8205/758/2/L29}, 758, L29

\bibitem[\protect\citeauthoryear{Gutermuth, Megeath, Myers, Allen, Pipher  \&
  Fazio}{Gutermuth et~al.}{2009}]{Gutermuth2009}
Gutermuth R.~A.,  Megeath S.~T.,  Myers P.~C.,  Allen L.~E.,  Pipher J.~L.,
  Fazio G.~G.,  2009, \mn@doi [Astrophys. Journal, Suppl. Ser.]
  {10.1088/0067-0049/184/1/18}, 184, 18

\bibitem[\protect\citeauthoryear{{Hansen}, {Klein}, {McKee}  \&
  {Fisher}}{{Hansen} et~al.}{2012}]{Hansen2012}
{Hansen} C.~E.,  {Klein} R.~I.,  {McKee} C.~F.,   {Fisher} R.~T.,  2012,
  \mn@doi [\apj] {10.1088/0004-637X/747/1/22}, \href
  {https://ui.adsabs.harvard.edu/abs/2012ApJ...747...22H} {747, 22}

\bibitem[\protect\citeauthoryear{Hobbs, Lorimer, Lyne  \& Kramer}{Hobbs
  et~al.}{2005}]{Hobbs2005}
Hobbs G.,  Lorimer D.~R.,  Lyne A.~G.,   Kramer M.,  2005, \mn@doi [Mon. Not.
  R. Astron. Soc.] {10.1111/j.1365-2966.2005.09087.x}, 360, 974

\bibitem[\protect\citeauthoryear{Hurley, Pols  \& Tout}{Hurley
  et~al.}{2000}]{Hurley2000}
Hurley J.~R.,  Pols O.~R.,   Tout C.~A.,  2000, \mn@doi [Mon. Not. R. Astron.
  Soc.] {10.1046/j.1365-8711.2000.03426.x}, 315, 543

\bibitem[\protect\citeauthoryear{Hurley, Tout  \& Pols}{Hurley
  et~al.}{2002}]{Hurley2002}
Hurley J.~R.,  Tout C.~A.,   Pols O.~R.,  2002, \mn@doi [Mon. Not. R. Astron.
  Soc.] {10.1046/j.1365-8711.2002.05038.x}, 329, 897

\bibitem[\protect\citeauthoryear{Kroupa}{Kroupa}{2001}]{KroupaIMF}
Kroupa P.,  2001, \mn@doi [Mon. Not. R. Astron. Soc.]
  {10.1046/j.1365-8711.2001.04022.x}, 322, 231

\bibitem[\protect\citeauthoryear{Kuhn, Hillenbrand, Sills, Feigelson  \&
  Getman}{Kuhn et~al.}{2019}]{Kuhn2019}
Kuhn M.~A.,  Hillenbrand L.~A.,  Sills A.,  Feigelson E.~D.,   Getman K.~V.,
  2019, \mn@doi [Astrophys. J.] {10.3847/1538-4357/aaef8c}, 870, 32

\bibitem[\protect\citeauthoryear{Lada \& Lada}{Lada \& Lada}{2003}]{Lada2003}
Lada C.~J.,  Lada E.~A.,  2003, \mn@doi [Annu. Rev. Astron. Astrophys.]
  {10.1146/annurev.astro.41.011802.094844}, 41, 57

\bibitem[\protect\citeauthoryear{Matzner \& McKee}{Matzner \&
  McKee}{2000}]{Matzner2000}
Matzner C.~D.,  McKee C.~F.,  2000, \mn@doi [The Astrophysical Journal]
  {10.1086/317785}, 545, 364

\bibitem[\protect\citeauthoryear{McKee \& Tan}{McKee \& Tan}{2003}]{mt03}
McKee C. C.~F.,  Tan J. J.~C.,  2003, \mn@doi [Astrophys. J.] {10.1086/346149},
  585, 850

\bibitem[\protect\citeauthoryear{{Mueller}, {Shirley}, {Evans}  \&
  {Jacobson}}{{Mueller} et~al.}{2002}]{Mueller2002}
{Mueller} K.~E.,  {Shirley} Y.~L.,  {Evans} Neal~J. I.,   {Jacobson} H.~R.,
  2002, \mn@doi [\apjs] {10.1086/342881}, \href
  {https://ui.adsabs.harvard.edu/abs/2002ApJS..143..469M} {143, 469}

\bibitem[\protect\citeauthoryear{Nakamura \& Li}{Nakamura \&
  Li}{2007}]{Nakamura2007}
Nakamura F.,  Li Z. Z.-Y. Z. Z.-Y.,  2007, \mn@doi [Astrophys. J.]
  {10.1086/517515}, 662, 395

\bibitem[\protect\citeauthoryear{Nakamura \& Li}{Nakamura \&
  Li}{2014}]{Nakamura2014}
Nakamura F.,  Li Z. Y. Z.-Y.,  2014, \mn@doi [Astrophys. J.]
  {10.1088/0004-637X/783/2/115}, 783, 115

\bibitem[\protect\citeauthoryear{Offner, Moe, Kratter, Sadavoy, Jensen  \&
  Tobin}{Offner et~al.}{2022}]{Offner2022}
Offner S.~S.,  Moe M.,  Kratter K.~M.,  Sadavoy S.~I.,  Jensen E.~L.,   Tobin
  J.~J.,  2022, \mn@doi [arXiv e-prints] {10.48550/arXiv.2203.10066}, p.
  arXiv:2203.10066

\bibitem[\protect\citeauthoryear{Oh \& Kroupa}{Oh \& Kroupa}{2016}]{Oh2016}
Oh S.,  Kroupa P.,  2016, \mn@doi [Astron. Astrophys.]
  {10.1051/0004-6361/201628233}, 590, A107

\bibitem[\protect\citeauthoryear{Perets \& {\v{S}}ubr}{Perets \&
  {\v{S}}ubr}{2012}]{Perets2012}
Perets H.~B.,  {\v{S}}ubr L.,  2012, \mn@doi [Astrophys. J.]
  {10.1088/0004-637X/751/2/133}, 751, 133

\bibitem[\protect\citeauthoryear{{Peters}, {Banerjee}, {Klessen}, {Mac Low},
  {Galv{\'a}n-Madrid}  \& {Keto}}{{Peters} et~al.}{2010}]{Peters2010}
{Peters} T.,  {Banerjee} R.,  {Klessen} R.~S.,  {Mac Low} M.-M.,
  {Galv{\'a}n-Madrid} R.,   {Keto} E.~R.,  2010, \mn@doi [\apj]
  {10.1088/0004-637X/711/2/1017}, \href
  {https://ui.adsabs.harvard.edu/abs/2010ApJ...711.1017P} {711, 1017}

\bibitem[\protect\citeauthoryear{{Peters}, {Banerjee}, {Klessen}  \& {Mac
  Low}}{{Peters} et~al.}{2011}]{Peters2011}
{Peters} T.,  {Banerjee} R.,  {Klessen} R.~S.,   {Mac Low} M.-M.,  2011,
  \mn@doi [\apj] {10.1088/0004-637X/729/1/72}, \href
  {https://ui.adsabs.harvard.edu/abs/2011ApJ...729...72P} {729, 72}

\bibitem[\protect\citeauthoryear{Proszkow \& Adams}{Proszkow \&
  Adams}{2009}]{Proszkow2009}
Proszkow E.-M. E.-M.,  Adams F. C.~F.,  2009, \mn@doi [Astrophys. J. Suppl.
  Ser.] {10.1088/0067-0049/185/2/486}, 185, 486

\bibitem[\protect\citeauthoryear{Qian, Cai, Zwart  \& Zhu}{Qian
  et~al.}{2017}]{Quian2017}
Qian P.~X.,  Cai M.~X.,  Zwart S.~P.,   Zhu M.,  2017, \mn@doi [Publ. Astron.
  Soc. Pacific] {10.1088/1538-3873/aa7c49}, 129, 094503

\bibitem[\protect\citeauthoryear{Raghavan et~al.,}{Raghavan
  et~al.}{2010}]{Raghavan2010}
Raghavan D.,  et~al., 2010, \mn@doi [Astrophys. J. Suppl. Ser.]
  {10.1088/0067-0049/190/1/1}, 190, 1

\bibitem[\protect\citeauthoryear{Reggiani \& Meyer}{Reggiani \&
  Meyer}{2011}]{Reggiani2011}
Reggiani M. M.~M.,  Meyer M. M.~R.,  2011, \mn@doi [Astrophys. J.]
  {10.1088/0004-637X/738/1/60}, 738, 60

\bibitem[\protect\citeauthoryear{{Rogers} \& {Pittard}}{{Rogers} \&
  {Pittard}}{2013}]{Rogers2013}
{Rogers} H.,  {Pittard} J.~M.,  2013, \mn@doi [\mnras] {10.1093/mnras/stt255},
  \href {https://ui.adsabs.harvard.edu/abs/2013MNRAS.431.1337R} {431, 1337}

\bibitem[\protect\citeauthoryear{{Stone}}{{Stone}}{1991}]{Stone1991}
{Stone} R.~C.,  1991, \mn@doi [\aj] {10.1086/115880}, \href
  {https://ui.adsabs.harvard.edu/abs/1991AJ....102..333S} {102, 333}

\bibitem[\protect\citeauthoryear{Tan}{Tan}{2006}]{Tan2006}
Tan J.~C.,  2006, \mn@doi [Proc. Int. Astron. Union]
  {10.1017/S1743921307001573}, 2, 258

\bibitem[\protect\citeauthoryear{Tan, Beltr{\'{a}}n, Caselli, Fontani, Fuente,
  Krumholz, McKee  \& Stolte}{Tan et~al.}{2014}]{Tan2014}
Tan J.~C.,  Beltr{\'{a}}n M.~T.,  Caselli P.,  Fontani F.,  Fuente A.,
  Krumholz M.~R.,  McKee C.~F.,   Stolte A.,  2014, in , Vol.~26, Protostars
  Planets VI.
University of Arizona Press, pp 18--21 (\mn@eprint {arXiv} {1402.0919}),
  \mn@doi{10.2458/azu_uapress_9780816531240-ch007}, \url
  {http://arxiv.org/abs/1402.0919{\%}0Ahttp://dx.doi.org/10.2458/azu{\_}uapress{\_}9780816531240-ch007
  http://muse.jhu.edu/books/9780816598762/9780816598762-13.pdf}

\bibitem[\protect\citeauthoryear{Tanaka, Tan  \& Zhang}{Tanaka
  et~al.}{2017}]{Tanaka2017}
Tanaka K. E.~I.,  Tan J.~C.,   Zhang Y.,  2017, \mn@doi [Astrophys. J.]
  {10.3847/1538-4357/835/1/32}, 835, 32

\bibitem[\protect\citeauthoryear{Tognelli, {Prada Moroni}  \&
  Degl'Innocenti}{Tognelli et~al.}{2011}]{Tognelli2011}
Tognelli E.,  {Prada Moroni} P.~G.,   Degl'Innocenti S.,  2011, \mn@doi
  [Astron. Astrophys.] {10.1051/0004-6361/200913913}, 533, 1

\bibitem[\protect\citeauthoryear{Wang, Spurzem, Aarseth, Nitadori, Berczik,
  Kouwenhoven  \& Naab}{Wang et~al.}{2015}]{Wang2015}
Wang L.,  Spurzem R.,  Aarseth S.,  Nitadori K.,  Berczik P.,  Kouwenhoven M.
  B.~N.,   Naab T.,  2015, \mn@doi [Mon. Not. R. Astron. Soc.]
  {10.1093/mnras/stv817}, 450, 4070

\bibitem[\protect\citeauthoryear{de Wit, Testi, Palla  \& Zinnecker}{de~Wit
  et~al.}{2005}]{Dewit2005}
de Wit W.~J.,  Testi L.,  Palla F.,   Zinnecker H.,  2005, \mn@doi [A{\&}A]
  {10.1051/0004-6361:20042489}, 437, 247

\makeatother
\end{thebibliography}

\appendix
\section{Ancillary results for the full set of binary properties}\label{ap:binaries}

Here we present results related to binary properties over the full range of $M_{\rm
cl}$, $\Sigma_{\rm cloud}$ and \sfeff\ explored in our grid of models.

\begin{figure*}
\includegraphics[width=\linewidth]{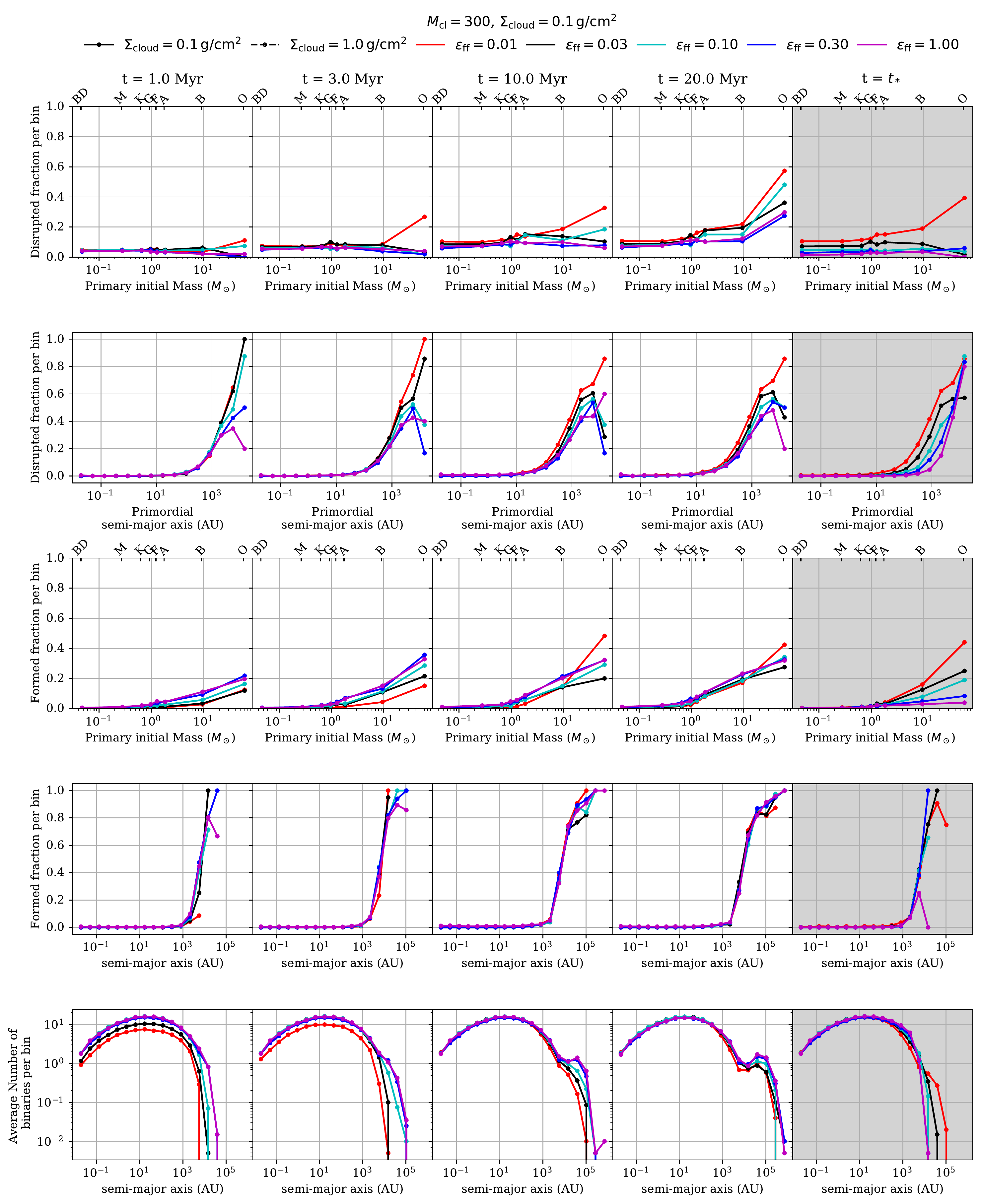} 
 \caption{
         Same as Figure~\ref{fig:Bdisrupted} but for models with $\mcl=300\Msun$
         and with $\Sigmacl=0.1\:{\rm g\:cm}^{-2}$  and $\sfeff = 0.01\:\gpcm$,
         0.03, 0.1, 0.3 and 1.0.
 }
 \label{fig:BdistM300}
\end{figure*}

\begin{figure*}
\includegraphics[width=\linewidth]{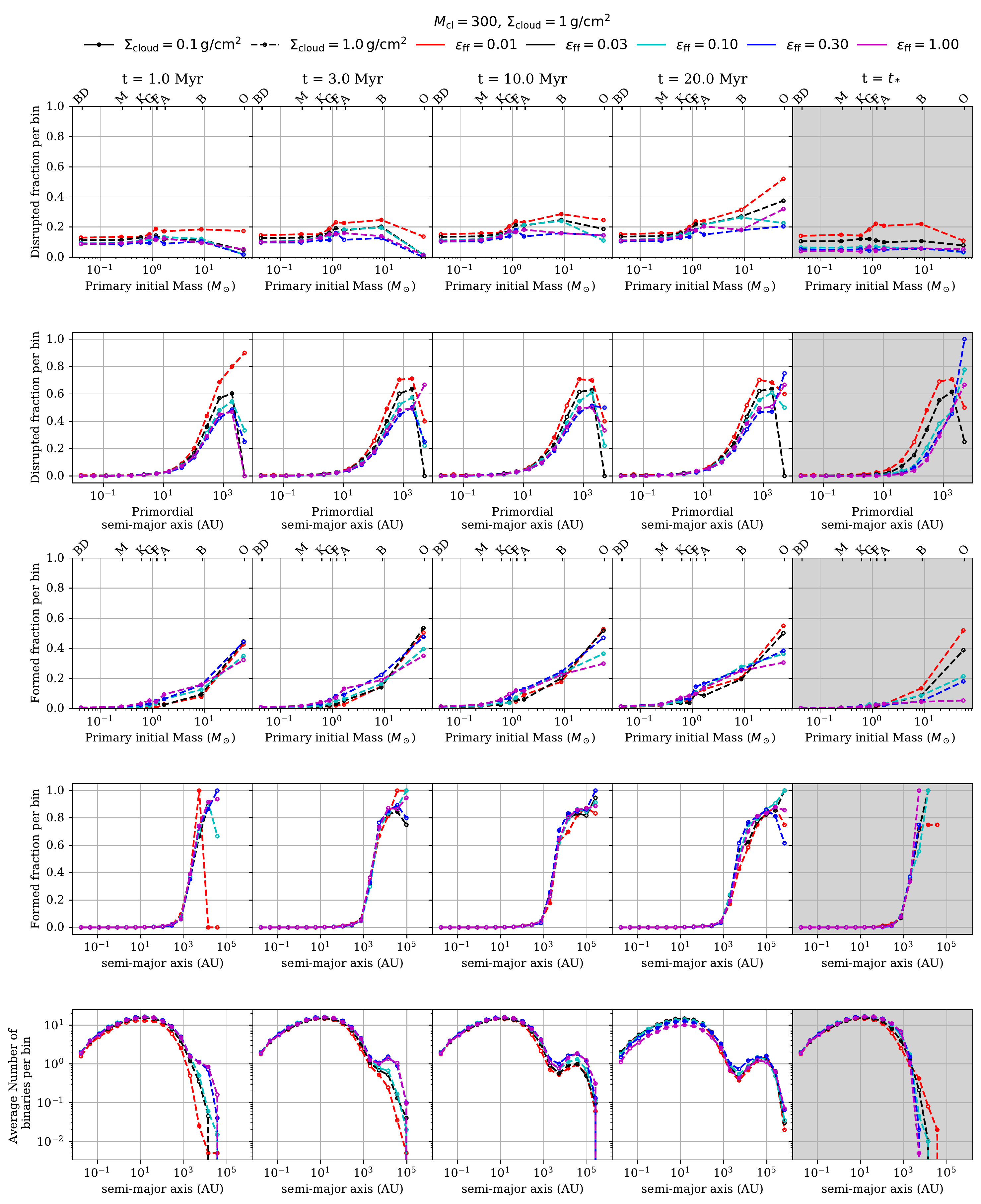} 
\caption{Same as Figure~\ref{fig:BdistM300} but for models with 
        $\Sigmacl=1.0\:\gpcm$  }
        \label{fig:BdistM300SS1}
\end{figure*}

\begin{figure*}
\includegraphics[width=\linewidth]{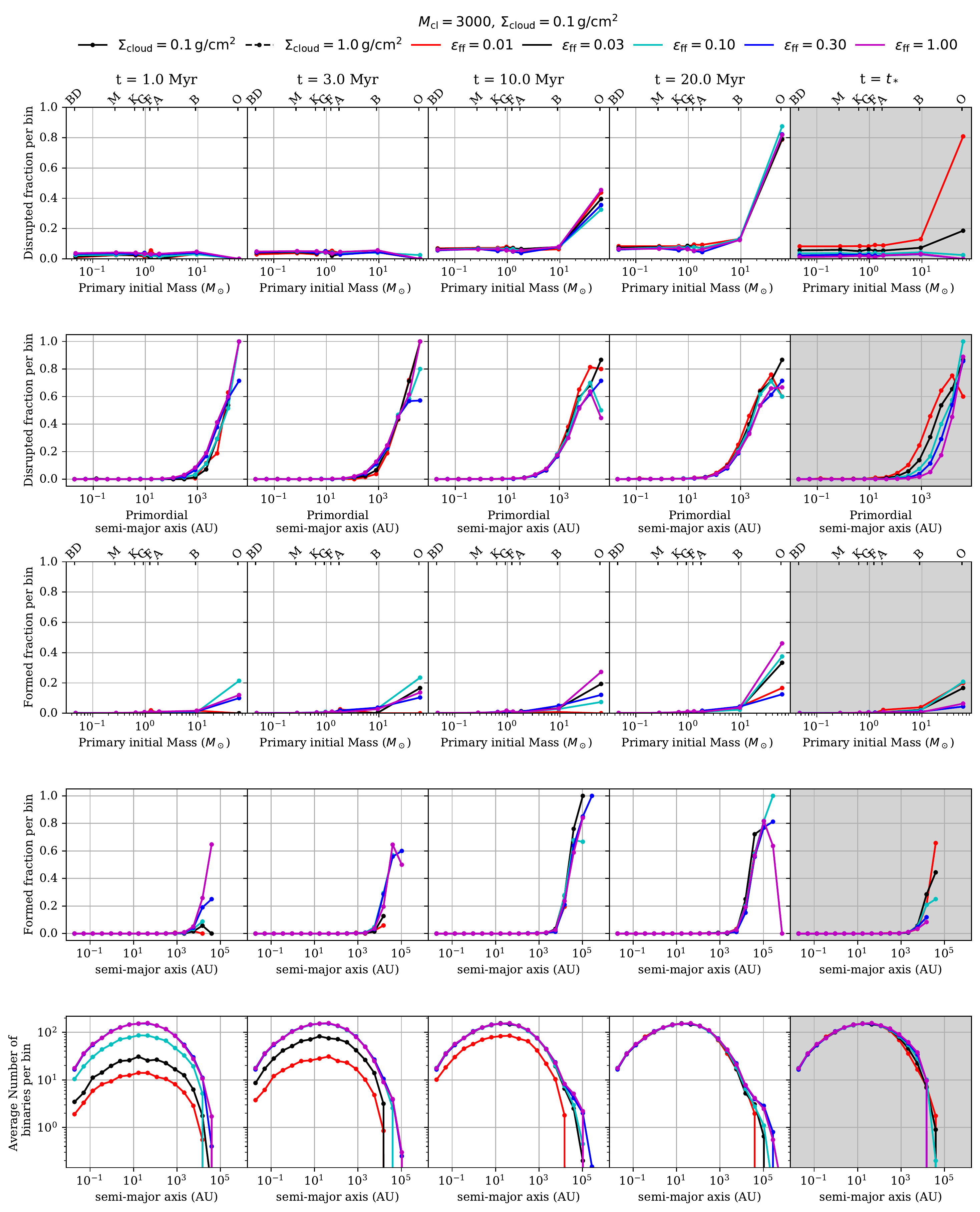} 
\caption{Same as Figure~\ref{fig:BdistM300} but for models with $\mcl=3,000\:\Msun$
        and $\Sigmacl=0.1\:\gpcm$ }
        \label{fig:BdistM3K}
\end{figure*}

\begin{figure*}
\includegraphics[width=\linewidth]{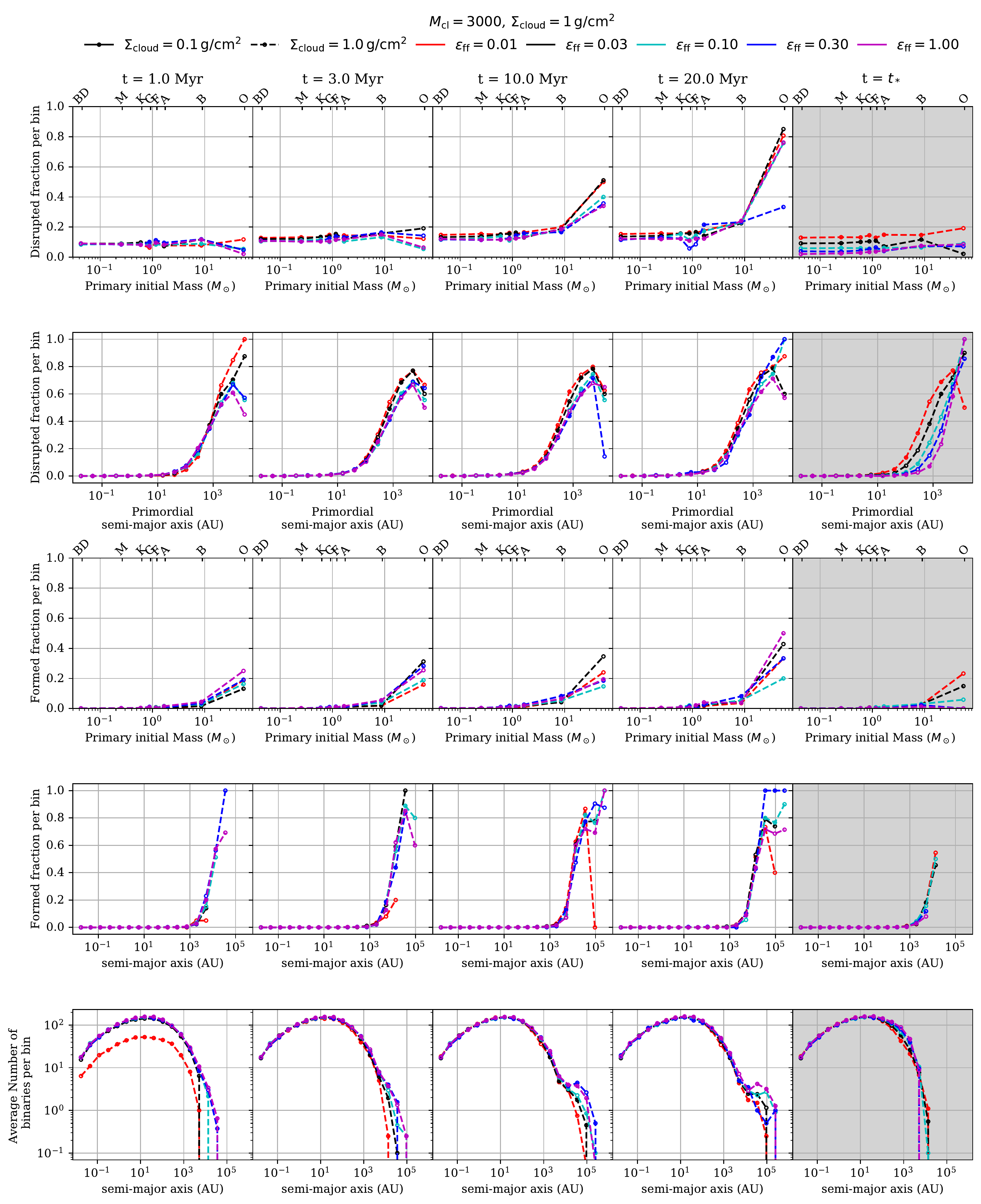} 
\caption{Same as Figure~\ref{fig:BdistM3K} but for models with  $\Sigmacl=1.0\:\gpcm$ }
        \label{fig:BdistM3KSS1}
\end{figure*}

\begin{figure*}
\includegraphics[width=\linewidth]{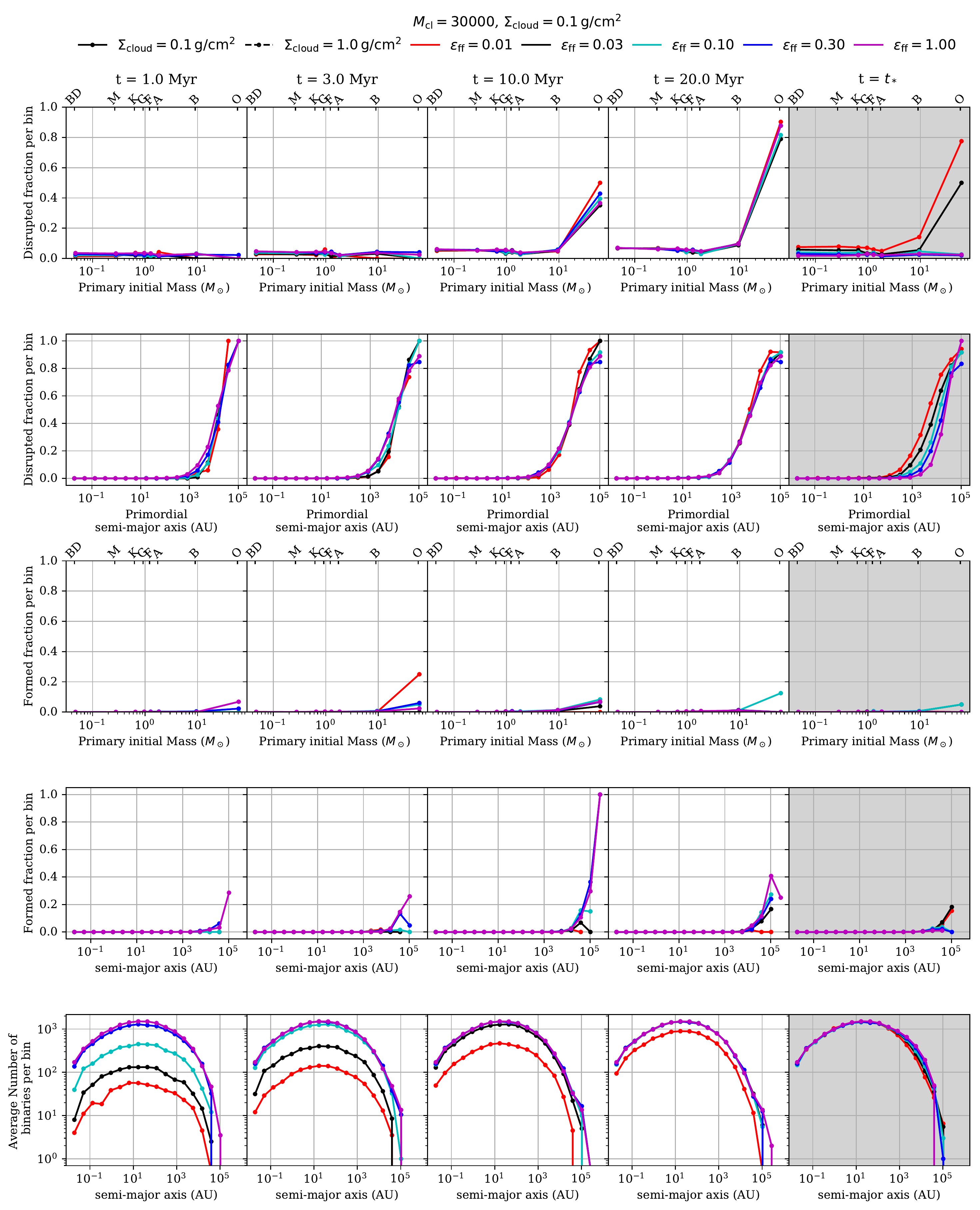} 
\caption{Same as Figure~\ref{fig:BdistM3K} but for models with $\mcl=30,000\:\Msun$
 and $\Sigmacl=0.1\:\gpcm$ }
        \label{fig:BdistM30K}
\end{figure*}

\begin{figure*}
\includegraphics[width=\linewidth]{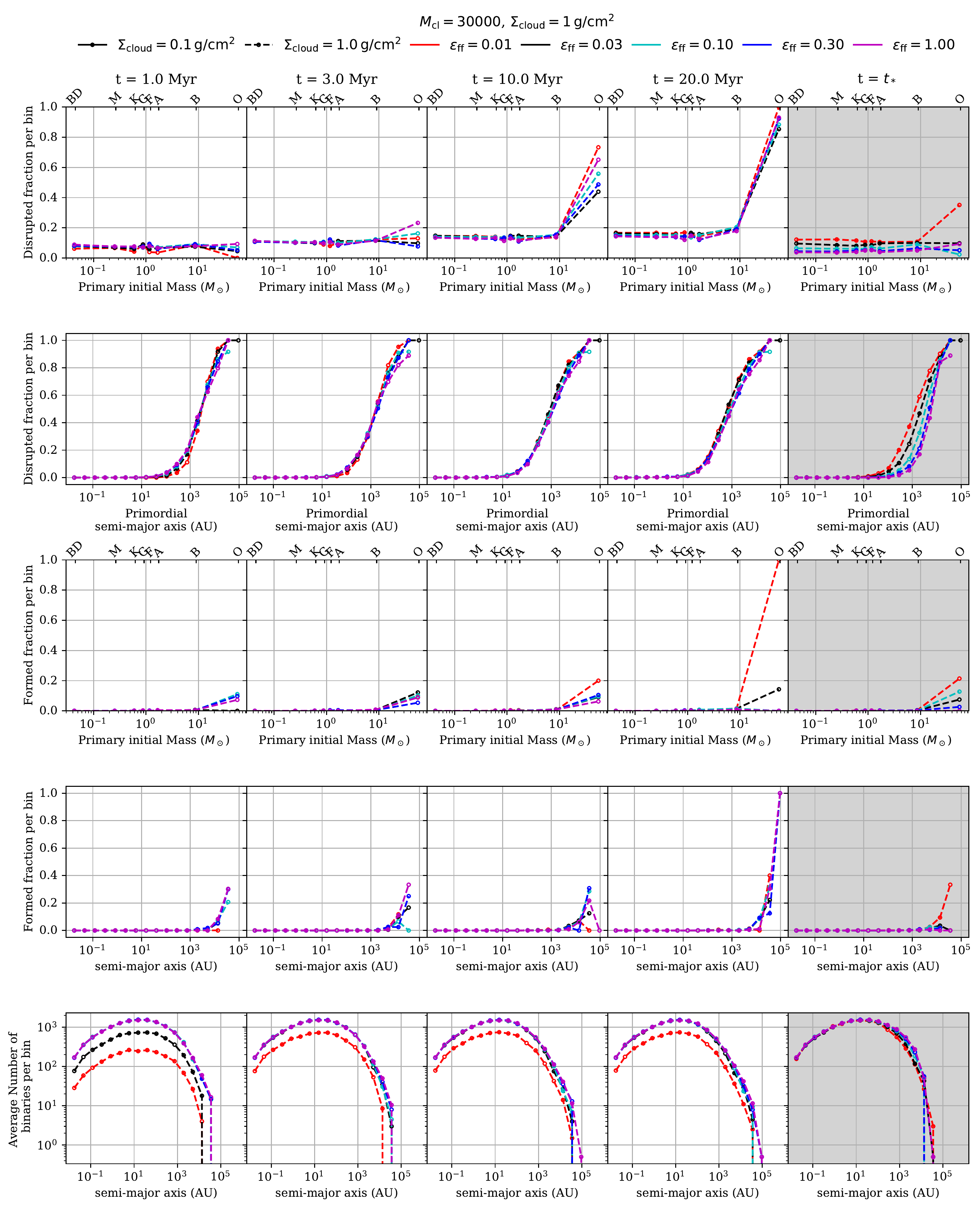} 
\caption{Same as Figure~\ref{fig:BdistM30K} but for models with $\Sigmacl=1.0\:\gpcm$ }
        \label{fig:BdistM30KSS1}
\end{figure*}

\bsp	%
\label{lastpage}
\end{document}